\documentclass[twocolumn,astrosymb,tighten,trackchanges,article]{aastex7}
\usepackage{comment}
\usepackage{CJK}
\usepackage{amsmath,amssymb,amsfonts}
\usepackage{appendix}
\usepackage{hyperref}
\usepackage{array,multirow}
\usepackage{wasysym}
\usepackage{float}
\usepackage{textcomp}
\usepackage{pifont}
\usepackage{stfloats}
\usepackage{wrapfig}
\usepackage[x11names]{xcolor}


\accepted{6 January 2026}
\submitjournal{The Astrophysical Journal}

\shorttitle{Parameter Constraints with Molecular Emission}
\shortauthors{Behrens et al.}

\graphicspath{{./}{Figures/}}

\begin{document}
\begin{CJK*}{UTF8}{gbsn}

\title{Testing the Physical Parameter Constraining Power of HCN and HNC with Neural Networks}

\author[0000-0002-2333-5474]{Erica Behrens}
\affiliation{Department of Astronomy, University of Virginia, P. O. Box 400325, 530 McCormick Road, Charlottesville, VA 22904-4325, USA}
\email{eb7he@virginia.edu}

\author[0000-0003-1183-9293]{Jeffrey G.~Mangum}
\affiliation{National Radio Astronomy Observatory, 520 Edgemont Road, Charlottesville, VA  22903-2475, USA}
\email{jmangum@nrao.edu}

\author[0000-0002-0370-8034]{Mathilde Bouvier}
\affiliation{Leiden Observatory, Leiden University, P.O. Box 9513, NL-2300 RA Leiden, The Netherlands}
\email{bouvier@strw.leidenuniv.nl}

\author[0000-0002-1185-2810]{Cosima Eibensteiner}
\altaffiliation{Jansky Fellow of the National Radio Astronomy Observatory}
\affiliation{National Radio Astronomy Observatory, 520 Edgemont Road, Charlottesville, VA  22903-2475, USA}
\email{ceibenst@nrao.edu}

\author[0000-0001-8504-8844]{Serena Viti}
\affiliation{Leiden Observatory, Leiden University, P.O. Box 9513, NL-2300 RA Leiden, The Netherlands}
\affiliation{Argelander Institut f\"{u}r Astronomie der Universit\"{a}t Bonn, Auf dem H\"{u}gel 71, 53121 Bonn Germany}
\affiliation{Department of Physics and Astronomy, University College London, Gower Street, London WC1E 6BT}
\email{viti@strw.leidenuniv.nl}

\correspondingauthor{Erica Behrens} \email{eb7he@virginia.edu}



\begin{abstract}
We quantify the utility of HCN and HNC to characterize gas conditions in the nearby starburst galaxy NGC\,253. We use measurements from the Atacama Large Millimeter/Submillimeter Array (ALMA) Large Program ALCHEMI: the ALMA Comprehensive High-resolution Molecular Inventory. Using different subsets of the eight total HCN and HNC transitions measured by ALCHEMI, we test the number and combinations of transitions necessary for constraining the temperature, H$_{2}$ volume and column densities, cosmic-ray ionization rate, and beam-filling factor in three representative regions within NGC\,253. We use these combinations of HCN and HNC transitions to constrain chemical and radiative transfer models and infer the gas conditions using a Bayesian nested sampling algorithm combined with neural network models for increased efficiency. By comparing the shapes of the resulting posterior distributions, as well as the medians and uncertainties for each gas parameter, from each test case to what we obtain with the full set of eight transitions (the control), we quantify how well each test reproduces the control. We find that multiple transitions each of both molecules are required to obtain a median parameter value within a factor of 2 of the control with an uncertainty less than 2--3 times that of the control. We also find that transition combinations that feature a range of upper-state energies are most effective. We show that single transitions, such as HCN $J=1-0$ or $3-2$, are among the worst-performing combinations and result in parameter values up to an order of magnitude different than the control.
\end{abstract}


\section{Introduction} \label{sec:intro}

Emission from molecular gas serves as a powerful tool for characterizing the physical conditions of gas in star-forming regions. The varying conditions (e.g. gas density, temperature, etc.) under which different molecules can be excited yields the opportunity to use individual or combinations of molecules as tracers of specific gas conditions or environments. In particular, molecular gas is chemically and physically influenced by the feedback processes from recently-formed nearby stars \citep[][and references therein]{Schinnerer2024}, such as shocks and cosmic-ray ionization from supernovae and their remnants \citep{Reach2019,Vaupre2014,Tu2025} and ultraviolet (UV) ionization from young stars \citep{Xu2019,Hernandez-Vera2023}. Understanding the impact of feedback processes on molecular gas helps us evaluate how the current generation of stars influences the material that will form the next generation. Some feedback mechanisms can have global effects on their host galaxies that influence the galaxies' abilities to form future stars \citep{Heckman1990,Lehnert1996,Ceverino2009,Hopkins2012,Agertz2013}. For instance, cosmic rays from supernova feedback have been shown to create large-scale pressure gradients that result in outflows, which eject star-forming material into the circumgalactic medium and thereby suppress the star-formation rate \citep{Peschken2023}. It is thus important to fully understand how to use molecular emission to infer gas conditions; like temperature, density, and cosmic-ray ionization rate (CRIR); in order to piece together the impact of star formation on galaxy evolution.

HCN in particular has been a popular molecule of study in extragalactic astronomy and is commonly used in ratios with other molecules to characterize dense, star-forming gas. In a seminal work regarding HCN's utility to study star formation, \cite{Gao2004} used the HCN($1-0$)/CO ratio as a dense gas tracer in a large sample of infrared galaxies. They additionally define an HCN-to-dense-gas conversion factor of $\alpha_{\text{HCN}} = 10$\,M$_{\odot}$\,pc$^{-2}$\,(K\,km\,s$^{-1}$)$^{-1}$. Since many extragalactic sources only feature measurements of single transitions due to sensitivity issues, many similar studies have since been performed that also use only single transitions, or ratios of single transitions, of HCN with CO and other molecules to identify correlations with infrared luminosity \citep[e.g.][]{Garcia-Burillo2012,Usero2015}, star-formation rate \citep[e.g.][]{Onus2018,Neumann2024}, and other global galaxy characteristics. However, more recent studies have begun to reveal that the relationship between HCN and these other quantities is not so straightforward. 

\cite{Patra2025} studied the dense gas conversion factor using HCN and HCO$^{+}$ $1-0$ and found that even within our own galaxy, this factor depends greatly on the environment's metallicity and location within the Galaxy. Other studies have examined HCN $3-2$/$1-0$ ratio in nearby galaxies and found that single transitions are not always reliable dense gas mass indicators \citep{Li2020} and that no ratio uniquely traces any single gas excitation parameter, like temperature or density \citep{Garcia-Rodriguez2023}.
\cite{Jones2023} studied the gas density traced by the HCN $1-0$ transition, which is typically considered to be a dense gas tracer, and found that in magnetohydrodynamical simulations, HCN emission can be found even when the cloud density is lower than its supposed characteristic density of $3\times10^4$\,cm$^{-3}$ \citep{Gao2004}. This result is expanded in \cite{Bemis2024}, which examined the HCN($1-0$)/CO($1-0$) ratio and determined that it will not always trace gravitationally bound star-forming gas. 
\cite{hacar20} explored the relationship between HCN and HNC in Milky Way star-forming environments, suggesting that the HCN/HNC $J=1-0$ intensity ratio can be used as a temperature probe. However, the use of these molecules as a temperature tracer is uncertain and has been found to be temperature independent for $T \gtrsim 50$\,K \citep{Behrens2022}. While \cite{hacar20} did compare their measurements to chemical models, the majority of these studies do not consider the chemical pathways related to HCN and HNC and their impact on observed intensities. Given these circumstances, more work is needed to understand how HCN can characterize molecular gas in star-forming regions and how this species is affected chemically by star-formation feedback mechanisms.

Studying HCN and HNC in extragalactic environments is essential for understanding the characteristics of dense molecular gas in a broader range of environments than is accessible within the Milky Way alone. An ideal target for an extragalactic study of HCN and HNC is the nearby \citep[$d \sim 3.5 \pm 0.2$\,Mpc,][]{Rekola2005} starburst galaxy NGC\,253. This nuclear starburst is forming stars at a rate of only $\sim 5$\,M$_{\odot}$\,yr$^{-1}$, but with about half of that star formation occurring in just the central kiloparsec \citep{leroy15}. The molecular diversity of NGC\,253's Central Molecular Zone (CMZ) has recently been studied in detail as a result of the ALMA Large Program ALCHEMI, the ALMA Comprehensive High-resolution Extragalactic Molecular Inventory \citep{ALCHEMI-ACA}. This program surveyed the molecular emission from $84.2-373.2$\,GHz, and the resulting measurements have been the focus of a wide range of studies analyzing the physical and chemical characteristics of this starburst CMZ \citep{holdship_c2h,harada21,Humire2022,Holdship2022,Harada2022,Behrens2022,Haasler2022,Huang2023,Harada2024ApJS,Bouvier2024,Tanaka2024,Bao2024,Behrens2024,Butterworth2024,Kishikawa2024,Gong2025arXiv,Bouvier2025arXiv}.

Of these studies, \cite{Behrens2022,Behrens2024} used the first four rotational transitions each of HCN and HNC ($J=1-0$, $2-1$, $3-2$, and $4-3$) to constrain the gas parameters in the NGC\,253 CMZ and found that HCN and HNC could effectively characterize the gas volume density and cosmic-ray ionization rate (CRIR). These studies found high volume densities and CRIRs in the center of the CMZ ($n_{\text{H}_{2}}\sim10^{5}$\,cm$^{-4}$, $\zeta\sim3\times10^{-13}$\,s$^{-1}$) that decreased towards the outskirts of the CMZ. \cite{Behrens2024} found that using multiple species and transitions of HCN and HNC to constrain chemical and radiative transfer models was necessary to accurately infer certain parameter values, particularly CRIR. However, very few galaxies feature a set of measurements as comprehensive as those from the ALCHEMI program, often making it difficult to robustly characterize the gas conditions with a high level of precision. Thus, it is not yet clear how many transitions of these frequently-observed molecules are necessary to infer the gas conditions in star-forming regions.

In this paper, we will test the ability and limits of different combinations of HCN and HNC transitions to constrain chemical and radiative transfer models and infer the gas conditions in starburst environments represented by NGC\,253. We will examine what number and combination of transitions yield the best results for constraining various physical parameters to provide guidance to those seeking to use these molecules to characterize gas in star-forming regions. In Section~\ref{sec:methods}, we will describe the Bayesian nested sampling algorithm we use to sample our gas parameter space and model observed intensities, as well as what regions within NGC\,253 and which combinations of transitions we will consider. We present the results of this analysis in Section~\ref{sec:results}, where we compare the results from each tested combination of HCN and HNC transitions to what can be obtained when using the full set available with ALCHEMI. In Section~\ref{sec:discussion}, we will discuss the implications of these results and provide guidance for choosing HCN and HNC transitions for constraining various gas parameters.

\section{Data} \label{sec:data}

We will use the first four rotational transitions each of HCN and HNC ($J=1-0$, $2-1$, $3-2$, and $4-3$), which were observed as part of the ALCHEMI Large Program during ALMA Cycles 5 (2017.1.00161.L) and 6 (2018.1.00162.S). These data feature a common resolution of 1.\!\!$^{\prime\prime}6$ ($\sim27$\,pc) and are sensitive to scales as large as 250 pc. Further details of the observations and measurements are described in \cite{ALCHEMI-ACA}, and the details of the spectral line signal extraction and moment map generation can be found in \cite{Behrens2022,Behrens2024}. We have provided the integrated intensities and root-mean-square (RMS) values for all eight HCN and HNC transitions measured toward each of the three regions (see Section~\ref{sec:regions}) we consider in this paper in Table~\ref{tab:obs}. Of particular importance is our use of 3$\sigma$ upper limits in place of measured intensities for some transitions in select regions that were not measured to have a signal-to-noise ratio (SNR) $\geq$ 3. As discussed in \cite{Behrens2024}, these limits are used so that our models are constrained by an equal number of HCN and HNC transitions in each region. The impacts of these limits on our results are discussed in Section~\ref{sec:results} and \ref{sec:SNR}.

\begin{deluxetable*}{>{\scriptsize}c>{\scriptsize}c>{\scriptsize}c>{\scriptsize}c>{\scriptsize}c>{\scriptsize}c>{\scriptsize}c>{\scriptsize}c>{\scriptsize}c>{\scriptsize}c>{\scriptsize}c>{\scriptsize}c>{\scriptsize}c}
\centering
\tablecolumns{9}
\tablecaption{Integrated intensity and RMS values for the regions studied in this work. \label{tab:obs}}
\tablehead{
 & &  &  & \multicolumn{2}{c}{Low-$n$-$\zeta$}  & & \multicolumn{2}{c}{High-$n$-$\zeta$} & & \multicolumn{2}{c}{Low-SNR}\\ 
 \cline{5-6} \cline{8-9} \cline{11-12}
 & \colhead{Transition} & \colhead{Rest Frequency} & \colhead{$E_{\text{u}}$} &\colhead{$\langle\int S_\nu d\nu \rangle$} &\colhead{RMS} & &\colhead{$\langle\int S_\nu d\nu \rangle$} & \colhead{RMS} & &\colhead{$\langle\int S_\nu d\nu \rangle$} & \colhead{RMS} \\ 
 &  & (GHz) & (K) & (K\,km\,s$^{-1}$) & (K\,km\,s$^{-1}$) & & (K\,km\,s$^{-1}$) & (K\,km\,s$^{-1}$) & & (K\,km\,s$^{-1}$) & (K\,km\,s$^{-1}$) 
}
\startdata
HCN     &  $1-0$  &  88.6316 & 4.25 & 286  &  43  & & 533 &  80 & & 51 & 8\\
        & $2-1$ &   177.2611 & 12.76 & 202  &  31   & &  509 &  76  & & 63  & 10\\
        & $3-2$ &   265.8864 & 25.52 & 94 &  14  & &  379 &   57  & &  5.6 & 1.0\\
        & $4-3$ &   354.5055 & 42.53 & 42 &  6  & &  261 &   39 & & 1.3\tablenotemark{\scriptsize a}  & 0.5 \\ \hline 
HNC     &  $1-0$ &  90.6636 & 4.35 & 126 &  19  & &  459 &  69  & & 28  & 8\\
        & $2-1$ &  181.3248 & 13.05 & 32 &  9   & &  288 &   44 & &  26 & 9 \\
        & $3-2$ &   271.9811 & 26.11 & 28&  4  & &  245 &  37  & &  1.9\tablenotemark{\scriptsize a} & 0.6\\
        & $4-3$ &   362.6303 & 43.51 & 9 &  1 & &  148 &   22 & & 1.8\tablenotemark{\scriptsize a}  & 0.6 \\
\enddata
\tablenotetext{\scriptstyle a}{\scriptsize 3$\sigma$ upper limit}
\end{deluxetable*}

\section{Bayesian Inference of Molecular Gas Conditions} \label{sec:methods}

We explore the ability and limitations of HCN and HNC to estimate molecular gas properties by constraining chemical and radiative transfer models with different combinations of measured HCN and HNC transitions toward the NGC\,253 CMZ. In the following, we describe the gas parameter space we explore, our Bayesian parameter sampling method, the regions within the CMZ we target, and the HCN and HNC transition combinations we test and molecular constraints for our models.

\begin{deluxetable*}{cccc|ccc}[]
\centering
\tablecolumns{3}
\tablewidth{0pt}
\tablecaption{Prior Distributions and Inferred Parameter Values\label{tab:priors}}
\tablehead{
\colhead{} &  \colhead{Parameter} & \colhead{Range} & \multicolumn{1}{c|}{Distribution Type} & \colhead{Low-$n$-$\zeta$}  & \colhead{High-$n$-$\zeta$} & \colhead{Low SNR}
}
\startdata
$T_\text{K}$  & Temperature (K) & 50--300 & Linear & $279^{+50}_{-84}$  & $179.47^{+100.56}_{-81.57}$ & $115^{+81}_{-47}$\\ [5pt]
$n_{\text{H}_2}$  & Volume Density (cm$^{-3}$) & 10$^3$--10$^7$ & Log & $4.1^{+0.4}_{-0.4}$\tablenotemark{a} & $4.93^{+0.37}_{-0.44}$\tablenotemark{a} & $3.9^{0.4}_{0.5}$\tablenotemark{a} \\ [5pt]
$\zeta$ & Cosmic-ray Ionization Rate (s$^{-1}$)& $10^{-17}$--$10^{-10}$ & Log & -$13.8^{+0.5}_{-0.4}$\tablenotemark{a}  & $-12.64^{+0.40}_{-0.48}$\tablenotemark{a} & $-13.6^{+0.5}_{-0.5}$\tablenotemark{a} \\[5pt]
$N_{\text{H}_2}$ & H$_{2}$ Column Density (cm$^{-2}$) & 10$^{22}$--$10^{25}$ & Log & $23.1^{+0.5}_{-0.3}\tablenotemark{a}$  & $24.43^{+0.38}_{-0.46}$\tablenotemark{a} & $22.7^{+0.9}_{-0.4}$\tablenotemark{a}\\[5pt]
\enddata
\tablenotetext{a}{Value is given in logspace}
\end{deluxetable*}

\subsection{Bayesian Inference Model Review}\label{sec:modelReview}

Here we provide an overview of our modeling algorithm, which has been used previously and described in more detail in \cite{Behrens2022,Behrens2024}. 
We are interested in constraining the following gas parameters in the NGC\,253 CMZ using HCN and HNC: kinetic temperature, $T_{\text{K}}$; H$_{2}$ volume density, $n_{\text{H}_{2}}$; cosmic-ray ionization rate (CRIR), $\zeta$; H$_{2}$ column density, $N_{\text{H}_{2}}$; and a beam-filling factor, $\eta_{\text{ff}}$. \cite{Behrens2022} and \cite{Behrens2024} constrained the gas conditions in the NGC\,253 CMZ using HCN and HNC and found that gas densities and CRIRs featured a distribution where these parameters peaked in the center of the CMZ, where most of the sources related to recent star formation are located, and decreased toward the outskirts. These studies found that H$_{2}$ volume and column densities, as traced by HCN and HNC ranges from $\sim 3\times10^{3}-10^{5}$\,cm$^{-3}$ and $\sim 2\times10^{22}-5\times10^{24}$\,cm$^{-2}$, respectively. The CRIR ranges from $\sim10^{-14}$\,s$^{-1}$ to $\sim 4\times 10^{-13}$\,s$^{-1}$. \cite{Behrens2024} also noted that the beam-filling factor $\eta_{\text{ff}}$ was lower in the center of the CMZ ($\sim0.1$) and higher near the outskirts ($\sim0.5$), indicating that HCN and HNC are emitting from smaller, denser components in the center of the CMZ while being more diffuse near the outskirts. Both studies were unable to constrain the kinetic temperature with HCN and HNC measurements, stating that the abundances of these two species were primarily influenced by cosmic-ray ionization, not mechanical heating, thus making it impossible to determine $T_{\text{K}}$ in this environment using HCN and HNC.

As in these previous studies, we use a forward model that includes both chemical and radiative transfer modeling, and we compare the results of these models to the measured HCN and HNC integrated intensities described in Section~\ref{sec:data}. Our chemical modeling component is derived from the two-phase gas-grain chemical code \texttt{UCLCHEM} \citep{Holdship2017}, which produces molecular abundances of HCN and HNC given a set of values for $T_{\text{K}}$, $n_{\text{H}_{2}}$, $\zeta$, and $N_{\text{H}_{2}}$. Note that cosmic-ray ionization is our primary excitation mechanism and that we do not consider UV ionization in this model. \cite{Behrens2024} demonstrated that cosmic-ray- versus UV-dominated regions can be identified by examining the HCN/HNC abundance ratios---when this abundance ratio is near unity, the chemistry of these two molecules is primarily influenced by cosmic rays, with little effect from UV radiation. Alternatively, if the HCN/HNC abundance ratios is greater than $\sim4$, HCN is being enhanced as a result of UV ionization. We have chosen to focus here on regions within NGC\,253 (further described in Section~\ref{sec:regions}) that are known to be cosmic-ray-dominated, and thus use a constant value of 1 Habing for the UV radiation field.

To improve the efficiency of our modeling algorithm, we have trained a neural network from the output of a grid of \texttt{UCLCHEM} models to replace the role of a chemical code in our algorithm. This neural network model \citep[details of which can be found in][]{Behrens2024} improves the speed of our algorithm by a factor of 10. After acquiring HCN and HNC abundance predictions from the neural network model, we convert these abundances to integrated intensities using the non-LTE radiative transfer model \texttt{RADEX} \citep{radex} via the Python wrapper \texttt{SpectralRadex} \citep{holdship_c2h}. With \texttt{RADEX}, we employ the default spherical geometry and assume a standard ortho:para H$_2$ ratio of 3:1. We then multiply the output integrated intensities by the beam-filling factor $\eta_{ff}$ and compare the modeled integrated intensities $F_{t}$ to our measured values $F_{d}$ and their uncertainties $\sigma_{F}$ for each transition $i$ using the following likelihood equation:
\begin{equation}
    P(\mathbf{F_d}|\boldsymbol{\theta}) = \exp\left(-\frac{\displaystyle 1}{\displaystyle 2} \sum\limits_{i} \frac{(F_{d,i} - F_{t,i})^2}{\sigma^2_{F,i}}\right)
    \label{eq:lik_fn}.
\end{equation}
where $\theta$ represents the gas parameters.

A lower likelihood value implies that the set of input parameters produced integrated intensities more similar to our measurements. Note that we assume all measured transitions are emitting from the same gas component, and thus we derive an average set of gas parameters for the bulk gas traced by HCN and HNC for each studied region \citep{Holdship2022,Behrens2022,Behrens2024}.

\subsection{Parameter Space and Sampling Method}\label{sec:nautilus}

To sample our parameter space and infer the gas conditions traced by HCN and HNC, we use a Bayesian nested sampling algorithm to constrain the chemical and radiative transfer models described in Section~\ref{sec:modelReview}. The main difference in this work is the use of a different nested sampling code, \texttt{Nautilus}\footnote{\url{https://nautilus-sampler.readthedocs.io/en/latest/index.html}} \citep{nautilus}. \texttt{Nautilus} is an importance nested sampling-based algorithm that uses neural networks to explore and sample the parameter space with increased efficiency. Both standard and importance nested sampling algorithms use a ``live set", which consists of the current set of parameter combinations, or live points, with the highest likelihoods. As new combinations within the parameter space are sampled, points are added to the live set if their likelihoods are higher than the lowest likelihood in the live set, which will then subsequently be removed. 

However, standard nested sampling algorithms, such as those used in \cite{Holdship2022, Behrens2022, Behrens2024}, are often inefficient when sampling the parameter space---sampling too few points can compromise the accuracy of the algorithm, so these algorithms often err on the side of sampling a great number of points, prioritizing accuracy over efficiency. As a result, many more sampled points are rejected from, rather than added to, the live set. 

\texttt{Nautilus} overcomes this limitation by (a) using importance nested sampling techniques \citep{Feroz2019} and (b) employing neural networks to apply more efficient bounds to the parameter space. Importance nested sampling allows the algorithm to use all previous likelihood calls, rather than just those in the live set, to determine which volume of the parameter space to sample next by associating a weight with each likelihood call. This fact, in combination with \texttt{Nautilus}'s use of neural networks to draw sampling bounds, results in a higher percentage of sampled points being added to the live set and thus makes fewer likelihood evaluations necessary to fully sample the parameter space.

\begin{deluxetable*}{>{\scriptsize}c|>{\scriptsize}c>{\scriptsize}c|>{\scriptsize}c>{\scriptsize}c|>{\scriptsize}c>{\scriptsize}c|>{\scriptsize}c>{\scriptsize}c}
\centering
\tablecolumns{9}
\tablewidth{0pt}
\vspace{-2mm}
\tablecaption{\scriptsize Molecular Transition Combinations \label{tab:mol_combos}}
\tablehead{
\multicolumn{1}{c|}{\scriptsize \# T} & \multicolumn{2}{c|}{\scriptsize$1-0$}  & \multicolumn{2}{c|}{\scriptsize $2-1$} &  \multicolumn{2}{c|}{\scriptsize$3-2$} & \multicolumn{2}{c}{\scriptsize$4-3$} \\ \hline
 &  HCN & HNC   & HCN & HNC & HCN & HNC &  HCN & HNC 
}
\startdata
1 & $\color{Sienna3} \blacklozenge$ & - & - & - & - & - & - & - \\ 
  & - & $\color{Sienna3} \blacklozenge$ & - & - & - & - & - & - \\ 
  & - & - & $\color{Sienna3} \blacklozenge$ & - & - & - & - & - \\
  & - & - & - & $\color{Sienna3} \blacklozenge$ & - & - & - & - \\
  & - & - & - & - & $\color{Sienna3} \blacklozenge$ & - & - & - \\
  & - & - & - & - & - & $\color{Sienna3} \blacklozenge$ & - & - \\
  & - & - & - & - & - & - & $\color{Sienna3} \blacklozenge$ & - \\
  & - & - & - & - & - & - & - & $\color{Sienna3} \blacklozenge$ \\ \hline
2 & \color{SpringGreen4}{\color{SpringGreen4}{\ding{53}}} & \color{SpringGreen4}{\ding{53}}  & - & - &  - & - &  - & -  \\ 
  & - & - & \color{SpringGreen4}{\ding{53}} & \color{SpringGreen4}{\ding{53}} &  - & - &  - & -  \\
  & - & - & - & - &  \color{SpringGreen4}{\ding{53}} & \color{SpringGreen4}{\ding{53}} &  - & -  \\
  & - & - & - & - &  - & - & \color{SpringGreen4}{\ding{53}} & \color{SpringGreen4}{\ding{53}}  \\ \hline
4 & \color{SpringGreen4}{\ding{53}} & \color{SpringGreen4}{\ding{53}} &  \color{SpringGreen4}{\ding{53}} & \color{SpringGreen4}{\ding{53}} &  - & - &  - & -  \\
  & \color{SpringGreen4}{\ding{53}} & \color{SpringGreen4}{\ding{53}} &  - & - &  \color{SpringGreen4}{\ding{53}} & \color{SpringGreen4}{\ding{53}} &  - & -  \\
  & \color{SpringGreen4}{\ding{53}} & \color{SpringGreen4}{\ding{53}} &  - & - & - & - &  \color{SpringGreen4}{\ding{53}} & \color{SpringGreen4}{\ding{53}}  \\
  & - & - & \color{SpringGreen4}{\ding{53}} & \color{SpringGreen4}{\ding{53}} &  \color{SpringGreen4}{\ding{53}} & \color{SpringGreen4}{\ding{53}} &  - & -  \\
  & - & - & \color{SpringGreen4}{\ding{53}} & \color{SpringGreen4}{\ding{53}} &  - & - &  \color{SpringGreen4}{\ding{53}} & \color{SpringGreen4}{\ding{53}}  \\
  & - & - & - & - &  \color{SpringGreen4}{\ding{53}} & \color{SpringGreen4}{\ding{53}} &  \color{SpringGreen4}{\ding{53}} & \color{SpringGreen4}{\ding{53}}  \\
  & \color{RoyalBlue3}$\blacksquare$ & - &   \color{RoyalBlue3}$\blacksquare$ & - &   \color{RoyalBlue3}$\blacksquare$ & - &   \color{RoyalBlue3}$\blacksquare$ & -  \\
  & - & \color{RoyalBlue3}$\blacksquare$ &  - & \color{RoyalBlue3}$\blacksquare$ &   - & \color{RoyalBlue3}$\blacksquare$ & - & \color{RoyalBlue3}$\blacksquare$  \\ \hline 
5 & \color{Purple4}\CIRCLE & \color{Purple4}\CIRCLE & \color{Purple4}\CIRCLE & - & \color{Purple4}\CIRCLE & - & \color{Purple4}\CIRCLE & - \\
  & \color{Purple4}\CIRCLE & \color{Purple4}\CIRCLE & - & \color{Purple4}\CIRCLE & - & \color{Purple4}\CIRCLE & - & \color{Purple4}\CIRCLE \\
  & \color{Purple4}\CIRCLE & - &  \color{Purple4}\CIRCLE & - &  \color{Purple4}\CIRCLE & - &  \color{Purple4}\CIRCLE & \color{Purple4}\CIRCLE \\
  & - & \color{Purple4}\CIRCLE &  - & \color{Purple4}\CIRCLE &  - & \color{Purple4}\CIRCLE &  \color{Purple4}\CIRCLE & \color{Purple4}\CIRCLE \\ \hline
6 & \color{SpringGreen4}{\ding{53}} & \color{SpringGreen4}{\ding{53}} &  \color{SpringGreen4}{\ding{53}} & \color{SpringGreen4}{\ding{53}} &  \color{SpringGreen4}{\ding{53}} & \color{SpringGreen4}{\ding{53}} &  - & - \\
  & \color{SpringGreen4}{\ding{53}} & \color{SpringGreen4}{\ding{53}} &  \color{SpringGreen4}{\ding{53}} & \color{SpringGreen4}{\ding{53}} &  - & - &  \color{SpringGreen4}{\ding{53}} & \color{SpringGreen4}{\ding{53}} \\
  & \color{SpringGreen4}{\ding{53}} & \color{SpringGreen4}{\ding{53}} &  - & - &  \color{SpringGreen4}{\ding{53}} & \color{SpringGreen4}{\ding{53}} &  \color{SpringGreen4}{\ding{53}} & \color{SpringGreen4}{\ding{53}} \\
  & - & - &  \color{SpringGreen4}{\ding{53}} & \color{SpringGreen4}{\ding{53}} &  \color{SpringGreen4}{\ding{53}} & \color{SpringGreen4}{\ding{53}} &  \color{SpringGreen4}{\ding{53}} & \color{SpringGreen4}{\ding{53}} \\ \hline
7 & - & $\color{VioletRed1}\blacktriangle$ & $\color{VioletRed1}\blacktriangle$ & $\color{VioletRed1}\blacktriangle$ & $\color{VioletRed1}\blacktriangle$ & $\color{VioletRed1}\blacktriangle$ & $\color{VioletRed1}\blacktriangle$ & $\color{VioletRed1}\blacktriangle$ \\ 
  & $\color{VioletRed1}\blacktriangle$ & - & $\color{VioletRed1}\blacktriangle$ & $\color{VioletRed1}\blacktriangle$ & $\color{VioletRed1}\blacktriangle$ & $\color{VioletRed1}\blacktriangle$ & $\color{VioletRed1}\blacktriangle$ & $\color{VioletRed1}\blacktriangle$ \\ 
  & $\color{VioletRed1}\blacktriangle$ & $\color{VioletRed1}\blacktriangle$ & - & $\color{VioletRed1}\blacktriangle$ & $\color{VioletRed1}\blacktriangle$ & $\color{VioletRed1}\blacktriangle$ & $\color{VioletRed1}\blacktriangle$ & $\color{VioletRed1}\blacktriangle$ \\
  & $\color{VioletRed1}\blacktriangle$ & $\color{VioletRed1}\blacktriangle$ & $\color{VioletRed1}\blacktriangle$ & - & $\color{VioletRed1}\blacktriangle$ & $\color{VioletRed1}\blacktriangle$ & $\color{VioletRed1}\blacktriangle$ & $\color{VioletRed1}\blacktriangle$ \\
  & $\color{VioletRed1}\blacktriangle$ & $\color{VioletRed1}\blacktriangle$ & $\color{VioletRed1}\blacktriangle$ & $\color{VioletRed1}\blacktriangle$ & - & $\color{VioletRed1}\blacktriangle$ & $\color{VioletRed1}\blacktriangle$ & $\color{VioletRed1}\blacktriangle$ \\ 
  & $\color{VioletRed1}\blacktriangle$ & $\color{VioletRed1}\blacktriangle$ & $\color{VioletRed1}\blacktriangle$ & $\color{VioletRed1}\blacktriangle$ & $\color{VioletRed1}\blacktriangle$ & - & $\color{VioletRed1}\blacktriangle$ & $\color{VioletRed1}\blacktriangle$ \\ 
  & $\color{VioletRed1}\blacktriangle$ & $\color{VioletRed1}\blacktriangle$ & $\color{VioletRed1}\blacktriangle$ & $\color{VioletRed1}\blacktriangle$ & $\color{VioletRed1}\blacktriangle$ & $\color{VioletRed1}\blacktriangle$ & - & $\color{VioletRed1}\blacktriangle$ \\ 
  & $\color{VioletRed1}\blacktriangle$ & $\color{VioletRed1}\blacktriangle$ & $\color{VioletRed1}\blacktriangle$ & $\color{VioletRed1}\blacktriangle$ & $\color{VioletRed1}\blacktriangle$ & $\color{VioletRed1}\blacktriangle$ & $\color{VioletRed1}\blacktriangle$ & - \\ \hline
8 & \color{SpringGreen4}{\ding{53}} & \color{SpringGreen4}{\ding{53}} &  \color{SpringGreen4}{\ding{53}} & \color{SpringGreen4}{\ding{53}} &  \color{SpringGreen4}{\ding{53}} & \color{SpringGreen4}{\ding{53}} &  \color{SpringGreen4}{\ding{53}} & \color{SpringGreen4}{\ding{53}} 
\enddata
\tablenotetext{\scriptstyle \color{Sienna3}\blacklozenge}{\scriptsize \space Single HCN OR HNC transition}
\vspace{-2mm}
\tablenotetext{\scriptstyle $\color{SpringGreen4}{\ding{53}}$}{\scriptsize \space Matched HCN and HNC transitions}
\vspace{-2mm}
\tablenotetext{\scriptstyle \color{RoyalBlue3}\blacksquare}{\scriptsize \space All HCN OR all HNC transitions}
\vspace{-2mm}
\tablenotetext{\scriptstyle \color{Purple4}\CIRCLE}{\scriptsize \space All HCN with one HNC transition OR all HNC with one HCN transition}
\vspace{-2mm}
\tablenotetext{\scriptstyle \color{VioletRed1}\blacktriangle}{\scriptsize \space Single HCN or HNC transition removed}
\end{deluxetable*}

\subsection{Region Selection} \label{sec:regions}

\begin{figure*}
\centering
    \includegraphics[trim = 3mm 2mm 0mm 0mm, width=\textwidth, clip=True]{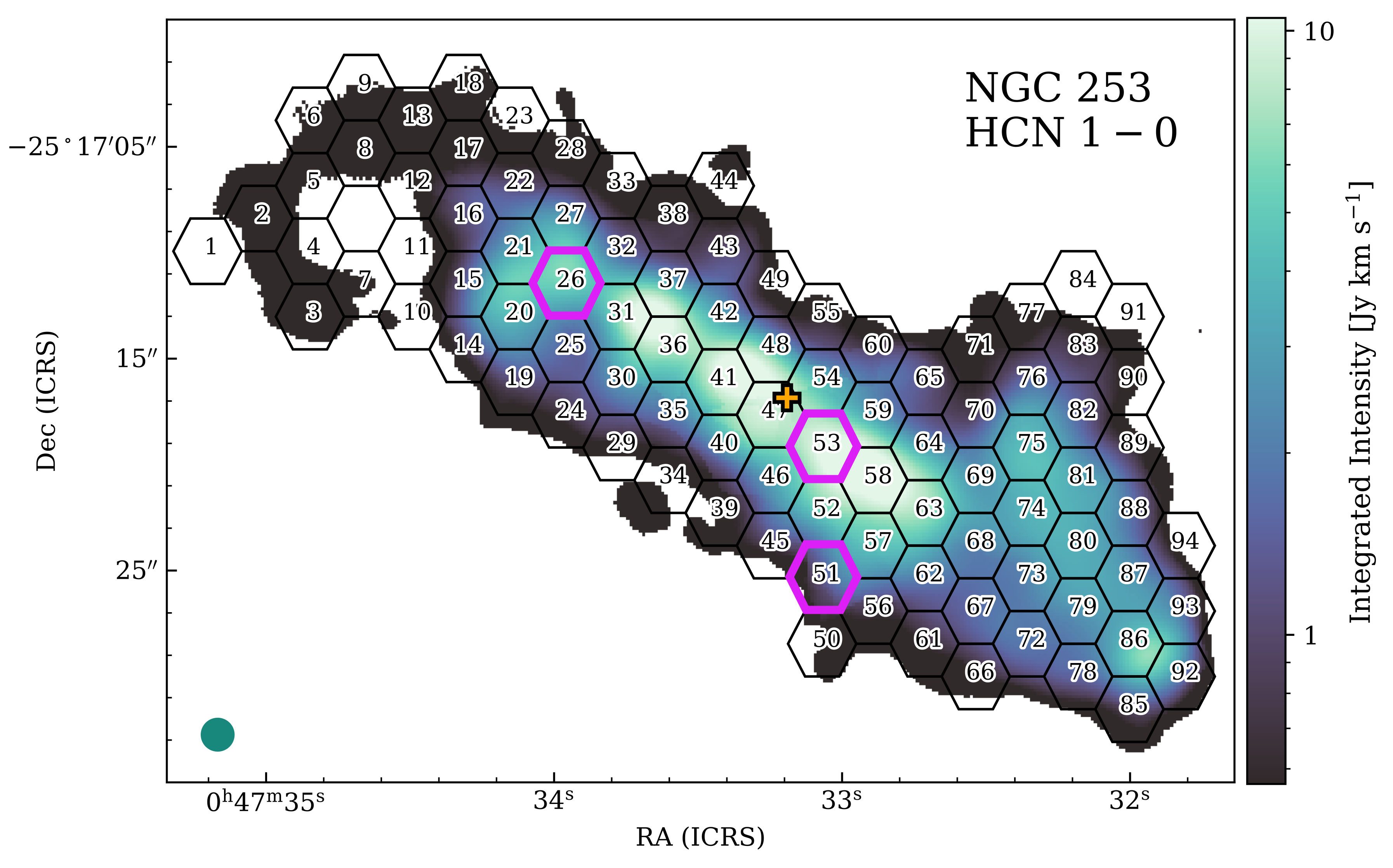}
    \caption{Map of NGC\,253 HCN 1--0 integrated emission with S/N $> 3$ overlaid with the regions analyzed in \cite{Behrens2024}. The three regions analyzed in this paper are highlighted in purple. The circle in the bottom left corner indicates the size of the 1.\!\!$^{\prime\prime}6$ ($\sim$27\,pc) ALCHEMI beam. The orange cross indicates the location of the galaxy's kinematic center \citep{Turner1985}.}
    \label{fig:ngc253}
\end{figure*}

We consider three regions in the NGC\,253 CMZ in which to test the constraining power of HCN and HNC in starburst environments. \cite{Behrens2024} divided the CMZ into 94 50-pc regions (Figure~\ref{fig:ngc253}) in order to sample GMC-scale structures and successfully constrained the gas conditions in many of these regions using 8 transitions of HCN and HNC. They identified a clear distribution in the gas conditions across the CMZ, with the highest volume densities and cosmic-ray ionization rates inferred toward the center and decreasing outward. Since other ALCHEMI studies \citep[e.g.][]{Holdship2022,Huang2023,Tanaka2024,Bouvier2024} have also found that the molecular gas properties vary spatially over the NGC\,253 CMZ, we select three regions that are representative of three different environments. The inferred parameter values for these 3 regions as determined by \cite{Behrens2024} are shown in Table~\ref{tab:priors}. 

We select region 53 (see Figure~\ref{fig:ngc253}), hereafter referred to as the ``high-$n$-$\zeta$ region", as an example of a central region with a high volume density ($n_{\text{H}_{2}} \sim 10^{5}$\,cm$^{-3}$) and cosmic-ray ionization rate (CRIR, $\zeta \sim 10^{-12.6}$\,s$^{-1}$). We choose region 26, hereafter referred to as the ``low-$n$-$\zeta$ region", as an example of an environment farther from the nucleus with more moderate gas conditions ($n_{\text{H}_{2}} \sim 10^{4}$\,cm$^{-3}$, $\zeta \sim 10^{-13.8}$\,s$^{-1}$). In both of these regions, we detected 4 transitions each of HCN and HNC with a signal-to-noise ratio (SNR) of at least 3. Though these regions are meant to distinctly represent more intense (high-$n$-$\zeta$) and more moderate (low-$n$-$\zeta$) conditions, they are neither the most extreme nor most quiescent regions within NGC\,253. They have instead been chosen to signify represent environments that could be generalized to many sources in the nearby universe.

We also chose to study region 51, hereafter referred to as the ``low-SNR region", which is located near the outskirts of the CMZ. \cite{Behrens2024} detected only 5 total HCN and HNC transitions toward the low-SNR region with a SNR $\ge 3$: HCN $J=1-0$, $2-1$, and $3-2$; and HNC $J=1-0$ and $2-1$. We chose this region due to its incomplete transition set as well as the fact that it is not strongly affected by UV, unlike many other low-signal regions on the outskirt of NGC\,253 \citep{Behrens2024}. Including the low-SNR region in our analysis will allow us to assess the impact of measurement SNRs on the constraining power of molecular transitions, while still employing the model framework described in Section~\ref{sec:modelReview}. As per \cite{Behrens2024}, we substitute 3$\sigma$ limits for the transitions that were not detected with SNR $\geq 3$ to use an equal number of constraining measurements per region---as indicated in Equation \ref{eq:lik_fn}, our log-likelihood function with which we compare our models and measurements does not account for the number of constraints. Thus we use upper limits to maintain common dimensionality across regions. Due to the low level of emission from these undetected transitions, and thus the low 3$\sigma$ limits, they do not significantly contribute to our inference of the gas parameters (see Figure~\ref{fig:reg51_onlyDet}). Nevertheless, using 3$\sigma$ upper limits prevents the model from invoking unrealistically high intensities for these transitions in the absence of constraints.
The transitions that are replaced with this intensity limit have the lowest SNRs of any used in this study and will provide useful information describing how marginally-detected lines influence gas parameter constraints.

\subsection{Molecular Constraints} \label{sec:mols}

We test several combinations of HCN and HNC transitions sampled from the first four HCN and HNC vibrational-ground rotational transitions: $J=1-0$, $2-1$, $3-2$, and $4-3$. Table \ref{tab:mol_combos} shows a comprehensive list of the transition combinations we test and categorizes them into several groupings, which we will discuss here. 

We first test the constraining power of single transitions when inferring gas parameters. These cases are marked with the $\color{Sienna3}\blacklozenge$ symbol in Table~\ref{tab:mol_combos} and utilize only one HCN or HNC transition to determine the gas conditions. In both nearby and high-redshift galaxies, oftentimes only one or two transitions, such as the HCN $J=1-0$ and $J=3-2$, are used to infer gas conditions such as volume density or dense gas mass \citep{Spilker2014,Oteo2017,Rybak2022}. We will consider the efficacy of using a single transition to constrain the molecular gas conditions and provide quantitative benchmarks that evaluate their accuracy in inferring gas parameters.

We then consider the case of matched HCN and HNC transitions (\textcolor{SpringGreen4}{\ding{53}}-marked combinations), i.e when we test the effects of constraining models with just the $J=1-0$ transition, we are using both the HCN and HNC $J=1-0$ transitions together. Previous studies \citep{Behrens2022,Behrens2024} have shown that the combination of HCN and HNC measurements, rather than just one species alone, is necessary in order to obtain clear constraints on certain model parameters, such as the cosmic-ray ionization rate. 

To further explore the necessity of using multiple species to constrain different gas parameters, we also test cases where all transitions from one species are used in combination with just a single transition of the other species (\textcolor{Purple4}{\CIRCLE}-marked combinations). We choose either the $J=1-0$ or $J=4-3$ transitions to supplement the full set of the other species in order to determine whether the excitation level of the single additional transition affects the efficacy of using two species to constrain the gas conditions. We will compare these tests to the cases where we use all the transitions of a single species ($\color{RoyalBlue3}\blacksquare$-marked combinations in Table~\ref{tab:mol_combos}) to determine both how well various gas conditions can be constrained with a single species as well as whether the addition of a single transition from another species improves these constraints.

We also test whether any one transition has a particularly enhanced ability to constrain the physical parameters. We explore how removing a single transition from the full set of 8 affects our results ($\color{VioletRed1}\blacktriangle$-marked combinations) to determine whether any one transition should be included in the constraints.

Using each combination of HCN and HNC transitions listed in Table~\ref{tab:mol_combos}, we run our Bayesian inference plus neural network algorithm (described in Section~\ref{sec:nautilus}) to obtain posterior distributions for each parameter listed in Table~\ref{tab:priors}. Example corner plots for each combination type can be found in Appendix \ref{sec:addFigs} for the low-$n$-$\zeta$ region. We compare the posterior distributions from each test case for each parameter to those obtained when using all eight transitions as constraints (see Table~\ref{tab:priors}), which we will hereafter refer to as the ``control distributions". Note that, as with all modeling frameworks, we cannot know with complete certainty that our model includes all of the physical and chemical processes occurring in a studied region, and thus we do not know if the parameters inferred from our set of control measurements are completely correct. However, our purpose with this study is instead to quantify how our understanding of the gas parameters may change when we use a different number of constraining measurements to infer said parameters.

To compare the test and control cases, we employ least-squares residuals. We do this by first binning the test and control distributions using the Freedman-Diaconis rule based on the data in the control distribution. This rule states that the bin width should be equal to $2 \times IQR \times N^{-1/3}$, where $IQR$ is the interquartile range (which is calculated in logarithmic space) and $N$ is the number of data points in the control distribution. We use identical bins for control and test distributions, which contain similar numbers of data points. We then calculate the residuals in each bin, sum them up, and normalize the final values to obtain a single residual $r$ for each test distribution using the following expression:

\begin{equation}
    r = \sum\limits_{i=1}^{n} (\dfrac{k_i}{N_{\text{control}}} - \dfrac{j_i}{N_{\text{test}}})^2, \label{eq:res}
\end{equation}
where $k$ and $j$ are the number of samples in the $i$th bin of the control and test distributions, respectively, and $N$ is the number of total samples in each distribution.

\begin{deluxetable*}{c|c|ccccc}[]
\centering
\tablecolumns{7}
\tablewidth{0pt}
\tablecaption{Inferred Parameter Values for Observationally and Synthetically Constrained Models\label{tab:synth}}
\tablehead{
 & \multicolumn{1}{c|}{Region} & \colhead{$T_{\mathrm{K}}$} & \colhead{$n_{\mathrm{H}_2}$} & \colhead{$\zeta$} & \colhead{$N_{\mathrm{H}_2}$} & \colhead{$\eta_{ff}$} \\
 & & (K) & (cm$^{-3}$) & (s$^{-1}$) & (cm$^{-2}$) & 
}
\startdata
\parbox[t]{2mm}{\multirow{3}{*}{\rotatebox[origin=c]{90}{Control}}} & High-$n$-$\zeta$ & $179.47^{+100.56}_{-81.57}$ & $4.93^{+0.37}_{-0.44}$\tablenotemark{a} & $-12.64^{+0.40}_{-0.48}$\tablenotemark{a} & $24.43^{+0.38}_{-0.46}$\tablenotemark{a} & $0.17^{+0.06}_{-0.04}$ \\
 & Synthetic, SNR = 3 & $160.92^{+117.47}_{-85.14}$ & $5.07^{+0.57}_{-0.74}$\tablenotemark{a} & $-12.80^{+0.80}_{-1.76}$\tablenotemark{a} & $24.07^{+0.65}_{-0.88}$\tablenotemark{a} & $0.21^{+0.26}_{-0.13}$ \\
 & Synthetic, SNR = 5 & $170.73^{+106.97}_{-83.98}$ & $4.99^{+0.40}_{-0.53}$\tablenotemark{a} & $-12.63^{+0.46}_{-0.64}$\tablenotemark{a} & $24.34^{+0.45}_{-0.59}$\tablenotemark{a} & $0.19^{0.10}_{-0.06}$ \\
 \hline
 \parbox[t]{2mm}{\multirow{3}{*}{\rotatebox[origin=c]{90}{HCN 1--0}}} & High-$n$-$\zeta$ & $199.09^{+101.42}_{-101.74}$ & $4.43^{+1.63}_{-0.92}$\tablenotemark{a} & $-12.91^{+1.42}_{-2.04}$\tablenotemark{a} & $23.28^{+1.06}_{-0.81}$\tablenotemark{a} & $0.46^{+0.36}_{-0.32}$ \\
 & Synthetic, SNR = 3 & $196.68^{+102.46}_{-102.14}$ & $4.38^{+1.61}_{-0.89}$\tablenotemark{a} & $-12.78^{+1.34}_{-2.05}$\tablenotemark{a} & $23.29^{+1.07}_{-0.85}$\tablenotemark{a} & $0.44^{+0.37}_{-0.31}$ \\ 
 & Synthetic, SNR = 5 & $199.97^{+101.23}_{-103.25}$ & $4.43^{+1.63}_{-0.90}$\tablenotemark{a} & $-12.91^{+1.41}_{-2.05}$\tablenotemark{a} & $23.31^{+1.08}_{-0.83}$\tablenotemark{a} & $0.44^{+0.36}_{-0.32}$ \\ 
 \hline
 \parbox[t]{2mm}{\multirow{3}{*}{\rotatebox[origin=c]{90}{HNC only}}} & High-$n$-$\zeta$ & $221.51^{+90.36}_{-110.47}$ & $4.95^{+0.36}_{-0.51}$\tablenotemark{a} & $-14.10^{+1.69}_{-1.29}$\tablenotemark{a} & $24.00^{+0.71}_{-0.89}$\tablenotemark{a} & $0.18^{+0.15}_{-0.06}$ \\
  & Synthetic, SNR = 3 & $210.58^{+97.05}_{-106.41}$ & $5.03^{+0.59}_{-0.75}$\tablenotemark{a} & $-13.80^{+1.92}_{-1.48}$\tablenotemark{a} & $23.77^{+0.83}_{-1.02}$\tablenotemark{a} & $0.37^{+0.37}_{-0.21}$ \\ 
& Synthetic, SNR = 5 & $217.11^{+92.70}_{-108.66}$ & $5.02^{+0.36}_{-0.58}$\tablenotemark{a} & $-13.97^{+1.69}_{-1.39}$\tablenotemark{a} & $23.89^{+0.77}_{-0.96}$\tablenotemark{a} & $0.25^{+0.24}_{-0.10}$ \\ 
\enddata
\tablenotetext{a}{Value is given in logspace}
\end{deluxetable*}


\subsection{Bayesian Inference Under Idealized Conditions} \label{sec:synth}

\begin{figure*}
\centering 
    \gridline{
    \leftfig{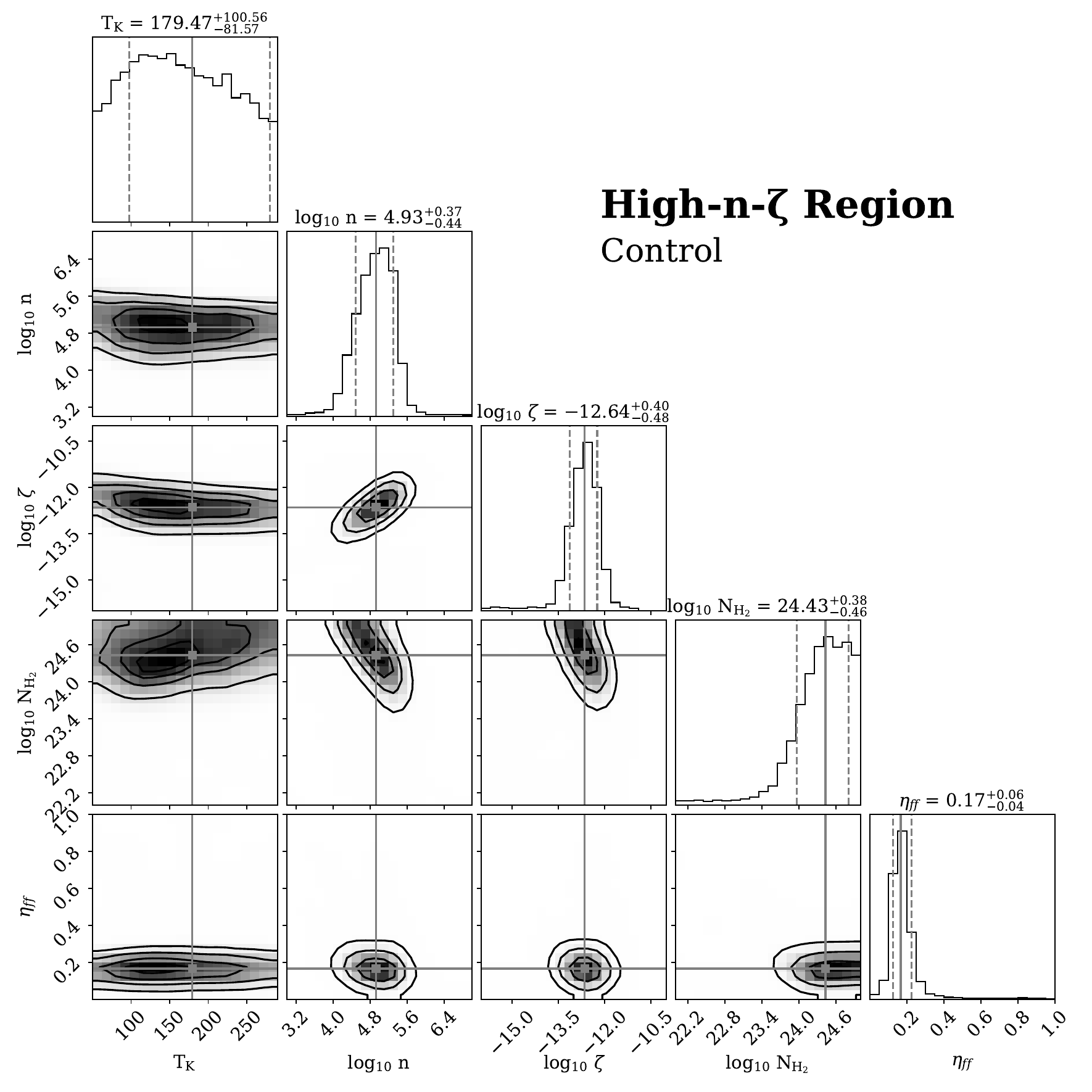}{0.5\textwidth}{}
    \rightfig{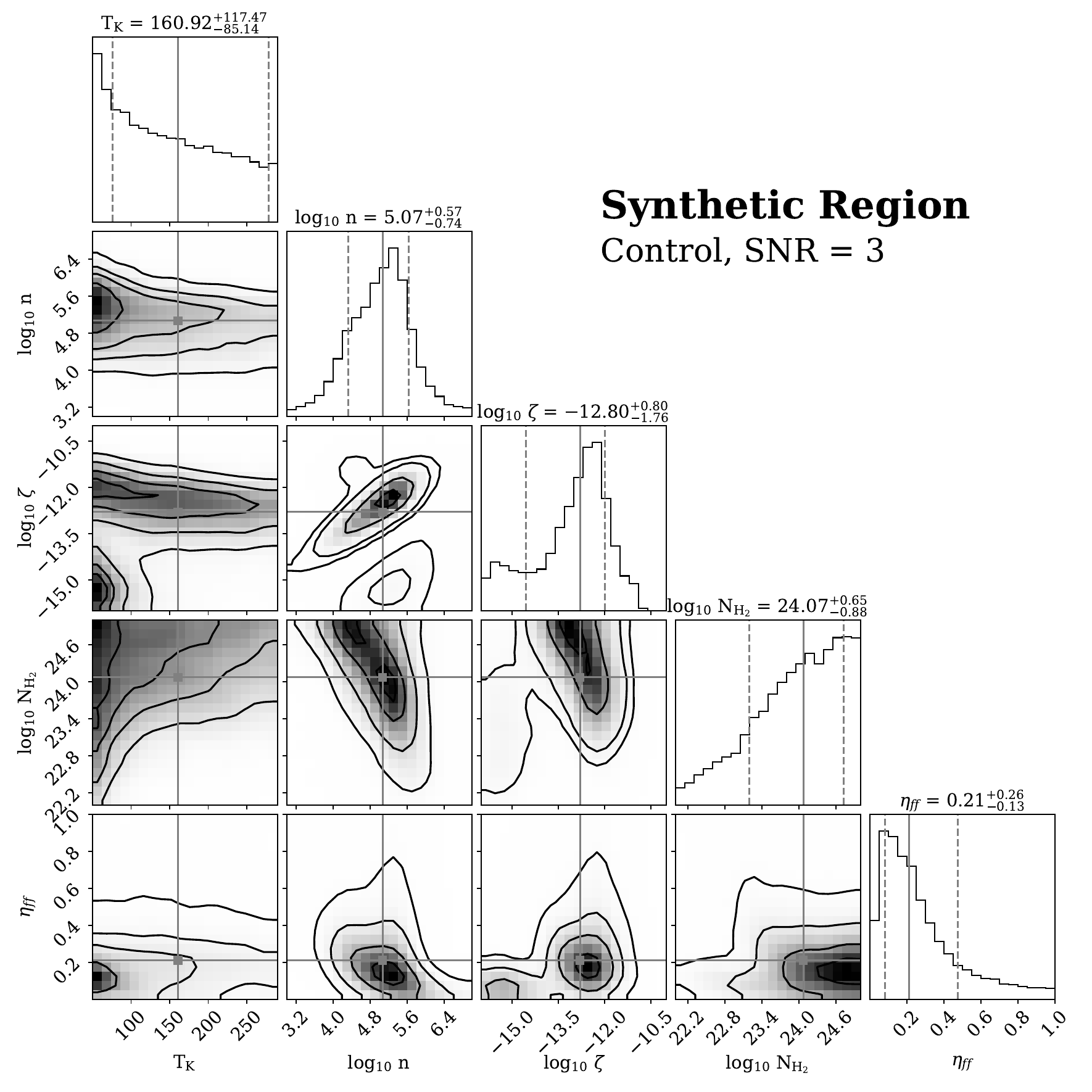}{0.5\textwidth}{}
    }
    \gridline{
    \fig{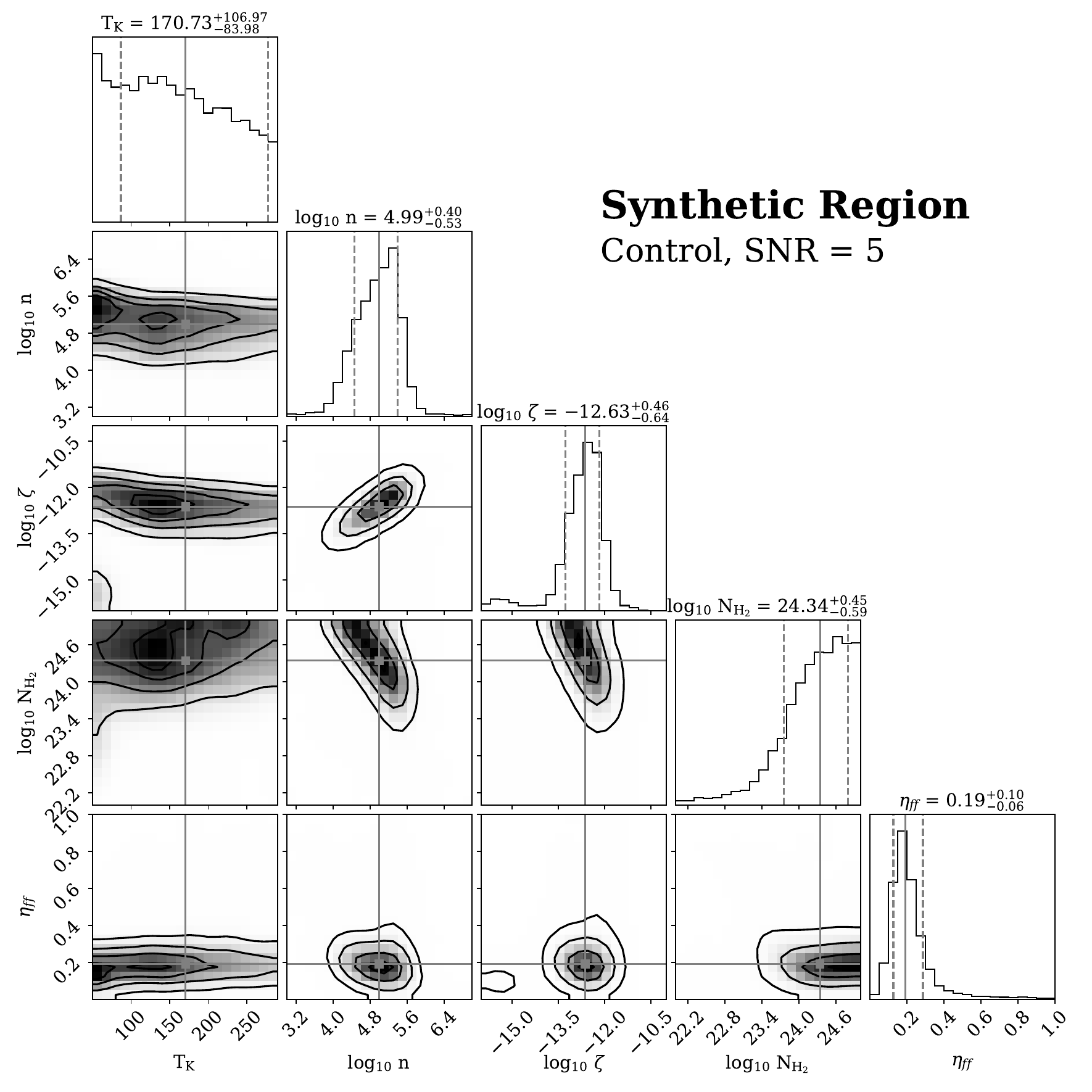}{0.5\textwidth}{}
    }
    \vspace{-8mm}
    \caption{Bayesian inference results for the controls of the high-$n$-$\zeta$ region (top left) and for the synthetic region where the synthetic data SNR = 3 (top right) and SNR = 5 (bottom).}
    \label{fig:synthControl}
\end{figure*}



To confirm that any differences in the inferred parameters arise solely from the constraining transitions and not from the Bayesian inference algorithm, we consider Bayesian inference results under idealized conditions with synthetic data. We generate synthetic data for a region with conditions mirroring those in the high-$n$-$\zeta$ region by taking the inferred parameters for the high-$n$-$\zeta$ region from Table \ref{tab:priors} and running them through our neural network and radiative transfer models to generate synthetic integrated intensities. We divide these integrated intensities by 3 and 5 to generate uncertainties that result in synthetic measurements with SNRs of 3 and 5, respectively. We then use various combinations of these synthetic integrated intensity and uncertainty values to constrain the Bayesian inference algorithm with the intention of recovering the same parameter values that were used to generate the synthetic data.

Figure \ref{fig:synthControl} shows the results of constraining models with the 8 HCN and HNC transitions (the control) measured toward the high-$n$-$\zeta$ region, as well as the results obtained from synthetic constraints with SNRs of 3 and 5. These results are also shown in Table \ref{tab:synth}, along with results for the same regions for models constrained with just HCN $1-0$ and just the four HNC transitions. Figures for the HCN $1-0$ and HNC only tests can be found in Appendix \ref{sec:addFigs}. Tests with the synthetic data show that the Bayesian inference algorithm successfully recovers the input high-$n$-$\zeta$ region gas parameters. Synthetic measurements with SNR = 5 result in better constrained gas parameters than in the SNR = 3 case, with smaller uncertainties and median parameter values that more closely match those obtained with the high-$n$-$\zeta$ measured transitions. However, all median parameter values inferred from both the SNR = 3 and SNR = 5 tests fall within the uncertainty range set by the high-$n$-$\zeta$ results. These tests indicate that any changes seen in the inferred gas parameters are a result of the combination and/or the SNR of the constraining transitions, not the Bayesian inference algorithm.

\section{Results} \label{sec:results}

\begin{figure*}
\centering
   \gridline{
   \leftfig{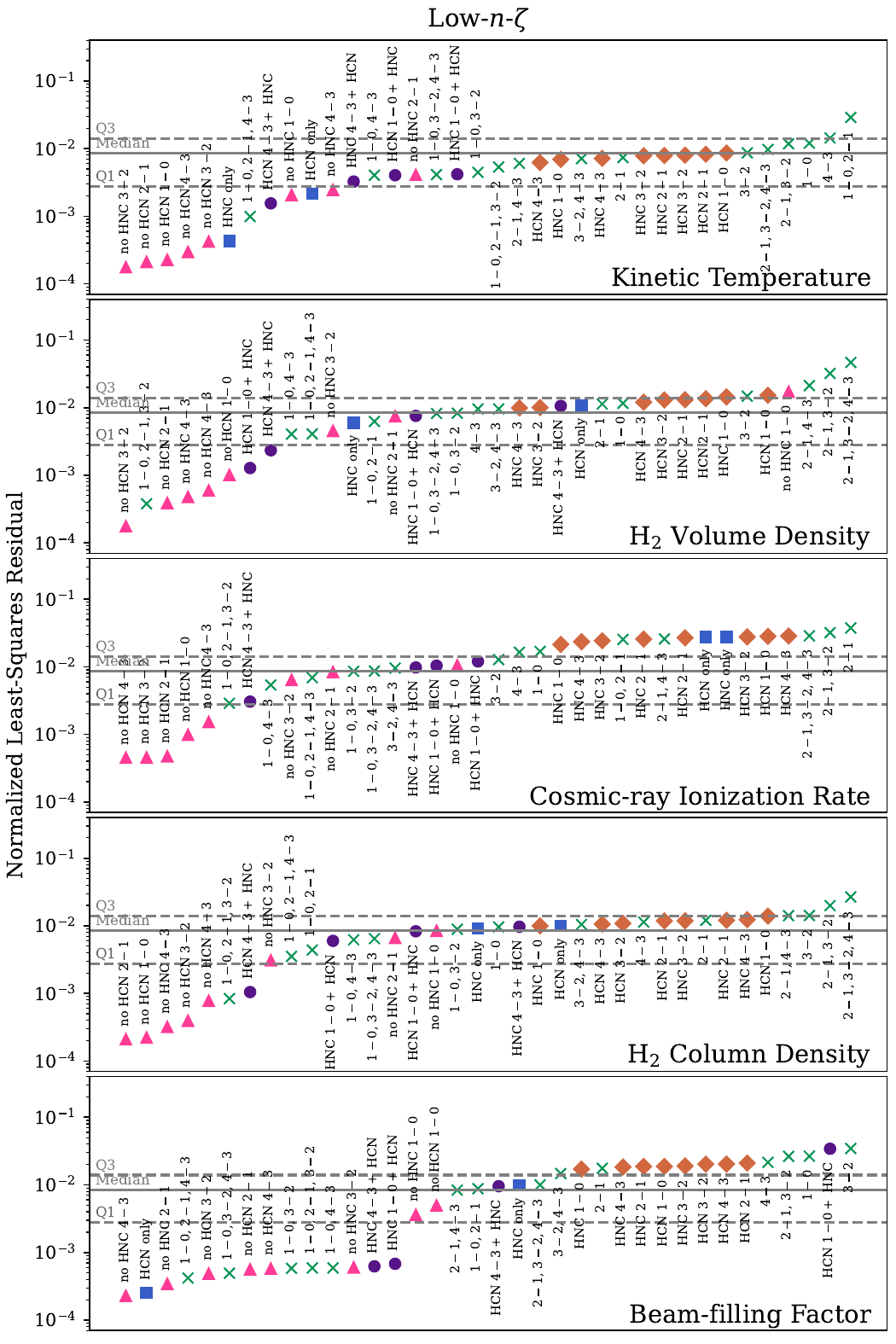}{0.77\textwidth}{}
   \rightfig{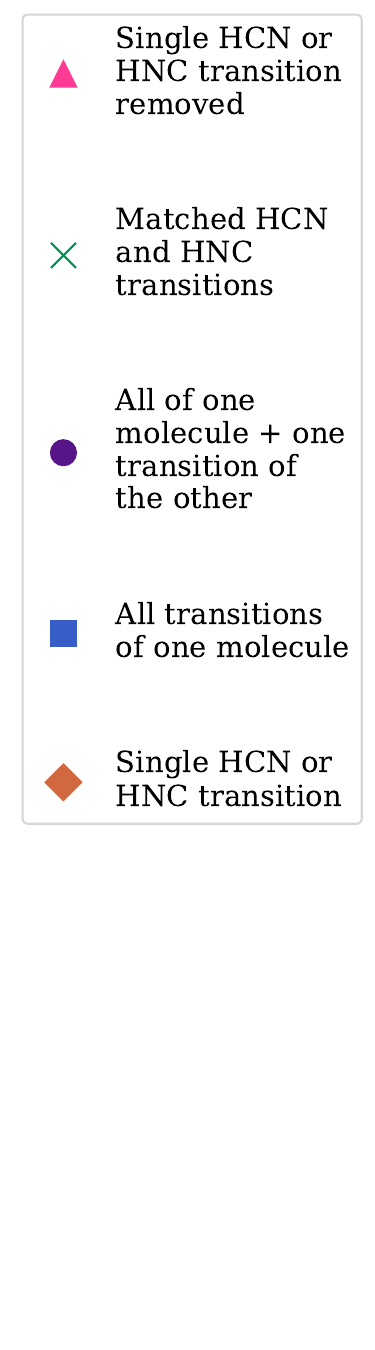}{0.23\textwidth}{}
   }
   \vspace{-8mm}
   \caption{Normalized least-squares residuals for each transition combination listed in Table~\ref{tab:mol_combos}, where residual values close to zero indicate better agreement between the test and control distributions. 
   The solid gray lines represent the median residual value across all parameters for the low-$n$-$\zeta$ region, and the dashed gray lines represent the first and third quartiles.}
   \label{fig:reg26_resids}
\end{figure*}
\begin{figure*}
\centering
    \gridline{
    \leftfig{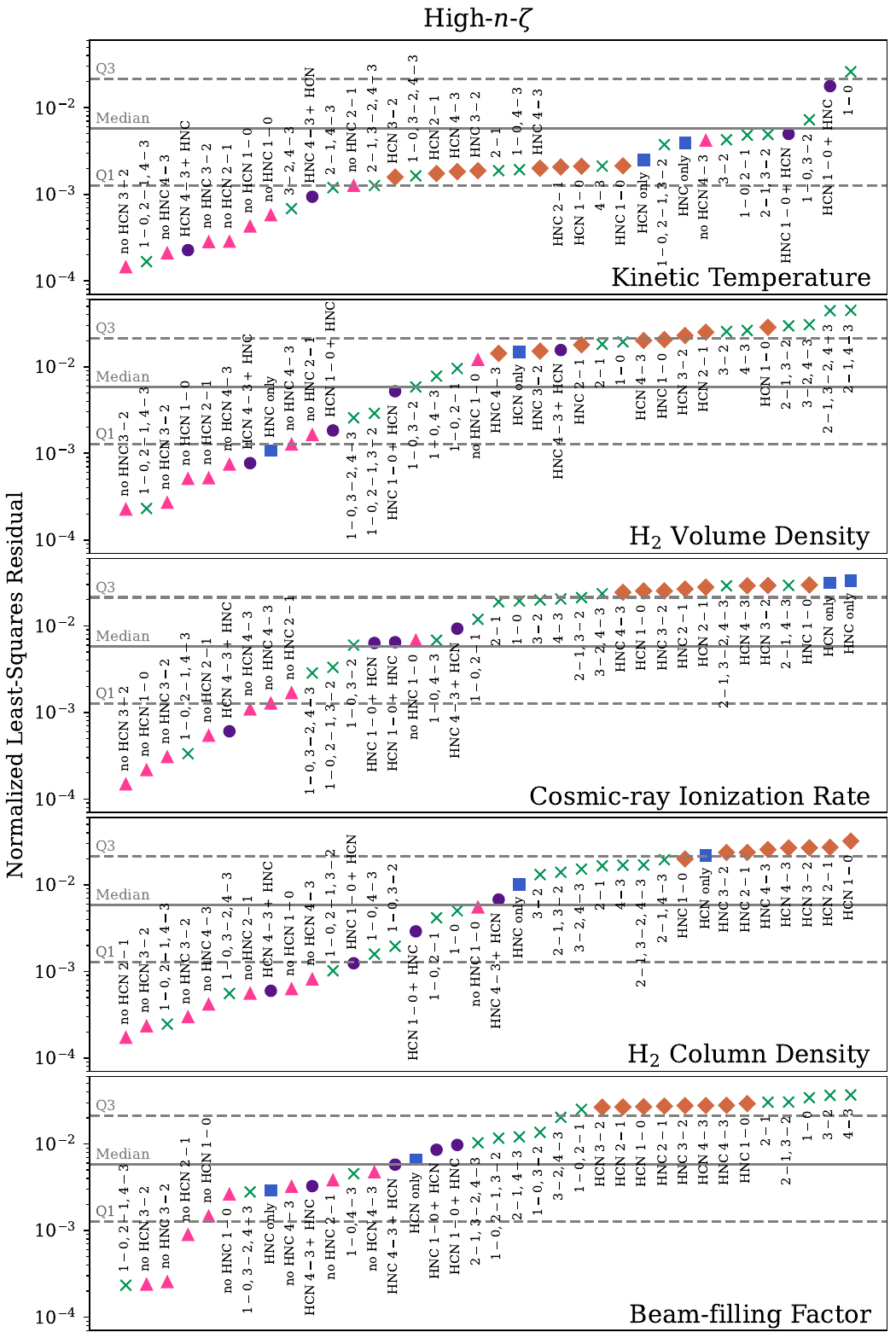}{0.77\textwidth}{}
    \rightfig{residuals_legend.pdf}{0.23\textwidth}{}
    }
    \vspace{-8mm}
    \caption{Same as in Figures~\ref{fig:reg26_resids} and \ref{fig:reg51_resids} but for the high-$n$-$\zeta$ region.}
    \label{fig:reg53_resids}
\end{figure*}
\begin{figure*}
\centering 
    \gridline{
    \leftfig{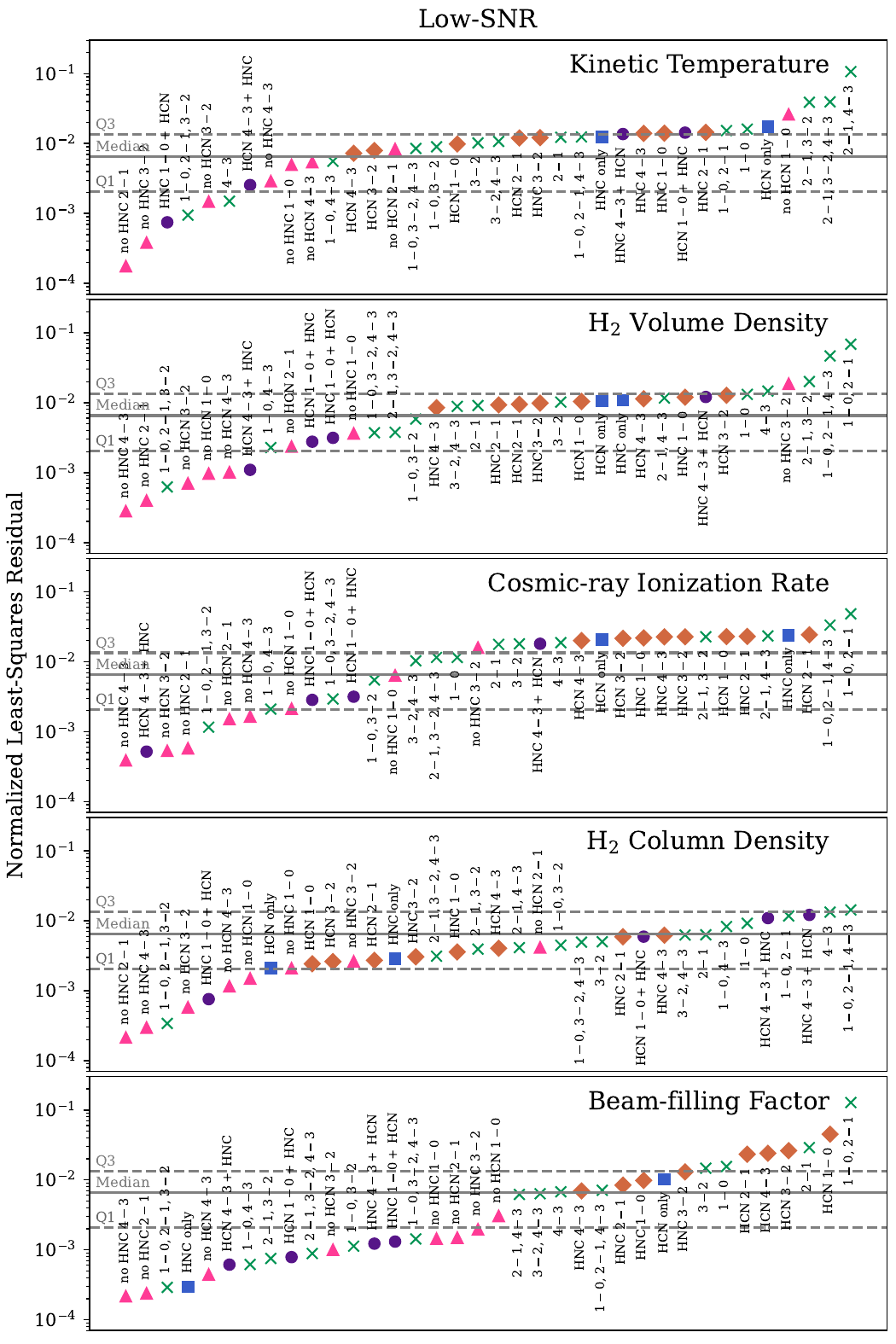}{0.77\textwidth}{}
    \rightfig{residuals_legend.pdf}{0.23\textwidth}{}
    }
    \vspace{-8mm}
    \caption{Same as in Figure~\ref{fig:reg26_resids} but for the low-SNR region.}
    \label{fig:reg51_resids}
\end{figure*}

In the following we will describe the results of our HCN and HNC measurement constraints investigation. The normalized residuals for each test case and parameter are shown in Figures~\ref{fig:reg26_resids}, \ref{fig:reg53_resids}, and \ref{fig:reg51_resids} for the low-$n$-$\zeta$, high-$n$-$\zeta$, and low-SNR regions, respectively. For each region, we collect the residuals across all parameters and calculate quartiles, which provide three thresholds (quartile 1, or Q1; the median, $r_{\text{med}}$; and quartile 3, or Q3) that separate the transition combinations into four groups based on the residuals. We will use these delineations in the following discussion to evaluate the efficacy of each combination's ability to reproduce the control distributions. 

\begin{figure*}[t!]
    \includegraphics[width=\linewidth]{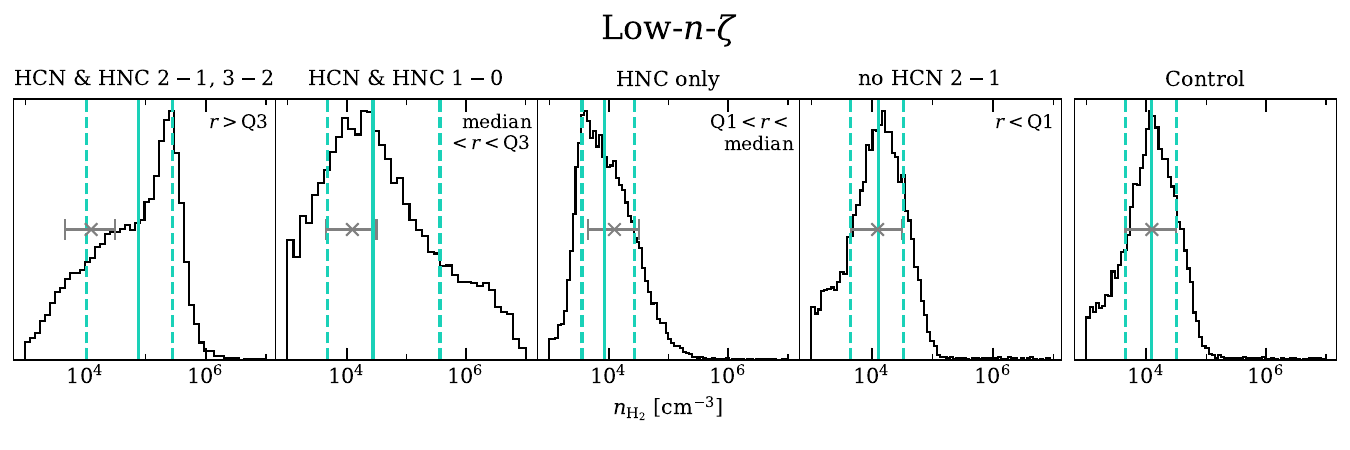}
    \caption{Example volume density posterior distributions for transition combinations that fall above, below, and between the three quartiles in the low-$n$-$\zeta$ region. Teal lines indicate the median (solid) and middle 66\% (dashed) of each distribution, while the gray X's and horizontal errors bars show the median and middle 66\%, respectively, of the control distribution (right panel).}
    \label{fig:benchmarks}
\end{figure*}

Examples of test combination posterior distributions that fall between each quartile are shown in Figure~\ref{fig:benchmarks} as compared to the control for volume density in the low-$n$-$\zeta$ region. Test combinations with residuals $r<$ Q1 are generally nearly indistinguishable from the control distributions. Transition combinations with Q1 $<r<r_{\text{med}}$  typically recover the same median value of the posterior distribution as the control and usually have similar uncertainties. It is thus up to the user to determine to what accuracy and precision threshold a given combination must rise in order to meet the requirements of their use case. Note, however, that these quartile values describe only where each transition combination falls with respect to the other combinations we describe in Section~\ref{sec:mols}, which are not necessarily representative of all possible combinations of the HCN and HNC $J=1-0$, $2-1$, $3-2$, and $4-3$ transitions. We will address each transition grouping, as described in Section~\ref{sec:mols} and Table~\ref{tab:mol_combos}, separately and then synthesize these results in Section~\ref{sec:discussion}.

\subsection{Single Transitions} \label{sec:single}

We first consider the test cases in which we used a single HCN or HNC transition to constrain the models used in our Bayesian inferencing algorithm. These test cases are denoted in Figures~\ref{fig:reg26_resids}, \ref{fig:reg53_resids}, and \ref{fig:reg51_resids} with the $\color{Sienna3}\blacklozenge$ symbol. In general, we caution readers against using single transitions to infer any physical parameters, as the degeneracy between key gas parameters (e.g. density and temperature) results in extremely uncertain or inaccurate parameter estimates. In nearly all cases, constraining the gas conditions with a single HCN or HNC transition resulted in residual values greater than the median residual for every parameter in each respective region. The only exceptions are for kinetic temperature in the low- and high-$n$-$\zeta$ regions and H$_{2}$ column density in the low-SNR region. As mentioned in Section~\ref{sec:nautilus}, we are unable to constrain the kinetic temperature in this environment with HCN and HNC measurements; thus we expect that any trends we note regarding the constraining power of HCN and HNC will likely not hold for kinetic temperature. We will further discuss any discrepancies in trends in the low-SNR region in Section~\ref{sec:discussion}.

In general, higher-$J$ ($J=3-2$ or $4-3$) single transitions are more effective at reproducing the control distributions across all regions than lower-$J$ transitions. Though the differences in residuals between the single-transition cases are often minimal, the most effective single-transition tests use the HCN or HNC $J=3-2$ or $4-3$ transitions in two-thirds of cases. HNC is also slightly more effective than HCN at reproducing control distributions when a single transition is used, with 60\% of cases featuring an HNC transition as the single-transition test with the lowest residual value. Alternatively, low-energy HCN and HNC transitions are among the least effective in reproducing the control posterior distributions. In particular, the HCN $J=1-0$ transition results in residual values that are above the third quartile in two-thirds of cases.

\subsection{Matched HCN and HNC Transitions} \label{sec:matched}

\begin{figure*}
\centering
    \includegraphics[width=0.85\textwidth]{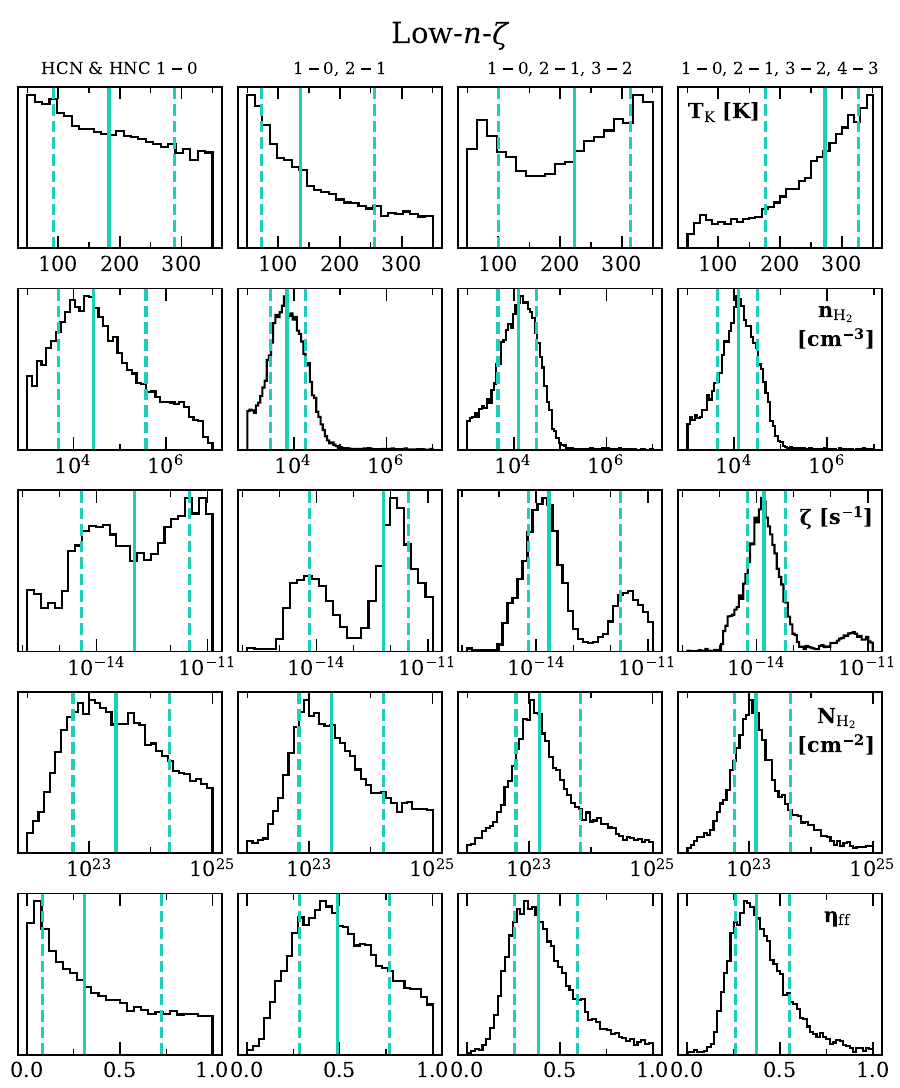}
    \caption{One-dimensional posterior distributions for the low-$n$-$\zeta$ region constrained by an increasing number of higher-$J$ HCN and HNC transitions. Median values for each distribution are denoted by the solid lines, and the middle 66\% of the distribution is indicated by the dashed lines.}
    \label{fig:reg26_1d}
\end{figure*}

Matched transitions, which use the same HCN and HNC transitions to constrain each model, are denoted in Figures~\ref{fig:reg26_resids}, \ref{fig:reg53_resids}, and \ref{fig:reg51_resids} with \textcolor{SpringGreen4}{\ding{53}} symbols. Since the single-transition tests (Section~\ref{sec:single}) indicate that higher-$J$ transitions are better at reproducing the control posterior distributions, we consider the same scenario for the matched-transition test cases. Across the three regions we test, the matched-transition test cases with the lowest residual values for each parameter contain either the HCN and HNC $J=3-2$ or $4-3$ transitions. 

To further investigate the impact of higher-energy transitions on our ability to constrain gas parameters, we show in Figure~\ref{fig:reg26_1d} how starting with just the HCN and HNC $J=1-0$ transitions and progressively adding matched sets of higher-energy transition constraints affects the parameter posterior distributions for each parameter in the low-$n$-$\zeta$ region. In the case of volume density, which our control indicates is $\sim10^4$\,cm$^{-3}$, a posterior distribution with a similar median and uncertainty range to our control can be obtained with just the HCN and HNC $J=1-0$ and $2-1$ transitions. Though using just the HCN and HNC $J=1-0$ transitions recovers a median density within the uncertainty of our control, the uncertainty on the test case is much larger, spanning nearly two orders of magnitude. On the other hand, Figure~\ref{fig:reg26_1d} shows that an accurate estimate of the CRIR with reasonable uncertainty cannot be achieved until the $J=4-3$ HCN and HNC transitions are included. This particular CRIR posterior distribution example features a bimodality \citep[discussed in][]{Behrens2022} that cannot be fully resolved without the addition of higher-$J$ transitions. 

However, we see from Figures~\ref{fig:reg26_resids}, \ref{fig:reg53_resids}, and \ref{fig:reg51_resids} that reasonable constraints that mirror the control CRIR posterior distributions can be obtained without all 8 HCN and HNC transitions. The matched-transition tests with the lowest residuals for CRIR across all regions contained the $J=1-0$, $2-1$, and either the $3-2$ or $4-3$ HCN and HNC transitions. These results show that a combination of low- and high-energy molecular transitions from both HCN and HNC can reliably reproduce the control posterior distributions without requiring all 8 of the transitions used in the control. 

The combinations of transitions that ultimately prove to provide the best constraints on the physical conditions may be dependent upon the specific conditions of the region under study. In the low-$n$-$\zeta$ region (Figure~\ref{fig:reg26_resids}), the most effective matched-transition combination to recover the volume density control posterior distribution ($n\sim10^4$\,cm$^{-3}$) is the $J=1-0$, $2-1$, and $3-2$ combination, which has a residual that falls well below the first quartile. Alternatively, replacing the $J=3-2$ transitions with the $J=4-3$ transitions yields a higher residual, between the first quartile and the median. However in the high-$n$-$\zeta$ region, which the control indicates has a higher volume density ($\sim10^5$\,cm$^{-3}$), the $J=1-0$, $2-1$, and $4-3$ transition combination is most effective at recovering the volume density with a residual below the first quartile, while swapping the $J=4-3$ transitions for the $J=3-2$ would result in a higher residual between the first quartile and median. Both combinations are generally effective, but the volume density posterior distribution is better recovered when using the highest-energy transitions in this higher-density region. We note that in the case of the low-SNR region, though the $J=1-0$, $2-1$, and $3-2$ combination performs quite well across all parameters, the $J=1-0$, $2-1$, and $4-3$ combination is actually one of the worst. This discrepancy may be a result of the lower SNR of the transitions measured in this region, which often results in less stable parameter constraints.

We also see that the inclusion of the $J=1-0$ HCN and HNC transitions is important for recovering the control distributions. Combinations that do not include the $J=1-0$ transitions, even when including the other 6 transitions, almost exclusively result in residuals higher than the median and rank behind almost all other tested combinations. When 6 transitions are used, we also see that including two sets of lower-energy transitions ($J=1-0$ and $2-1$) and one higher-energy transition ($J=3-2$ or $4-3$) is more effective than including only one lower-energy and two higher-energy pairs. We will discuss these results further in Section~\ref{sec:discussion}.

\subsection{Single Species} \label{sec:one_species}
Models that we constrain with all four transitions of a single species are shown in Figures~\ref{fig:reg26_resids}, \ref{fig:reg53_resids}, and \ref{fig:reg51_resids} with $\color{RoyalBlue3}\blacksquare$ symbols. Single-species constraints generally only yield posterior distributions that resemble the control in the case of volume density, and only for a specific subset of cases. In the low-$n$-$\zeta$ region, the HNC-only test yields a residual value between the first quartile (Q1) and the median for H$_{2}$ volume density, and in the high-$n$-$\zeta$ region, the same test results in a residual value below Q1. Alternatively, the HCN-only test case results in residual values greater than the median in all regions for volume density, indicating that HNC by itself may be more useful for constraining the gas density than HCN. Using a single species to constrain our models results in very inaccurate and ineffective constraints on the cosmic-ray ionization rate and inconsistent results for all other parameters across all three regions.

\subsection{Single Species with One Opposing Species Transition Added} \label{sec:oneSpecies_plus1}

The test cases where we have used all the transitions from one species with one additional high- or low-energy transition from the opposing species are shown with \textcolor{Purple4}{\CIRCLE} symbols in Figures~\ref{fig:reg26_resids}, \ref{fig:reg53_resids}, and \ref{fig:reg51_resids}. In nearly all cases, the \textcolor{Purple4}{\CIRCLE} test with the lowest residual value for each parameter includes all HNC transitions with one additional HCN transition. In the high-$n$-$\zeta$ region, supplementing the HNC transitions with a single HCN transition results in posterior distributions with residual values below Q1 for all parameters except the beam-filling factor. In the low- and high-$n$-$\zeta$ regions, using the HCN $J=4-3$ transition to supplement the HNC constraints, rather than the HCN $J=1-0$, is more effective at recovering the control distributions across all parameters, except the beam-filling factor in the low-$n$-$\zeta$ region. The results are less consistent in the low-SNR region, which shows a mix of favoring the HNC $+$ HCN $J=4-3$ combination as well as HCN $+$ HNC $J=1-0$ for the most effective test cases in this category. In general, adding one transition from a different species results in significant improvements in the posterior distributions compared to using only transitions from a single species. Using a higher-energy transition as the supplementary constraint may further improve results.

\subsection{Removing a Single Transition} \label{sec:one_transition_removed}

In order to assess whether any single transition has a significant impact on the gas parameter constraints, we try removing one transition at a time from our measurement constraints, which are denoted by $\color{VioletRed1}\blacktriangle$ symbols in Figures~\ref{fig:reg26_resids}, \ref{fig:reg53_resids}, and \ref{fig:reg51_resids}. These results show that the HNC $J=1-0$ transition has the largest impact on our parameter inferencing results. While most tests in this category result in residuals below Q1, removing the HNC $J=1-0$ transition from our constraints often leads to residual values greater than the median. This result is especially evident for volume density in the low- and high-$n$-$\zeta$ regions, where significant gaps exist between this case and all other $\color{VioletRed1}\blacktriangle$ test cases. The low-SNR region is less consistent, with a variety of other transitions appearing to be the most impactful on the parameter inferencing results. Other parameters do not show significant preferences toward any particular transition.

\section{Discussion} \label{sec:discussion}


\begin{figure*}
    \gridline{
    \leftfig{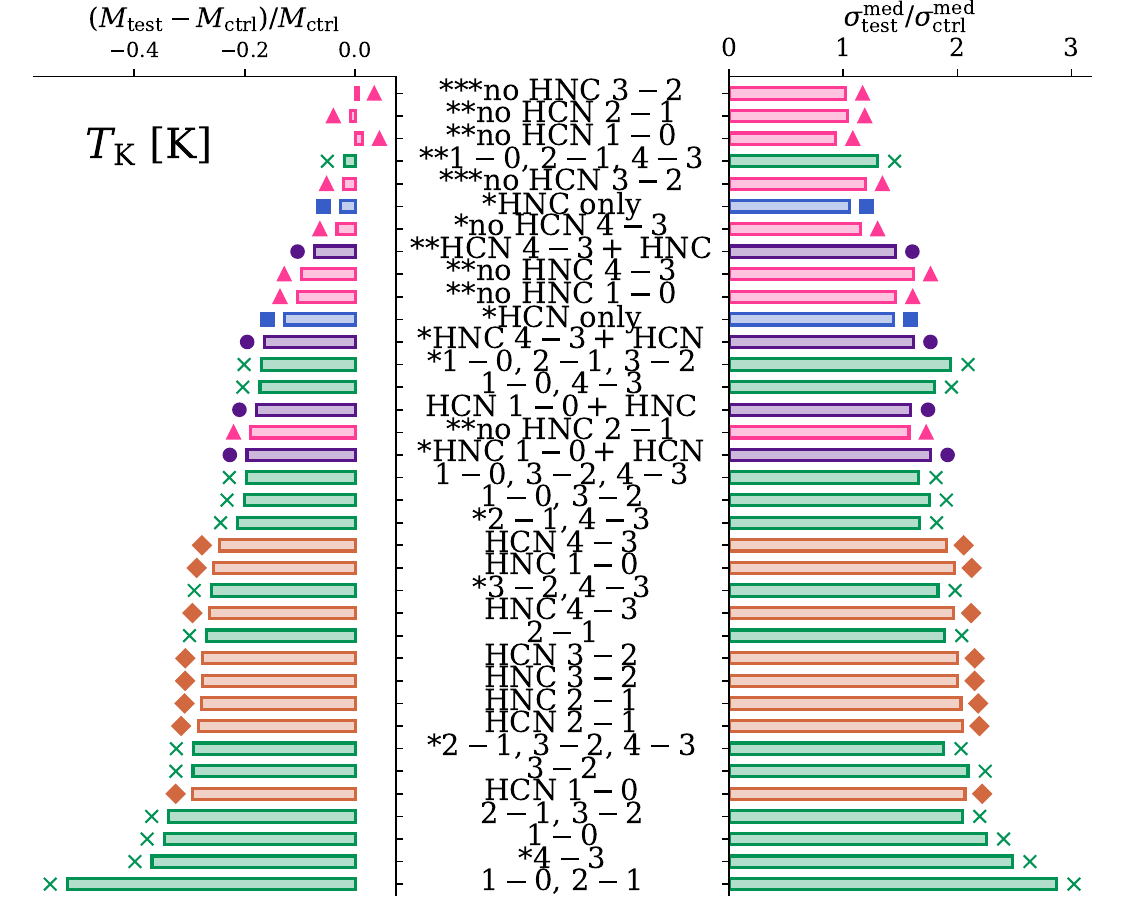}{0.5\textwidth}{}
    \rightfig{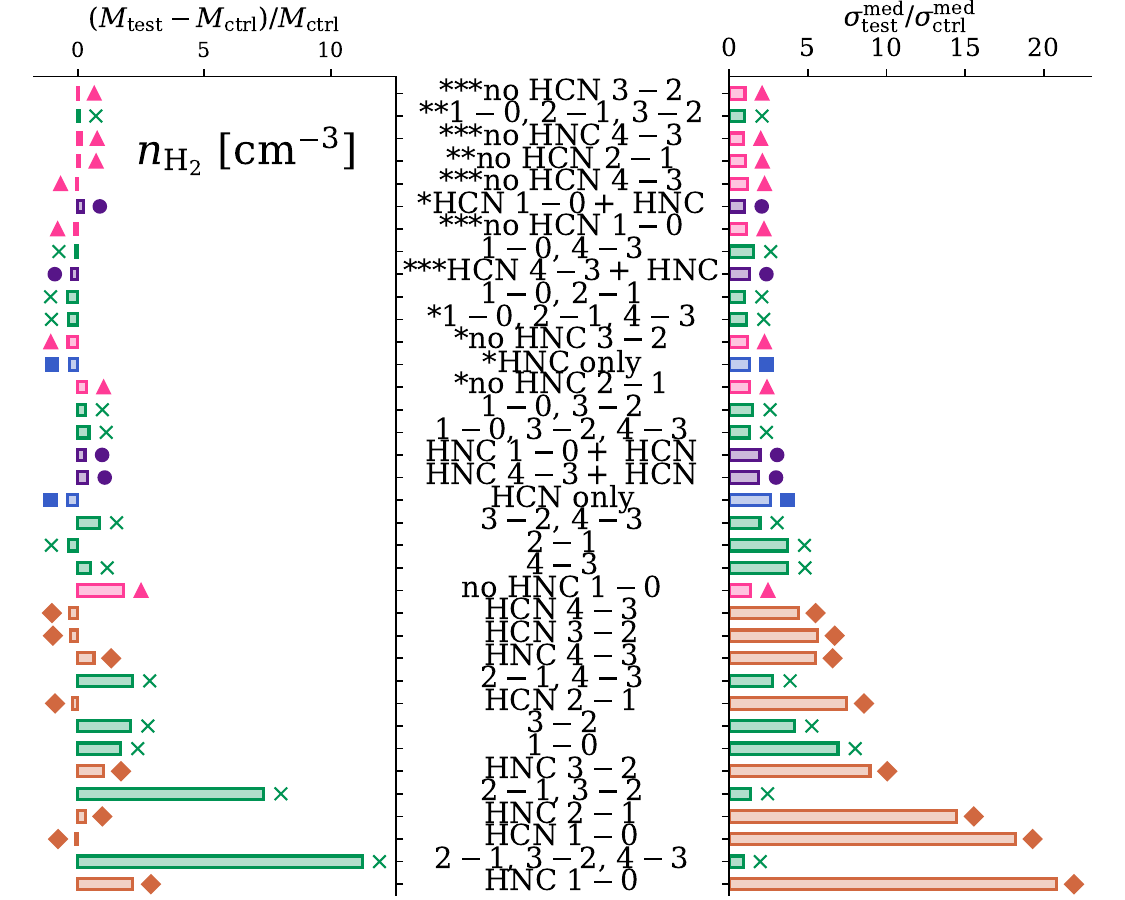}{0.5\textwidth}{}
    }
    \vspace{-9mm}
    \gridline{
    \leftfig{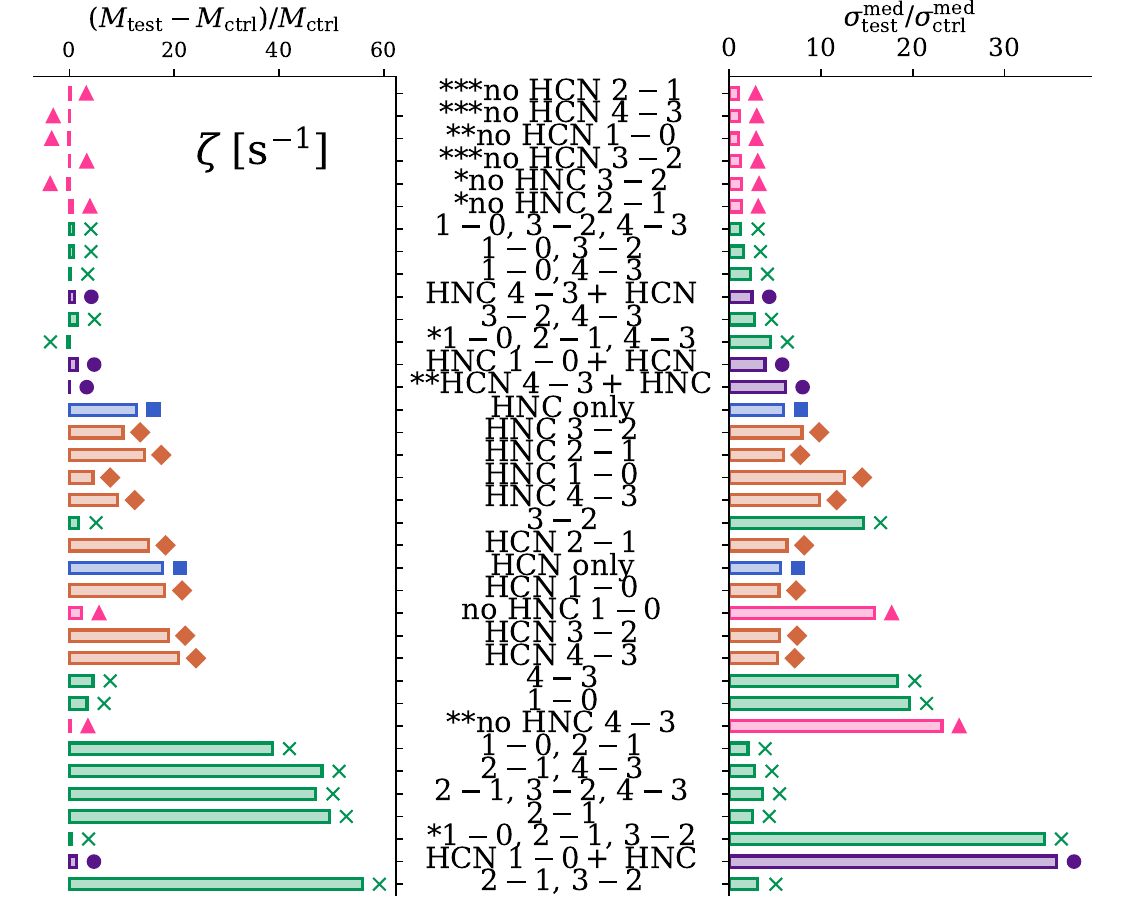}{0.5\textwidth}{}
    \rightfig{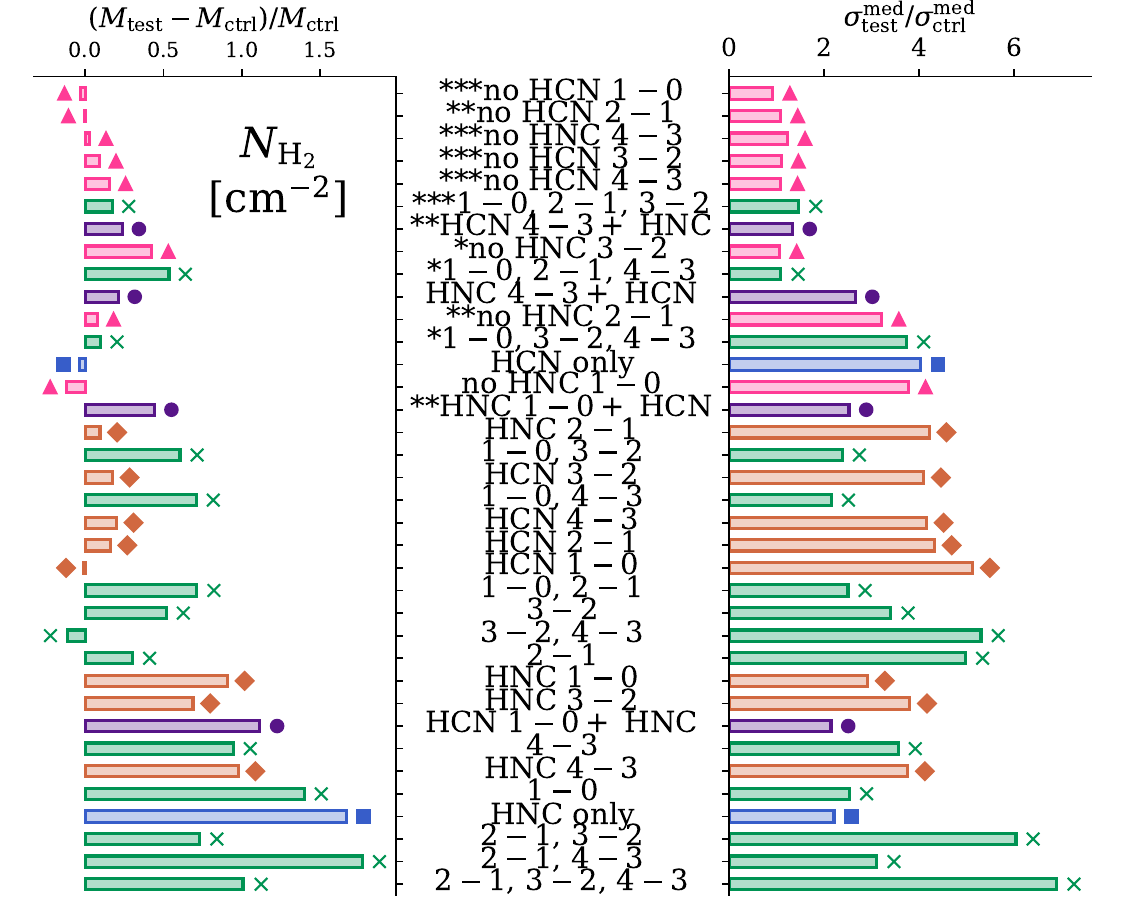}{0.5\textwidth}{}
    }
    \vspace{-9mm}
    \gridline{
    \leftfig{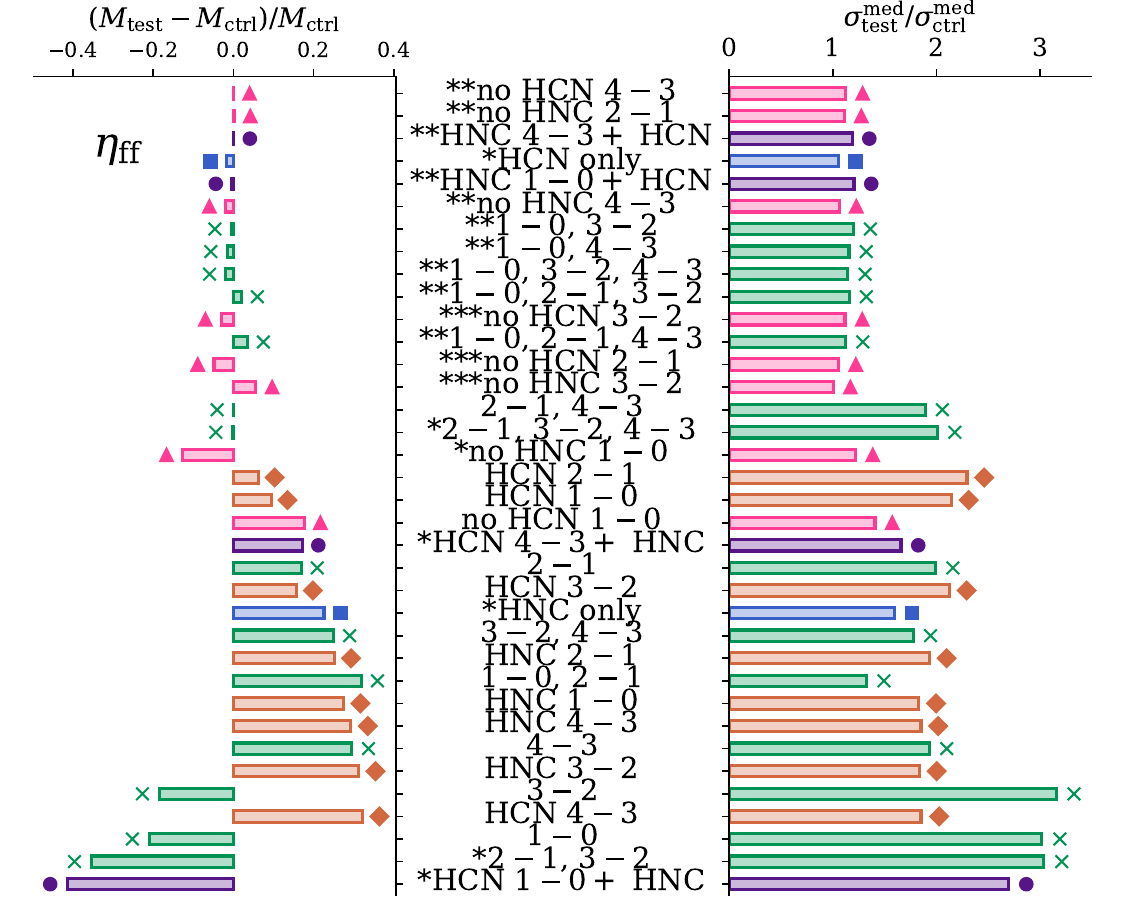}{0.5\textwidth}{}
    \rightfig{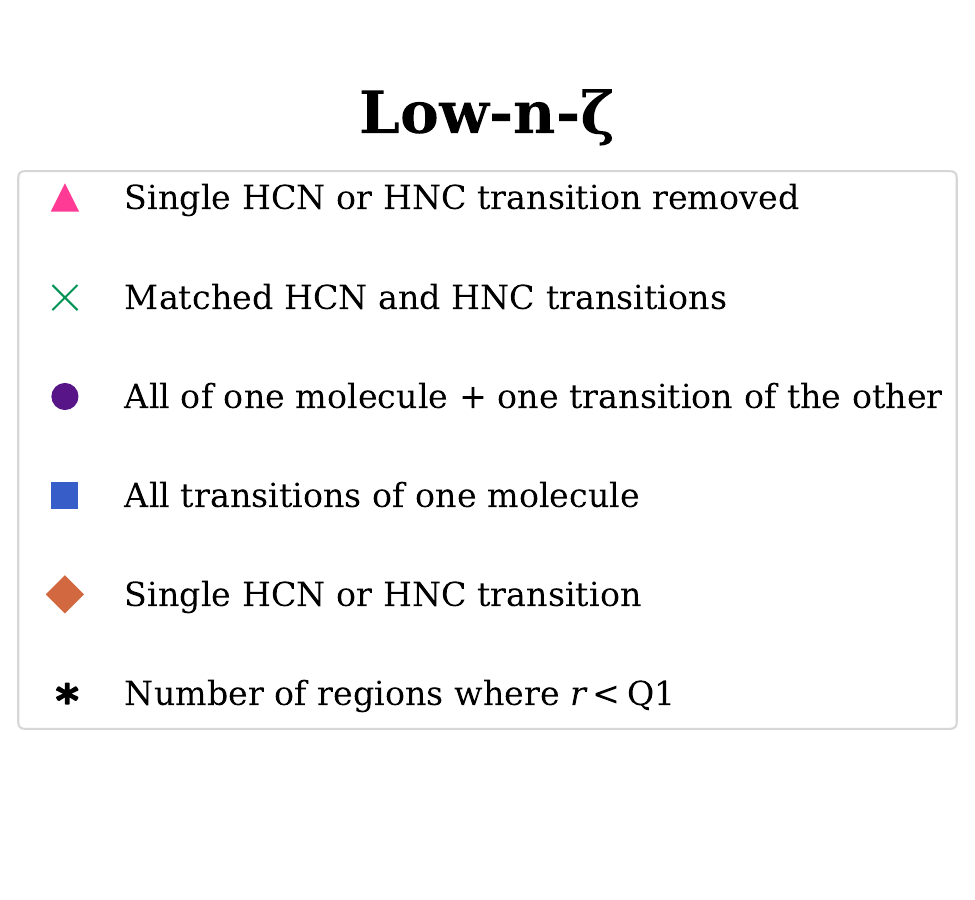}{0.4\textwidth}{}
    }
    \vspace{-8mm}
    \caption{Fractional difference between the test and control medians for each test combination as well as ratio of test to control uncertainty in the low-$n$-$\zeta$ region for $T_{\text{K}}$  (top left), $n_{\text{H}_{2}}$ (top right), $\zeta$ (middle left), $N_{\text{H}_2}$ (middle right), and $\eta_{\text{ff}}$ (bottom left). Number of asterisks (*) next to each test combination represents the number of regions for which that combination's calculated residual value was below Q1.}
    \label{fig:bars26}
\end{figure*}

\begin{figure*}
    \gridline{
    \leftfig{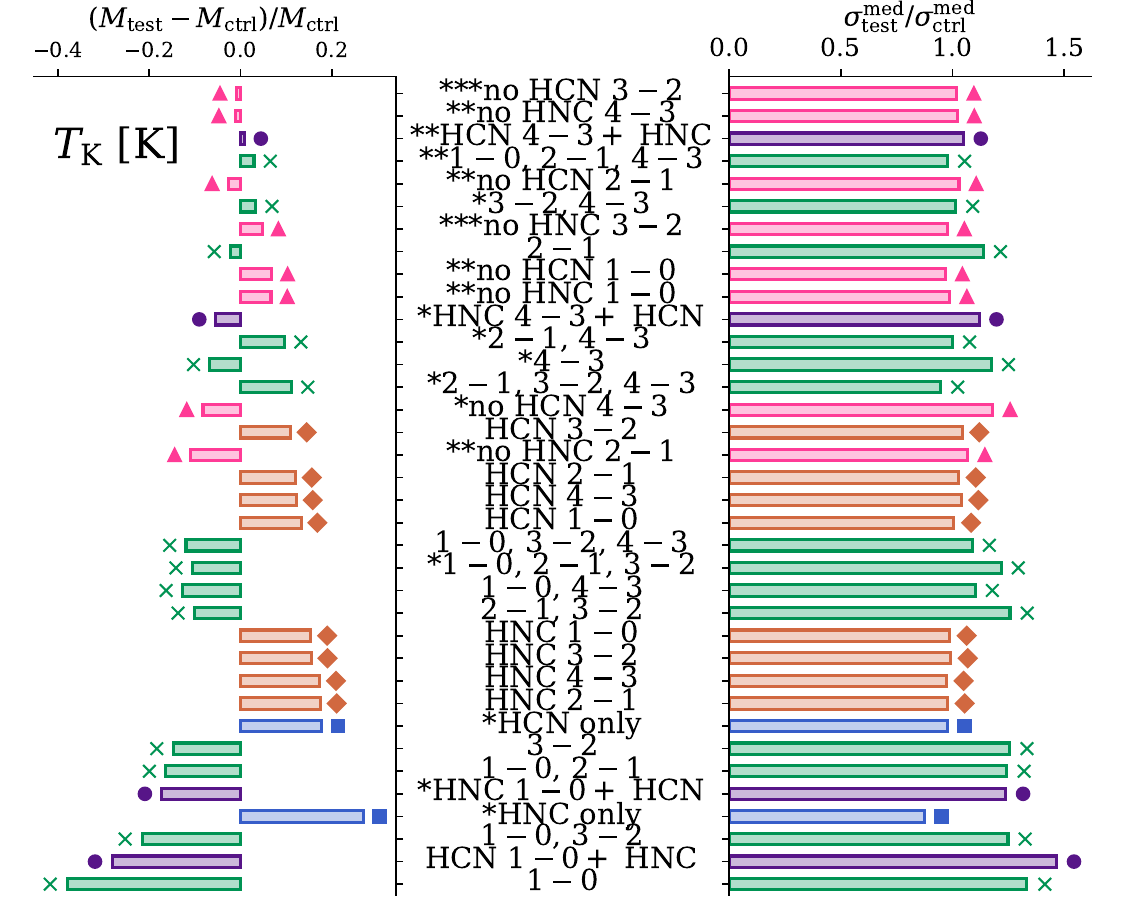}{0.5\textwidth}{}
    \rightfig{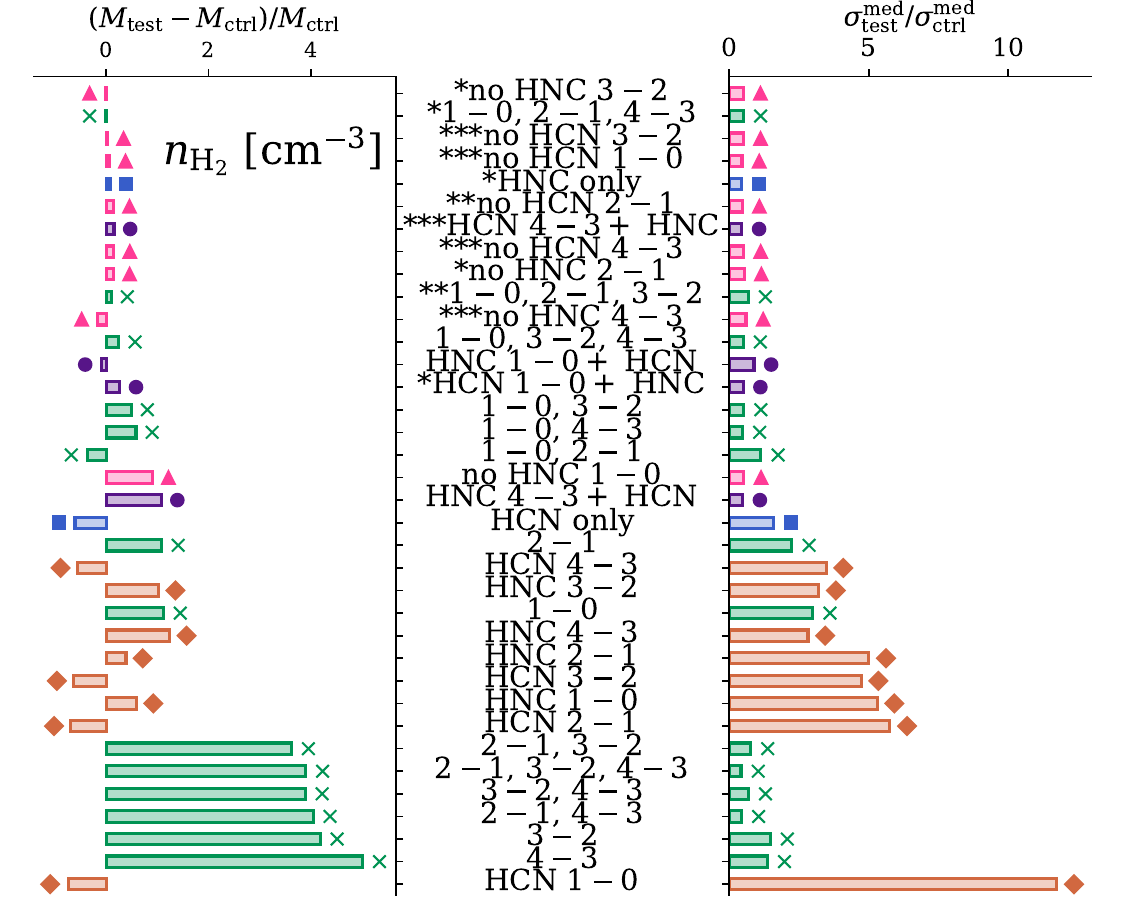}{0.5\textwidth}{}
    }
    \vspace{-9mm}
    \gridline{
    \leftfig{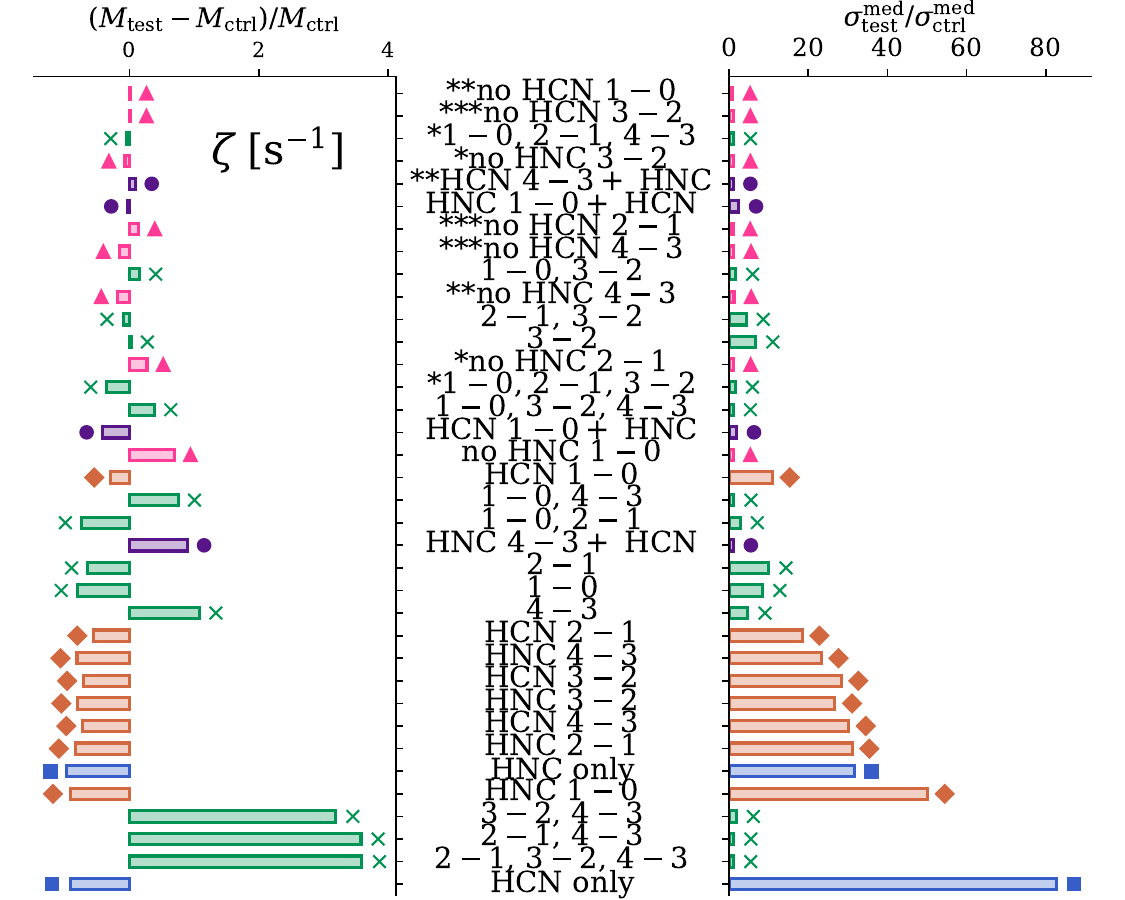}{0.5\textwidth}{}
    \rightfig{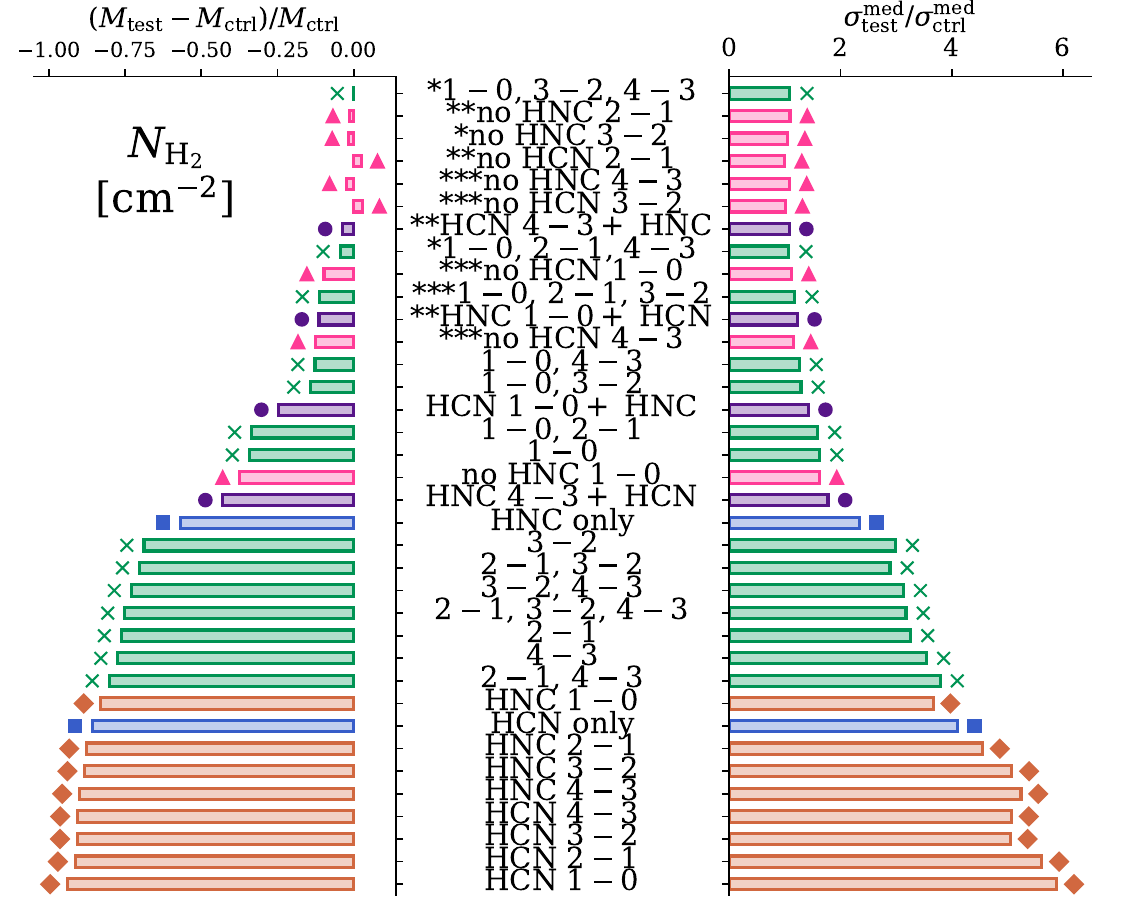}{0.5\textwidth}{}
    }
    \vspace{-9mm}
    \gridline{
    \leftfig{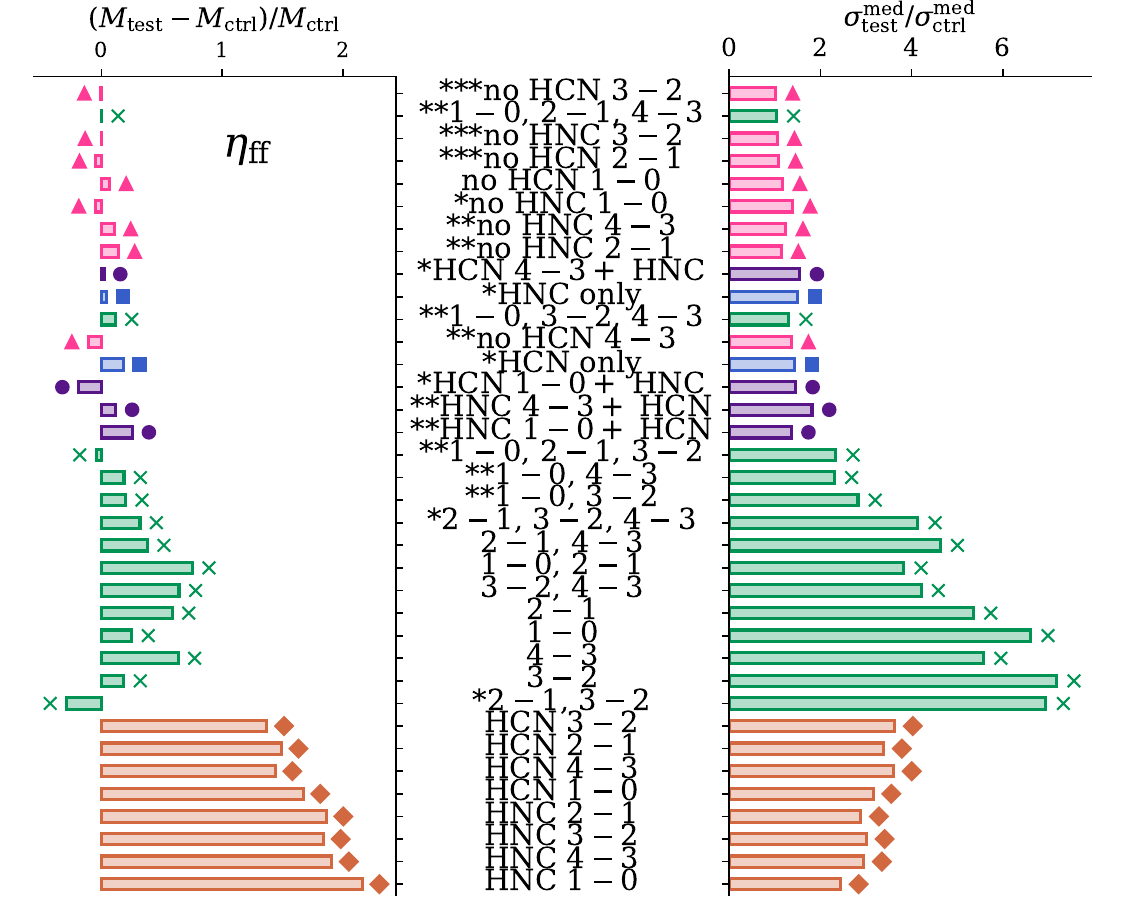}{0.5\textwidth}{}
    \rightfig{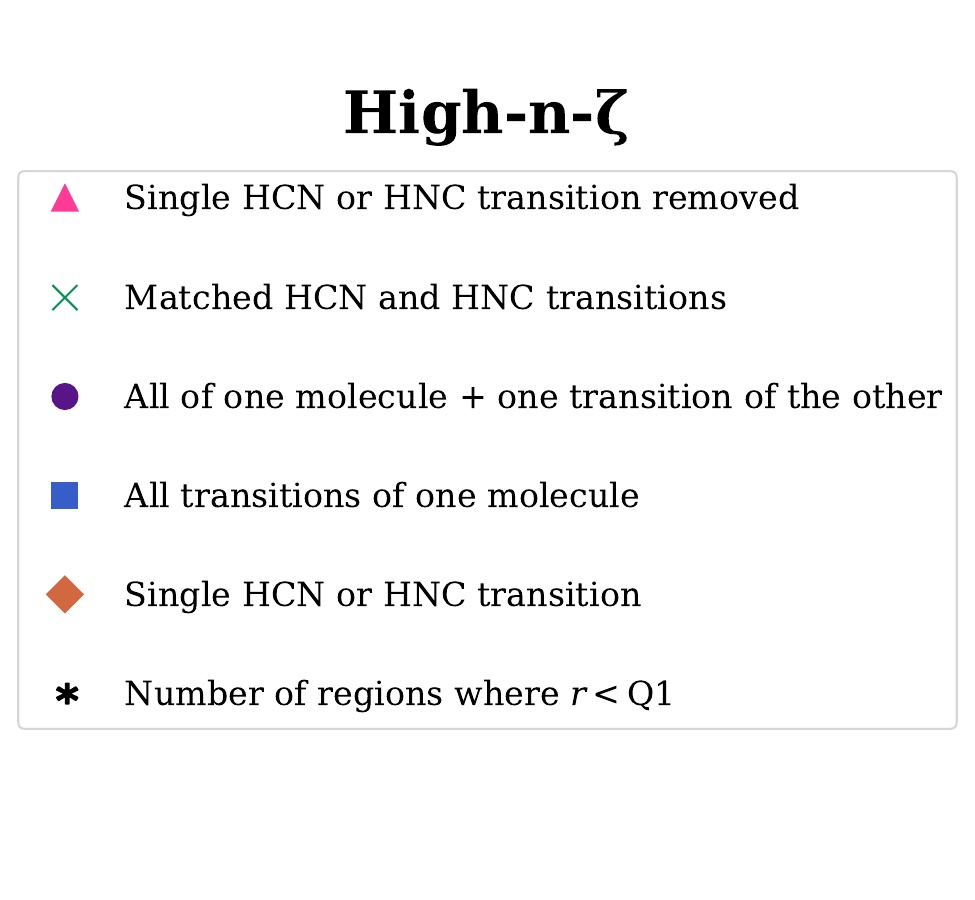}{0.4\textwidth}{}
    }
    \vspace{-8mm}
    \caption{Same as in Figure~\ref{fig:bars26} but for the high-$n$-$\zeta$ region.}
    \label{fig:bars53}
\end{figure*}

\begin{figure*}
    \gridline{
    \leftfig{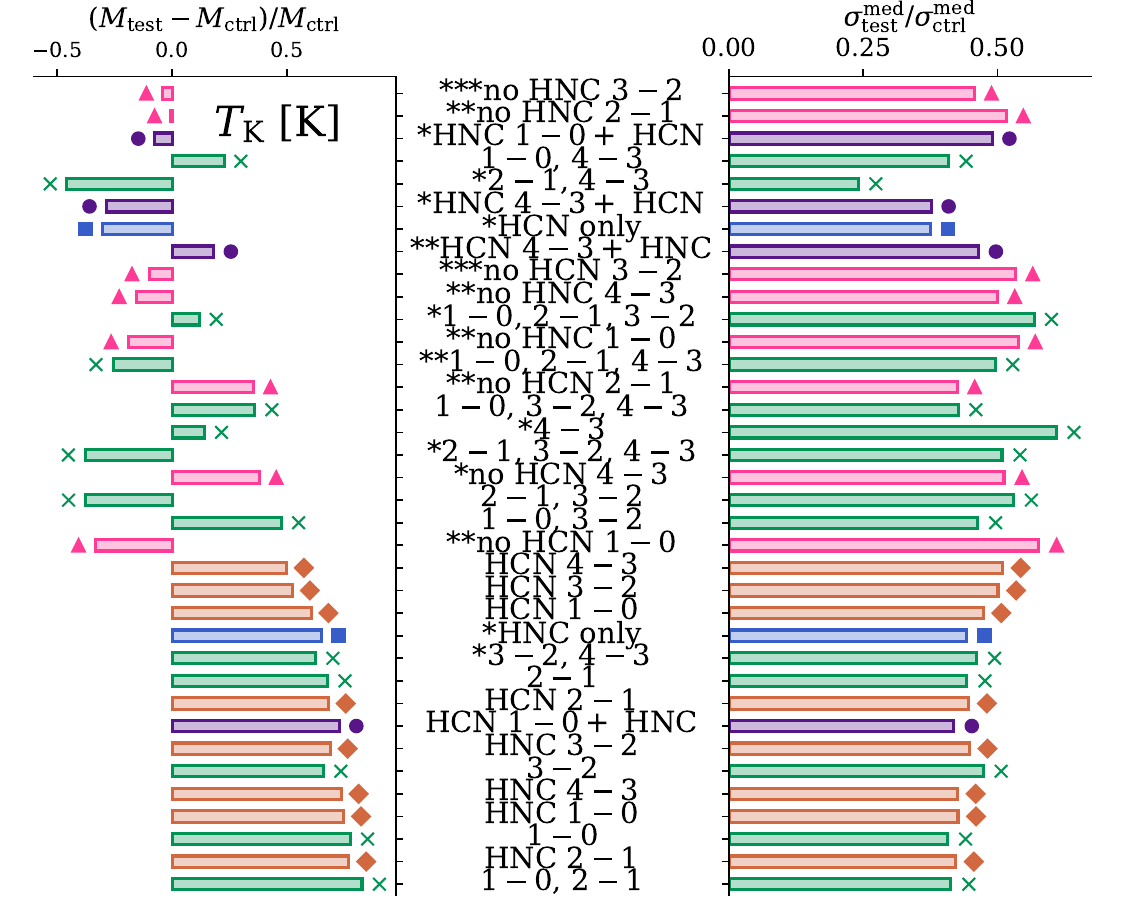}{0.5\textwidth}{}
    \rightfig{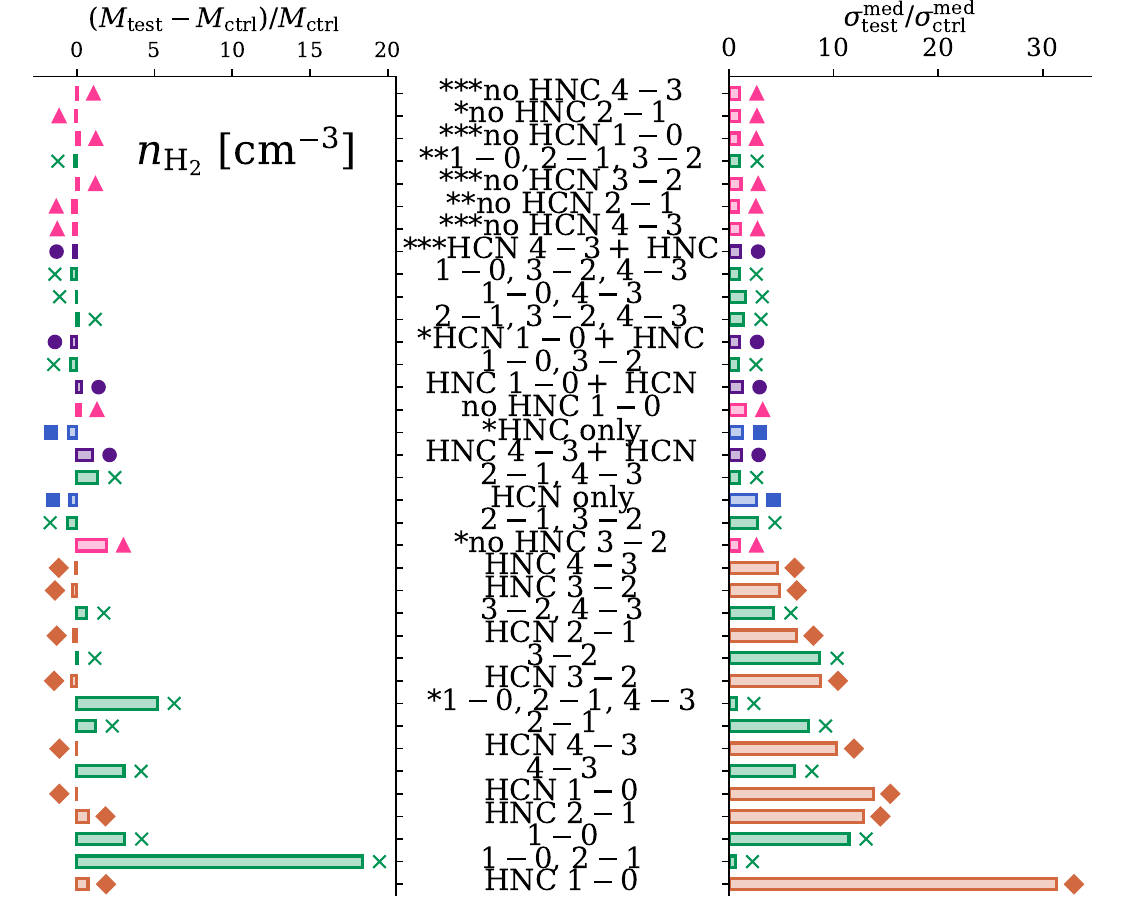}{0.5\textwidth}{}
    }
    \vspace{-9mm}
    \gridline{
    \leftfig{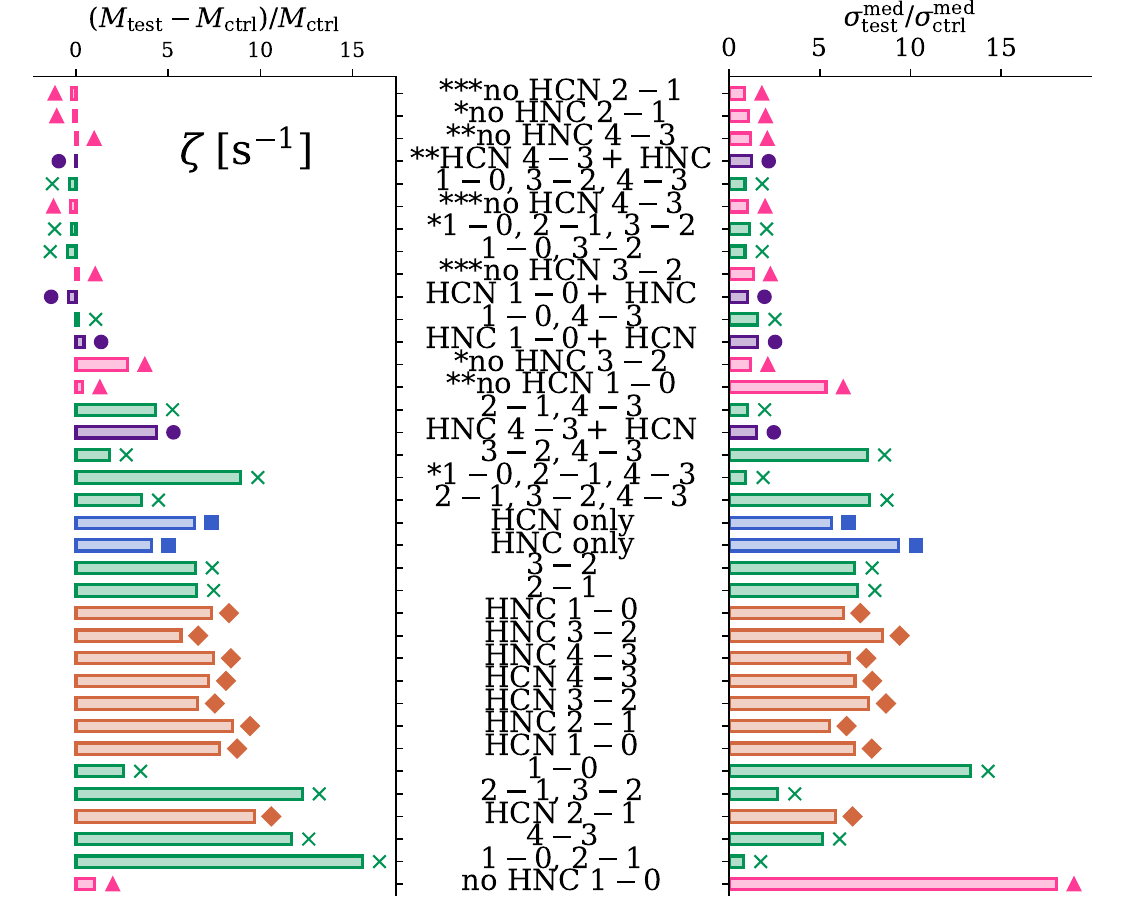}{0.5\textwidth}{}
    \rightfig{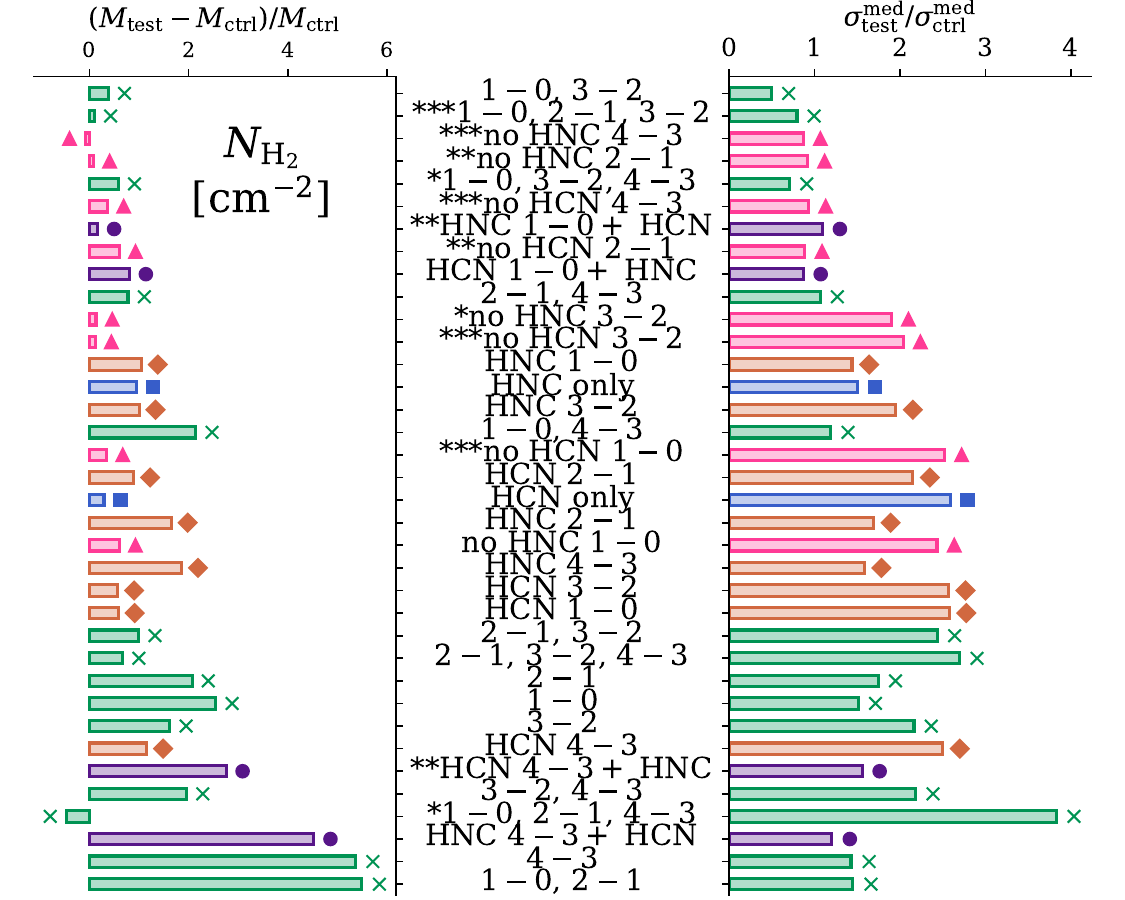}{0.5\textwidth}{}
    }
    \vspace{-9mm}
    \gridline{
    \leftfig{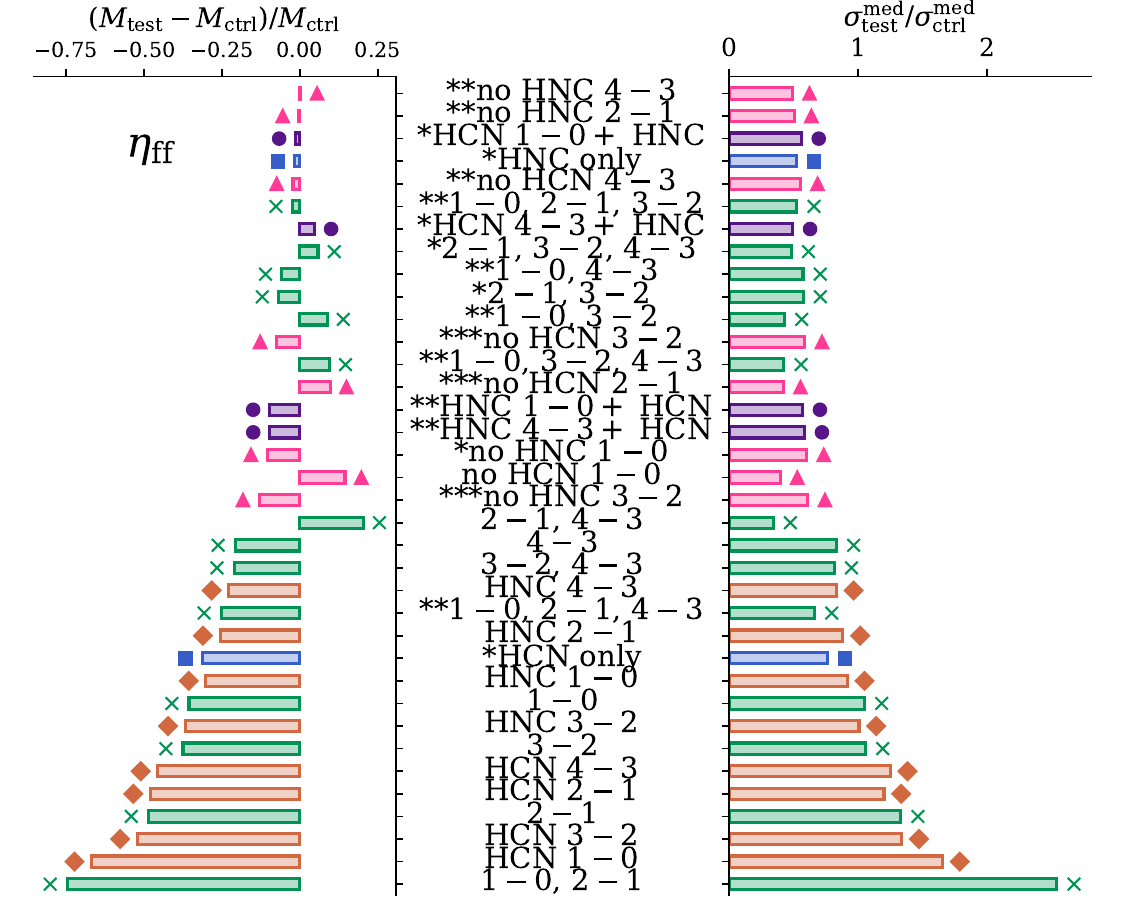}{0.5\textwidth}{}
    \rightfig{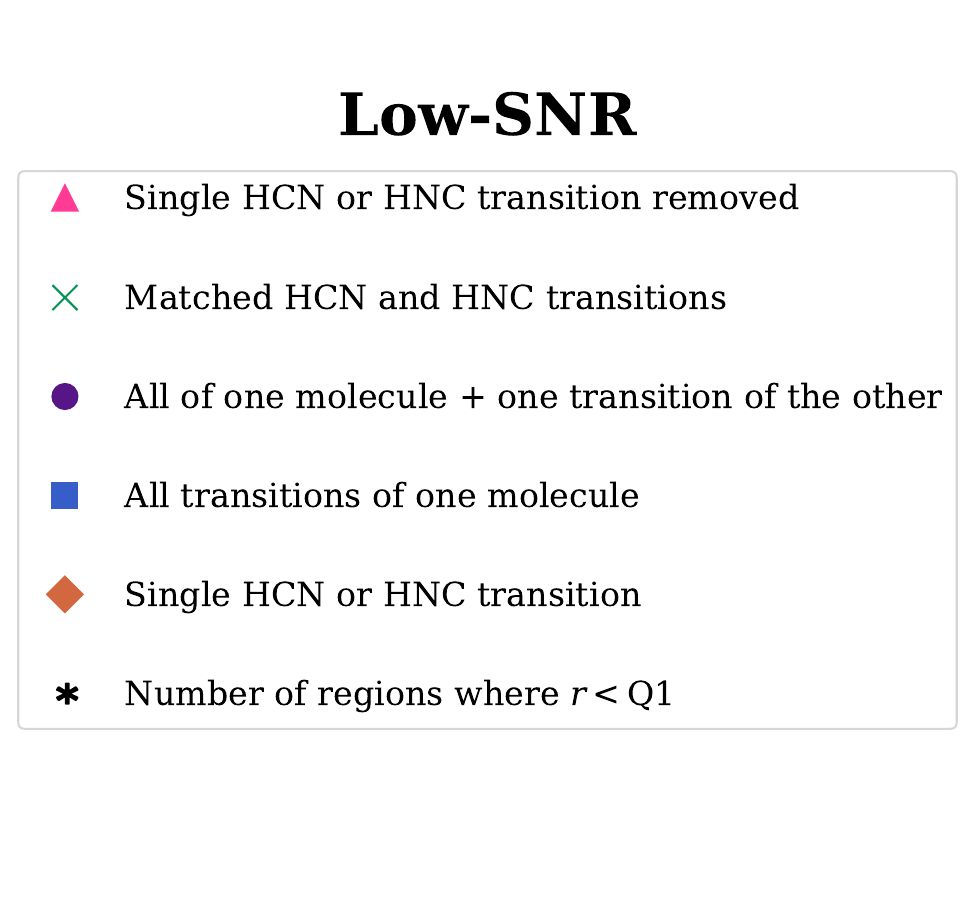}{0.4\textwidth}{}
    }
    \vspace{-8mm}
    \caption{Same as in Figures~\ref{fig:bars26} and \ref{fig:bars53} but for the low-SNR region.}
    \label{fig:bars51}
\end{figure*}

\subsection{Quantifying the Efficacy of Transition Constraints Tests} \label{sec:quant}

In Section~\ref{sec:results} we described how well each category of transition constraints performed in reproducing the control gas parameter posterior distributions that utilize all eight HCN and HNC transitions measured toward NGC\,253 by the ALCHEMI Large Program. To aid observers in choosing the transition combinations that suit their needs, we consider an additional metric for evaluating the performance of each test case and compare this metric to the results obtained by calculating posterior distribution residuals.

This metric separates the accuracy (how close a test median is to the control median) and precision (how much larger the test uncertainty is than the control) of the posterior distribution results for each test case so that observers may choose the HCN and HNC combinations most suited to their science case. We first calculate the fractional difference between the median values of each gas parameter derived from each test case, $M_{\text{test}}$, to the median gas parameter values derived from the control distributions, $M_{\text{ctrl}}$, for each of the three regions. We calculate the fractional difference between $M_{\text{test}}$ and $M_{\text{ctrl}}$, which is given by $(M_{\text{test}} - M_{\text{ctrl}}) / M_{\text{ctrl}}$. Note that we calculate these values in linear space, even for the three gas parameters for which we use log-uniform distributions in our modeling ($n_{\text{H}_2}$, $\zeta$, and $N_{\text{H}_2}$). 

To evaluate the precision of each test case in constraining our five gas parameters, we consider the ratio of the median-weighted test uncertainty to the median-weighted control uncertainty, $\sigma^{\text{med}}_{\text{test}}/\sigma^{\text{med}}_{\text{ctrl}}$, where $\sigma^{\text{med}} \equiv \sigma / M$ for both the control and test fit uncertainties. As with the median fractional difference, we calculate the median-weighted uncertainty ratio values within each of the three regions for each gas parameter and test case. The results of both the accuracy and precision metrics are shown in Figures~\ref{fig:bars26}, \ref{fig:bars53}, and \ref{fig:bars51}. We have ordered the transition combination test cases according to their total bar length (see Figures \ref{fig:bars26}, \ref{fig:bars53}, and \ref{fig:bars51}), which is given by

\begin{equation}
    \text{Tot. bar length} = \frac{\Delta M / M_{\text{ctrl}}}{(\Delta M / M_{\text{ctrl}})_{\text{max}}} + \frac{\sigma^{\text{med}}_{\text{test}}/\sigma^{\text{med}}_{\text{ctrl}}}{(\sigma^{\text{med}}_{\text{test}}/\sigma^{\text{med}}_{\text{ctrl}})_\text{max}},
\end{equation}

\noindent where $\Delta M$ is the difference between the test and control medians, and $(\Delta M / M_{\text{ctrl}})_{\text{max}}$ and $(\sigma^{\text{med}}_{\text{test}}/\sigma^{\text{med}}_{\text{ctrl}})_\text{max}$ are the maximum values for the median and uncertainty metrics for a given gas parameter and region. The test cases with the smallest total bar lengths, and thus the closest analogs to the control distributions according to these metrics, are at the top of each panel, while the worst cases are located at the bottom.

These metrics, presented in Figures~\ref{fig:bars26}, \ref{fig:bars53}, and \ref{fig:bars51}, largely echo the results presented in Section~\ref{sec:results}. 
As noted earlier, we do not constrain kinetic temperature well in any of these regions even with the control. Nevertheless, the results from the temperature panels reflect the results seen in other parameters. The best analogs to the controls are unsurprisingly the combinations where only a single transition has been removed. In the case of volume density, the top half of transition combinations will generally yield a median parameter value within a factor of two and an uncertainty less than five times that of the control. The worst cases for volume density constraint consist mostly of matched transitions (\textcolor{SpringGreen4}{\ding{53}}) that either did not include the $J=1-0$ transitions or poorly sampled the upper-state energy ladder and single transitions (\textcolor{Sienna3}{$\blacklozenge$}). These combinations may still yield a median value within a factor of 2--5 of the control, but the uncertainties are much larger, 5--10 times larger than the control uncertainties. For the CRIR, the results worsen significantly when only one molecule is used to constrain our models. Combinations that use a single transition (\textcolor{Sienna3}{$\blacklozenge$}) or single molecule (\textcolor{RoyalBlue3}{$\blacksquare$}) either overestimate the median by a factor of $\sim20$, in the case of the low-$n$-$\zeta$ region (Figure~\ref{fig:bars26}), or have an uncertainty 20--50 times greater than that of the control, as is the case in the high-$n$-$\zeta$ region (Figure~\ref{fig:bars53}). Most transition combinations yield a column density within a factor of two of the control median with uncertainties 1--6 times that of the control. Finally, we see in the beam-filling factor plots that across all regions, using single transitions or matched transition combinations that do not appropriately sample the transition energy ladder result in poorer estimations of the beam-filling factor but still yield median values within a factor of 2--3 of the control median. 

To tie in the results of Section~\ref{sec:results}, we also show in Figures~\ref{fig:bars26}, \ref{fig:bars53}, and \ref{fig:bars51} the number of regions for which a given test combination yielded residuals below the Q1 threshold for each parameter using asterisks (*). For instance, in the $\zeta$ panel of Figure~\ref{fig:bars26}, the test combination that used all HCN and HNC transitions excluding the HCN $3-2$ yielded a residual below Q1 in all three regions for the CRIR, while the combination that used the $1-0$, $2-1$, and $4-3$ HCN and HNC transitions only resulted in residuals below the Q1 threshold in one region. Combinations near the top of a given panel in Figures~\ref{fig:bars26}, \ref{fig:bars53}, and \ref{fig:bars51} with more than one asterisk are likely to be the most consistently useful across different environments when constraining a given parameter.

\subsection{Signal-to-Noise Ratio} \label{sec:SNR}

Our results demonstrate that parameter values inferred by the HCN and HNC measurements toward the low-SNR region are significantly less stable than those inferred toward the low- and high-$n$-$\zeta$ regions. As noted in Section~\ref{sec:regions}, we included the low-SNR region in our analysis to determine the impacts of molecular constraints with low SNRs on our parameter inferencing results. Five HCN and HNC transitions were detected toward the low-SNR region in NGC\,253 with SNR $\geq$ 3: HCN $J=1-0$, $2-1$, and $3-2$ and HNC $J=2-1$, and $3-2$. These transitions were detected with SNRs of $\sim 6.6$, $6.1$, and $5.9$ for HCN, respectively, and $\sim6.6$ and 3.0 for HNC. Note that the maximum SNR of 6.6 across all transitions studied here is set by our flux calibration uncertainty of 15\% $=1/6.6$. 

We note in Section~\ref{sec:results} that the trends seen in the low- and high-$n$-$\zeta$ regions are not always consistent with those seen in the low-SNR region. When discussing the matched HCN and HNC transition combinations in Section~\ref{sec:matched}, we note that some of the combinations that prove to be the best in the other regions do not always yield low residuals in the low-SNR region. While matched transitions of $J=1-0$, $2-1$, and $3-2$ perform well, $J=1-0$, $2-1$, and $4-3$ perform quite poorly. This result is likely due to the fact that in the low-SNR region, this combination contains no transitions with higher energies than the $J=2-1$ that have SNR $>$ 3, as the HCN and HNC $J=4-3$ transitions were not detected at this threshold and thus were replaced with SNR limits. In the case of the HCN and HNC $J=1-0$, $2-1$, and $3-2$ combination, the HCN $J=3-2$ transition was detected with a SNR of $\sim6$ and thus provides the parameter inferencing algorithm with some reliable information on the intensity of higher-energy transitions.

\subsection{HCN versus HNC as gas condition tracers} \label{sec:HCNvHNC}

Our results indicate that the necessity for using HCN, HNC, or the two species together to infer gas conditions in star-forming environments may be dependent on the specific parameter one desires to constrain. The results presented in Section~\ref{sec:results} and Figures~\ref{fig:bars26}, \ref{fig:bars53}, and \ref{fig:bars51} indicate that for volume density, HNC may be a more useful constraint than HCN. Figures~\ref{fig:bars26}, \ref{fig:bars53}, and \ref{fig:bars51} show that for volume density, HNC by itself (\textcolor{RoyalBlue3}{$\blacksquare$}) ranks higher than HCN by itself across all three regions. However, both combinations will result in a median volume density within a factor of two of the control median with comparable uncertainty relative to the control. Additionally, in the cases where all transitions of one species are supplemented with a single transition of the opposing species (\textcolor{Purple4}{\CIRCLE}), using majority HNC transitions is more effective for constraining the gas volume density. Figure~\ref{fig:bars26} shows that for the low-$n$-$\zeta$ region, HNC supplemented with the HCN $J=1-0$ and then the $J=4-3$ rank sixth and ninth, respectively, overall for gas density. Furthermore, the combination with four HNC transitions + HCN $4-3$ results in a residual below Q1 in all three regions for volume density (Figure~\ref{fig:reg26_resids}). On the other hand, HCN supplemented with the $1-0$ and $4-3$ HNC transitions rank seventeenth and eighteenth, respectively, in the low-$n$-$\zeta$ region. These results indicate that when using an unequal number of constraints from HCN and HNC, having more HNC transitions is beneficial for inferring the gas density. Note, however, that supplementing one species' four transitions with one transition of the other species almost always improves the density constraints compared to only using a single species.

HNC's utility for constraining gas conditions may lie in its chemistry. While HCN has been frequently used as a dense gas tracer due to its high critical density \citep{Gao2004,Shirley2015}, some studies have found that the lower-$J$ HCN transitions may actually trace more diffuse gas \citep[$n\lesssim10^4$\,cm$^{-3}$][]{Jones2023}. Additionally, several recent studies have suggested that the HCN abundance can be influenced by star-formation feedback mechanisms such as UV \citep{Santa-Maria2023} and cosmic-ray \citep{Behrens2022,Behrens2024} ionization. {While HCN is largely formed at low CRIR via neutral-neutral reactions with temperature barriers of at least 50\,K, higher CRIRs cause HCN to form via recombination reactions involving cosmic-ray electrons, resulting in feedback-dependent formation paths \citep{Behrens2022}. Due to these additional factors, HCN's role as a dense gas tracer may not be as straightforward. While HNC is also known as a dense gas tracer \citep{Aalto2002,Eibensteiner2022,Imanishi2023}, it may not be affected as strongly by feedback mechanisms. The routes to form HNC at low CRIRs are far less temperature-dependent than those that form HCN \citep{Behrens2022}, suggesting that HNC will be less affected by feedback mechanisms. Assuming that the reaction networks in our chemical model \citep[derived from the UMIST12 database,][]{McElroy2013} are complete, our results indicate that HNC is a more reliable tracer of the gas density than HCN.



\subsection{Sampling the Energy Ladder} \label{sec:energyLadder}

Our results indicate that constraining chemical and radiative transfer models with molecular transitions that span a range of upper-state energies $E_{\text{u}}$ generally improves the constraints as compared to only using either low- or high-energy transitions. The HCN and HNC transitions range in $E_{u}$ from $\sim4-43$\,K \citep[Table~\ref{tab:obs}; ][]{Saykally1976,Ahrens2002}. When constraining gas volume density, CRIR, and the beam-filling factor, Figures~\ref{fig:bars26}, \ref{fig:bars53}, and \ref{fig:bars51} indicate that a combination of the $J=1-0$ HCN and HNC transitions with either or both of the $J=3-2$ and $J=4-3$ transitions are very effective and, in the majority of cases, constrain the median value of these gas parameters within a factor of two. As discussed in Section~\ref{sec:matched}, regions with lower gas densities (relative to the high gas densities of $\sim10^5$\,cm$^{-3}$ that we see in the NGC 253 starburst nucleus) may be more easily characterized with the $J=3-2$ versus the $J=4-3$ transitions due to their critical densities. Because critical densities assume a fully optically thin environment, which is likely not the case for HCN and HNC in NGC\,253, we will adopt the nomenclature used in \cite{Shirley2015} and instead refer to the effective excitation density $n_{\text{eff}}$, which takes into account radiative trapping. $n_{\text{eff}}$ is dependent on the gas temperature, and though we cannot constrain kinetic temperature with HCN and HNC, other studies suggest that the temperature is $\sim80-100$\,K at distances greater than $\sim100$\,pc from the center of the CMZ and up to $200$\,K at the nucleus \citep{mangum19,Tanaka2024}. Under these conditions, the maximum $n_{\text{eff}}$ for HCN $J=1-0$, $2-1$, $3-2$, and $4-3$ would be $2.6\times10^{3}$, $7.3\times10^{3}$, $2.5\times10^4$, and $8.1\times10^4$\,cm$^{-3}$, respectively, with similar $n_{\text{eff}}$ for HNC \citep{Shirley2015}. 

Given that the $n_{\text{H}_{2}}$ control median for the low-$n$-$\zeta$ region is $\sim10^4$\,cm$^{-3}$, it is not surprising that using the $J=3-2$ HCN and HNC transitions instead of the $J=4-3$ provides more information on the gas density in this region. Similarly, the $J=4-3$ transitions are more useful in the high-$n$-$\zeta$ region due to its higher median density of $\sim10^5$\,cm$^{-3}$. However, it is important to note that these transitions by themselves may not be very useful for determining the density. Though single transitions appear to constrain the gas density within, at times, a factor of 2 or 3 from the control median in Figures~\ref{fig:bars26}, \ref{fig:bars53}, and \ref{fig:bars51}, the uncertainty on these estimates is often large, 4--10 times the control uncertainty. With such a large uncertainty range, the density remains essentially unconstrained. Supplementary corner plots in the Appendix (Figure~\ref{fig:singleTcorners}) also show that the posterior distributions for gas density when constrained with single transitions often peak at one end of our prior distribution range, which casts some doubt on the validity of these constraints. 

Single transitions may not be the most effective to characterize gas parameters because the intensity balance of low- and high-$E_{\text{u}}$ transitions is important and informs the models how each energy level is populated relative to other levels, which can be reflective of the gas density. Additionally, the gas we study is not a single homogeneous structure. Though we derive a single value for the gas density in a given region, this number is simply an average over what is likely several gas components. Lower-$E_{\text{u}}$ transitions may trace a more diffuse gas component, possibly being emitted from a larger area, whereas the higher-$E_{\text{u}}$ transitions are associated with smaller, denser components. In extragalactic astrochemistry, where even the highest resolution measurements are averaging over several gas components, it is critical to obtain measurements of multiple transitions that span the $J$-ladder in order to fully characterize the gas.

We find that the gas we are probing has an important contribution from the $J=1-0$ transitions. As mentioned in Section~\ref{sec:matched}, matched pairs of HCN and HNC transitions that do not include the HCN or HNC $1-0$ transitions perform quite poorly. Figures~\ref{fig:bars26}, \ref{fig:bars53}, and \ref{fig:bars51} show that matched transition combinations without the $1-0$ almost exclusively fall in the bottom half of the tested combinations. For example, the matched case (\textcolor{SpringGreen4}{\ding{53}}) that includes the HCN and HNC $2-1$, $3-2$, and $4-3$ transitions is one of the worst tested combinations for volume density and CRIR, despite containing six transitions. We find that models that are not constrained by the $1-0$ transitions often severely underestimate the $1-0$ emission. In this case, the model estimates the HCN $1-0$ integrated intensity to be 190 K\,km\,s$^{-1}$, while the measured value is 285\,$\pm$\,43\,K\,km\,s$^{-1}$. This underestimation clearly impacts the inferred parameter results, causing the algorithm to infer higher volume densities and CRIRs than it would otherwise. In the regions studied here, the HCN and HNC $J=1-0$ transitions clearly contribute significantly to the overall gas emission.

\section{Conclusion}

We examine the efficacy of various combinations of HCN and HNC transitions in constraining kinetic temperature, H$_2$ volume density, cosmic-ray ionization rate (CRIR), H$_2$ column density, and a beam-filling factor in three representative 50-pc regions in the NGC\,253 CMZ. We calculate residuals of the the posterior distributions for all the test combinations of HCN and HNC transitions as compared to controls, or the posterior distributions obtained when using the $J=1-0$, $2-1$, $3-2$, and $4-3$ transitions of both HCN and HNC. Our conclusions are as follows.

\begin{itemize}
    \item Single transitions are not effective at constraining gas parameters and generally fall into the worst quartile of combinations we study here. While they may yield volume median densities within a factor of 2-5 of the control, uncertainties on these estimates can be larger by a factor of 5-10. 
    \item Using single transitions or a single molecule (HCN or HNC) to constrain the CRIR results in median CRIRs different from the control by up to an order of magnitude. Single-transition or single-molecule constraints also result in CRIR uncertainties that are 10-50 times greater than that of our control models.
    \item Including multiple transitions that span a range of upper-state energies is important when inferring gas parameters, especially volume density and CRIR. The intensity balance of these transitions provides constraints for the models, especially when measurements  average over a large area that likely contains multiple gas components. In particular, combinations that include the $J=1-0$ transitions and either the $J=3-2$ or $J=4-3$ transitions usually constrain the gas parameters within a factor of 2 of the control medians.
    \item Most transition combinations yield molecular hydrogen column densities within a factor of two of the control median, with uncertainties 1--6 times that of the control. 
    \item For the beam-filling factor, most models derive median values within a factor of 2--3 of the control median.  Uncertainties on the beam-filling factor derivations are generally less than 3 times the control model uncertainty, but can be as large as 7 times the control uncertainty for single or matched HCN and HNC transition models.
    \item In the case of HCN and HNC, transitions from both species with a range of upper-state energies are required to constrain the CRIR. Using a single species or transitions from multiple species with a limited range in upper-state energies results in poor constraints on the CRIR, with the median CRIR different from the control by up to an order of magnitude.
    \item HNC is a more direct tracer of gas density than HCN, as HCN emission can be influenced by factors besides volume density (e.g. UV and cosmic-ray ionization). On the other hand, gas density is the main factor contributing to HNC abundances.
    \item We recommend that the quality metrics provided in Figures~\ref{fig:bars26}, \ref{fig:bars53}, and \ref{fig:bars51} be used by observers to design experiments which provide physical parameter constraints commensurate with the specific needs of their research projects.
\end{itemize}

These results show that in starburst environments that feature complex chemistry as a result of competing star-formation feedback mechanisms, multiple transitions of multiple molecular species are required in order to accurately and precisely constrain the conditions of dense gas. More work is needed to understand how these results might extend to non-starburst environments or those dominated by different chemical and feedback mechanisms, such as shocked or photodissociation regions.

\section*{Acknowledgments}

The authors thank R. Indebetouw for encouraging the pursuit of this project. The authors also thank K. D. Foster for productive conversations that have improved the quality of this work. This paper makes use of the following ALMA data: ADS/JAO.ALMA \#2017.1.00161.L and 2018.1.00162.S. ALMA is a partnership of ESO (representing its member states), NSF (USA) and NINS (Japan), together with NRC (Canada), MOST and ASIAA (Taiwan), and KASI (Republic of Korea), in cooperation with the Republic of Chile. The Joint ALMA Observatory is operated by ESO, AUI/NRAO and NAOJ. The National Radio Astronomy Observatory and Green Bank Observatory are facilities of the U.S. National Science Foundation operated under cooperative agreement by Associated Universities, Inc. Support for this work was provided by the NSF through the Grote Reber Fellowship Program administered by Associated Universities, Inc./National Radio Astronomy Observatory.

\begin{contribution}
    EB led the development of the modeling, analysis, interpretation of the results, development of effective presentation of those results, and writing of the article. JGM provided guidance regarding the interpretation of the results and overall project direction. MB, CE, and SV contributed to the interpretation and presentation of the results. SV provided guidance regarding the astrochemistry presented in this article.
\end{contribution}

\facility{ALMA}

\software{\texttt{Astropy} \citep{astropy:2013,astropy:2018,astropy:2022}, \texttt{CASA} \citep{CASATeam2022PASP}, \texttt{corner} \citep{corner}, \texttt{Nautilus} \citep{nautilus}, \texttt{SpectralRadex}\footnote{\url{https://spectralradex.readthedocs.io/en/latest/}}, \texttt{TensorFlow} \citep{tensorflow2015-whitepaper}, \texttt{UCLCHEM} \citep{Holdship2017}.}

\newpage
\bibliography{Behrens}{}

@ARTICLE{ALCHEMI-ACA,bibcode={2021A&A...656A..46M},
       author = {{Mart{\'\i}n}, S. and {Mangum}, J.~G. and {Harada}, N. and {Costagliola}, F. and {Sakamoto}, K. and {Muller}, S. and {Aladro}, R. and {Tanaka}, K. and {Yoshimura}, Y. and {Nakanishi}, K. and {Herrero-Illana}, R. and {M{\"u}hle}, S. and {Aalto}, S. and {Behrens}, E. and {Colzi}, L. and {Emig}, K.~L. and {Fuller}, G.~A. and {Garc{\'\i}a-Burillo}, S. and {Greve}, T.~R. and {Henkel}, C. and {Holdship}, J. and {Humire}, P. and {Hunt}, L. and {Izumi}, T. and {Kohno}, K. and {K{\"o}nig}, S. and {Meier}, D.~S. and {Nakajima}, T. and {Nishimura}, Y. and {Padovani}, M. and {Rivilla}, V.~M. and {Takano}, S. and {van der Werf}, P.~P. and {Viti}, S. and {Yan}, Y.~T.},
        title = "{ALCHEMI, an ALMA Comprehensive High-resolution Extragalactic Molecular Inventory. Survey presentation and first results from the ACA array}",
      journal = {\aap},
         year = 2021,
        month = dec,
       volume = {656},
          eid = {A46},
        pages = {A46},
          doi = {10.1051/0004-6361/202141567},
archivePrefix = {arXiv},
       eprint = {2109.08638},
 primaryClass = {astro-ph.GA},
       adsurl = {https://ui.adsabs.harvard.edu/abs/2021A&A...656A..46M}
}

@ARTICLE{leroy15,
       author = {{Leroy}, Adam K. and {Bolatto}, Alberto D. and {Ostriker}, Eve C. and {Rosolowsky}, Erik and {Walter}, Fabian and {Warren}, Steven R. and {Donovan Meyer}, Jennifer and {Hodge}, Jacqueline and {Meier}, David S. and {Ott}, J{\"u}rgen and {Sandstrom}, Karin and {Schruba}, Andreas and {Veilleux}, Sylvain and {Zwaan}, Martin},
        title = "{ALMA Reveals the Molecular Medium Fueling the Nearest Nuclear Starburst}",
      journal = {\apj},
         year = 2015,
        month = mar,
       volume = {801},
       number = {1},
          eid = {25},
        pages = {25},
          doi = {10.1088/0004-637X/801/1/25},
archivePrefix = {arXiv},
       eprint = {1411.2836},
 primaryClass = {astro-ph.GA},
       adsurl = {https://ui.adsabs.harvard.edu/abs/2015ApJ...801...25L}
}

@ARTICLE{mangum19,
       author = {{Mangum}, Jeffrey G. and {Ginsburg}, Adam G. and {Henkel}, Christian and {Menten}, Karl M. and {Aalto}, Susanne and {van der Werf}, Paul},
        title = "{Fire in the Heart: A Characterization of the High Kinetic Temperatures and Heating Sources in the Nucleus of NGC 253}",
      journal = {\apj},
         year = 2019,
        month = feb,
       volume = {871},
       number = {2},
          eid = {170},
        pages = {170},
          doi = {10.3847/1538-4357/aafa15},
archivePrefix = {arXiv},
       eprint = {1812.09219},
 primaryClass = {astro-ph.GA},
       adsurl = {https://ui.adsabs.harvard.edu/abs/2019ApJ...871..170M}
}

@ARTICLE{radex,
       author = {{van der Tak}, F.~F.~S. and {Black}, J.~H. and {Sch{\"o}ier}, F.~L. and {Jansen}, D.~J. and {van Dishoeck}, E.~F.},
        title = "{A computer program for fast non-LTE analysis of interstellar line spectra. With diagnostic plots to interpret observed line intensity ratios}",
      journal = {\aap},
         year = 2007,
        month = jun,
       volume = {468},
       number = {2},
        pages = {627-635},
          doi = {10.1051/0004-6361:20066820},
archivePrefix = {arXiv},
       eprint = {0704.0155},
 primaryClass = {astro-ph},
       adsurl = {https://ui.adsabs.harvard.edu/abs/2007A&A...468..627V}
}

@ARTICLE{uclchem,
       author = {{Holdship}, J. and {Viti}, S. and {Jim{\'e}nez-Serra}, I. and {Makrymallis}, A. and {Priestley}, F.},
        title = "{UCLCHEM: A Gas-grain Chemical Code for Clouds, Cores, and C-Shocks}",
      journal = {\aj},
         year = 2017,
        month = jul,
       volume = {154},
       number = {1},
          eid = {38},
        pages = {38},
          doi = {10.3847/1538-3881/aa773f},
archivePrefix = {arXiv},
       eprint = {1705.10677},
 primaryClass = {astro-ph.GA},
       adsurl = {https://ui.adsabs.harvard.edu/abs/2017AJ....154...38H}
}

@ARTICLE{Turner1985,
       author = {{Turner}, J.~L. and {Ho}, P.~T.~P.},
        title = "{The 1 parsec radio core and possible nuclear ejection in NGC 253.}",
      journal = {\apjl},
         year = 1985,
        month = dec,
       volume = {299},
        pages = {L77-L81},
          doi = {10.1086/184584},
       adsurl = {https://ui.adsabs.harvard.edu/abs/1985ApJ...299L..77T}
}

@ARTICLE{Rekola2005,
       author = {{Rekola}, R. and {Richer}, M.~G. and {McCall}, Marshall L. and {Valtonen}, M.~J. and {Kotilainen}, J.~K. and {Flynn}, Chris},
        title = "{Distance to NGC 253 based on the planetary nebula luminosity function}",
      journal = {\mnras},
         year = 2005,
        month = jul,
       volume = {361},
       number = {1},
        pages = {330-336},
          doi = {10.1111/j.1365-2966.2005.09166.x},
       adsurl = {https://ui.adsabs.harvard.edu/abs/2005MNRAS.361..330R}
}

@ARTICLE{hacar20,
       author = {{Hacar}, A. and {Bosman}, A.~D. and {van Dishoeck}, E.~F.},
        title = "{HCN-to-HNC intensity ratio: a new chemical thermometer for the molecular ISM}",
      journal = {\aap},
         year = 2020,
        month = mar,
       volume = {635},
          eid = {A4},
        pages = {A4},
          doi = {10.1051/0004-6361/201936516},
archivePrefix = {arXiv},
       eprint = {1910.13754},
 primaryClass = {astro-ph.GA},
       adsurl = {https://ui.adsabs.harvard.edu/abs/2020A&A...635A...4H}
}

@ARTICLE{holdship_c2h,
       author = {{Holdship}, J. and {Viti}, S. and {Mart{\'\i}n}, S. and {Harada}, N. and {Mangum}, J. and {Sakamoto}, K. and {Muller}, S. and {Tanaka}, K. and {Yoshimura}, Y. and {Nakanishi}, K. and {Herrero-Illana}, R. and {M{\"u}hle}, S. and {Aladro}, R. and {Colzi}, L. and {Emig}, K.~L. and {Garc{\'\i}a-Burillo}, S. and {Henkel}, C. and {Humire}, P. and {Meier}, D.~S. and {Rivilla}, V.~M. and {van der Werf}, P.},
        title = "{The distribution and origin of C$_{2}$H in NGC 253 from ALCHEMI}",
      journal = {\aap},
         year = 2021,
        month = oct,
       volume = {654},
          eid = {A55},
        pages = {A55},
          doi = {10.1051/0004-6361/202141233},
archivePrefix = {arXiv},
       eprint = {2107.04580},
 primaryClass = {astro-ph.GA},
       adsurl = {https://ui.adsabs.harvard.edu/abs/2021A&A...654A..55H}
}

@ARTICLE{harada21,
       author = {{Harada}, Nanase and {Mart{\'\i}n}, Sergio and {Mangum}, Jeffrey G. and {Sakamoto}, Kazushi and {Muller}, Sebastien and {Tanaka}, Kunihiko and {Nakanishi}, Kouichiro and {Herrero-Illana}, Rub{\'e}n and {Yoshimura}, Yuki and {M{\"u}hle}, Stefanie and {Aladro}, Rebeca and {Colzi}, Laura and {Rivilla}, V{\'\i}ctor M. and {Aalto}, Susanne and {Behrens}, Erica and {Henkel}, Christian and {Holdship}, Jonathan and {Humire}, P.~K. and {Meier}, David S. and {Nishimura}, Yuri and {van der Werf}, Paul P. and {Viti}, Serena},
        title = "{Starburst Energy Feedback Seen through HCO$^{+}$/HOC$^{+}$ Emission in NGC 253 from ALCHEMI}",
      journal = {\apj},
         year = 2021,
        month = dec,
       volume = {923},
       number = {1},
          eid = {24},
        pages = {24},
          doi = {10.3847/1538-4357/ac26b8},
archivePrefix = {arXiv},
       eprint = {2109.06476},
 primaryClass = {astro-ph.GA},
       adsurl = {https://ui.adsabs.harvard.edu/abs/2021ApJ...923...24H}
}

@ARTICLE{McElroy2013,
       author = {{McElroy}, D. and {Walsh}, C. and {Markwick}, A.~J. and {Cordiner}, M.~A. and {Smith}, K. and {Millar}, T.~J.},
        title = "{The UMIST database for astrochemistry 2012}",
      journal = {\aap},
         year = 2013,
        month = feb,
       volume = {550},
          eid = {A36},
        pages = {A36},
          doi = {10.1051/0004-6361/201220465},
archivePrefix = {arXiv},
       eprint = {1212.6362},
 primaryClass = {astro-ph.SR},
       adsurl = {https://ui.adsabs.harvard.edu/abs/2013A&A...550A..36M}
}

@ARTICLE{CASATeam2022PASP,
       author = {{CASA Team} and {Bean}, Ben and {Bhatnagar}, Sanjay and {Castro}, Sandra and {Donovan Meyer}, Jennifer and {Emonts}, Bjorn and {Garcia}, Enrique and {Garwood}, Robert and {Golap}, Kumar and {Gonzalez Villalba}, Justo and {Harris}, Pamela and {Hayashi}, Yohei and {Hoskins}, Josh and {Hsieh}, Mingyu and {Jagannathan}, Preshanth and {Kawasaki}, Wataru and {Keimpema}, Aard and {Kettenis}, Mark and {Lopez}, Jorge and {Marvil}, Joshua and {Masters}, Joseph and {McNichols}, Andrew and {Mehringer}, David and {Miel}, Renaud and {Moellenbrock}, George and {Montesino}, Federico and {Nakazato}, Takeshi and {Ott}, Juergen and {Petry}, Dirk and {Pokorny}, Martin and {Raba}, Ryan and {Rau}, Urvashi and {Schiebel}, Darrell and {Schweighart}, Neal and {Sekhar}, Srikrishna and {Shimada}, Kazuhiko and {Small}, Des and {Steeb}, Jan-Willem and {Sugimoto}, Kanako and {Suoranta}, Ville and {Tsutsumi}, Takahiro and {van Bemmel}, Ilse M. and {Verkouter}, Marjolein and {Wells}, Akeem and {Xiong}, Wei and {Szomoru}, Arpad and {Griffith}, Morgan and {Glendenning}, Brian and {Kern}, Jeff},
        title = "{CASA, the Common Astronomy Software Applications for Radio Astronomy}",
      journal = {\pasp},
         year = 2022,
        month = nov,
       volume = {134},
       number = {1041},
          eid = {114501},
        pages = {114501},
          doi = {10.1088/1538-3873/ac9642},
archivePrefix = {arXiv},
       eprint = {2210.02276},
 primaryClass = {astro-ph.IM},
       adsurl = {https://ui.adsabs.harvard.edu/abs/2022PASP..134k4501C}
}

@ARTICLE{Holdship2022,
       author = {{Holdship}, Jonathan and {Mangum}, Jeffrey G. and {Viti}, Serena and {Behrens}, Erica and {Harada}, Nanase and {Mart{\'\i}n}, Sergio and {Sakamoto}, Kazushi and {Muller}, Sebastien and {Tanaka}, Kunihiko and {Nakanishi}, Kouichiro and {Herrero-Illana}, Rub{\'e}n and {Yoshimura}, Yuki and {Aladro}, Rebeca and {Colzi}, Laura and {Emig}, Kimberly L. and {Henkel}, Christian and {Nishimura}, Yuri and {Rivilla}, V{\'\i}ctor M. and {van der Werf}, Paul P. and {Alma Comprehensive High-Resolution Extragalactic Molecular Inventory (Alchemi) Collaboration}},
        title = "{Energizing Star Formation: The Cosmic-Ray Ionization Rate in NGC 253 Derived from ALCHEMI Measurements of H$_{3}$O$^{+}$ and SO}",
      journal = {ApJ},
         year = 2022,
        month = jun,
       volume = {931},
       number = {2},
          eid = {89},
        pages = {89},
          doi = {10.3847/1538-4357/ac6753},
archivePrefix = {arXiv},
       eprint = {2204.03668},
 primaryClass = {astro-ph.GA},
       adsurl = {https://ui.adsabs.harvard.edu/abs/2022ApJ...931...89H}
}

@ARTICLE{Humire2022,
      author = {{Humire}, P.~K. and {Henkel}, C. and {Hern{\'a}ndez-G{\'o}mez}, A. and {Mart{\'\i}n}, S. and {Mangum}, J. and {Harada}, N. and {Muller}, S. and {Sakamoto}, K. and {Tanaka}, K. and {Yoshimura}, Y. and {Nakanishi}, K. and {M{\"u}hle}, S. and {Herrero-Illana}, R. and {Meier}, D.~S. and {Caux}, E. and {Aladro}, R. and {Mauersberger}, R. and {Viti}, S. and {Colzi}, L. and {Rivilla}, V.~M. and {Gorski}, M. and {Menten}, K.~M. and {Huang}, K. -Y. and {Aalto}, S. and {van der Werf}, P.~P. and {Emig}, K.~L.},
       title = "{Methanol masers in NGC 253 with ALCHEMI}",
     journal = {\aap},
        year = 2022,
       month = jul,
      volume = {663},
         eid = {A33},
       pages = {A33},
         doi = {10.1051/0004-6361/202243384},
archivePrefix = {arXiv},
      eprint = {2205.03281},
primaryClass = {astro-ph.GA},
      adsurl = {https://ui.adsabs.harvard.edu/abs/2022A&A...663A..33H}
}

@ARTICLE{Behrens2022,
       author = {{Behrens}, Erica and {Mangum}, Jeffrey G. and {Holdship}, Jonathan and {Viti}, Serena and {Harada}, Nanase and {Mart{\'\i}n}, Sergio and {Sakamoto}, Kazushi and {Muller}, Sebastien and {Tanaka}, Kunihiko and {Nakanishi}, Kouichiro and {Herrero-Illana}, Rub{\'e}n and {Yoshimura}, Yuki and {Aladro}, Rebeca and {Colzi}, Laura and {Emig}, Kimberly L. and {Henkel}, Christian and {Huang}, Ko-Yun and {Humire}, P.~K. and {Meier}, David S. and {Rivilla}, V{\'\i}ctor M. and {van der Werf}, Paul P. and {Alma Comprehensive High-Resolution Extragalactic Molecular Inventory (Alchemi) Collaboration}},
        title = "{Tracing Interstellar Heating: An ALCHEMI Measurement of the HCN Isomers in NGC 253}",
      journal = {ApJ},
         year = 2022,
        month = nov,
       volume = {939},
       number = {2},
          eid = {119},
        pages = {119},
          doi = {10.3847/1538-4357/ac91ce},
archivePrefix = {arXiv},
       eprint = {2209.06244},
 primaryClass = {astro-ph.GA},
       adsurl = {https://ui.adsabs.harvard.edu/abs/2022ApJ...939..119B}
}

@ARTICLE{Holdship2017,
       author = {{Holdship}, J. and {Viti}, S. and {Jim{\'e}nez-Serra}, I. and {Makrymallis}, A. and {Priestley}, F.},
        title = "{UCLCHEM: A Gas-grain Chemical Code for Clouds, Cores, and C-Shocks}",
      journal = {AJ},
         year = 2017,
        month = jul,
       volume = {154},
       number = {1},
          eid = {38},
        pages = {38},
          doi = {10.3847/1538-3881/aa773f},
archivePrefix = {arXiv},
       eprint = {1705.10677},
 primaryClass = {astro-ph.GA},
       adsurl = {https://ui.adsabs.harvard.edu/abs/2017AJ....154...38H}
}

@ARTICLE{Huang2023,
       author = {{Huang}, K. -Y. and {Viti}, S. and {Holdship}, J. and {Mangum}, J.~G. and {Mart{\'\i}n}, S. and {Harada}, N. and {Muller}, S. and {Sakamoto}, K. and {Tanaka}, K. and {Yoshimura}, Y. and {Herrero-Illana}, R. and {Meier}, D.~S. and {Behrens}, E. and {van der Werf}, P.~P. and {Henkel}, C. and {Garc{\'\i}a-Burillo}, S. and {Rivilla}, V.~M. and {Emig}, K.~L. and {Colzi}, L. and {Humire}, P.~K. and {Aladro}, R. and {Bouvier}, M.},
        title = "{Reconstructing the shock history in the CMZ of NGC 253 with ALCHEMI}",
      journal = {A\&A},
         year = 2023,
        month = jul,
       volume = {675},
          eid = {A151},
        pages = {A151},
          doi = {10.1051/0004-6361/202245659},
archivePrefix = {arXiv},
       eprint = {2303.12685},
 primaryClass = {astro-ph.GA},
       adsurl = {https://ui.adsabs.harvard.edu/abs/2023A&A...675A.151H}
}

@ARTICLE{Haasler2022,
       author = {{Haasler}, D. and {Rivilla}, V.~M. and {Mart{\'\i}n}, S. and {Holdship}, J. and {Viti}, S. and {Harada}, N. and {Mangum}, J. and {Sakamoto}, K. and {Muller}, S. and {Tanaka}, K. and {Yoshimura}, Y. and {Nakanishi}, K. and {Colzi}, L. and {Hunt}, L. and {Emig}, K.~L. and {Aladro}, R. and {Humire}, P. and {Henkel}, C. and {van der Werf}, P.},
        title = "{First extragalactic detection of a phosphorus-bearing molecule with ALCHEMI: Phosphorus nitride (PN)}",
      journal = {A\&A},
         year = 2022,
        month = mar,
       volume = {659},
          eid = {A158},
        pages = {A158},
          doi = {10.1051/0004-6361/202142032},
archivePrefix = {arXiv},
       eprint = {2112.04849},
 primaryClass = {astro-ph.GA},
       adsurl = {https://ui.adsabs.harvard.edu/abs/2022A&A...659A.158H}
}

@ARTICLE{Harada2022,
       author = {{Harada}, Nanase and {Mart{\'\i}n}, Sergio and {Mangum}, Jeffrey G. and {Sakamoto}, Kazushi and {Muller}, Sebastien and {Rivilla}, V{\'\i}ctor M. and {Henkel}, Christian and {Meier}, David S. and {Colzi}, Laura and {Yamagishi}, Mitsuyoshi and {Tanaka}, Kunihiko and {Nakanishi}, Kouichiro and {Herrero-Illana}, Rub{\'e}n and {Yoshimura}, Yuki and {Humire}, P.~K. and {Aladro}, Rebeca and {van der Werf}, Paul P. and {Emig}, Kimberly L.},
        title = "{ALCHEMI Finds a ``Shocking'' Carbon Footprint in the Starburst Galaxy NGC 253}",
      journal = {ApJ},
         year = 2022,
        month = oct,
       volume = {938},
       number = {1},
          eid = {80},
        pages = {80},
          doi = {10.3847/1538-4357/ac8dfc},
archivePrefix = {arXiv},
       eprint = {2208.13983},
 primaryClass = {astro-ph.GA},
       adsurl = {https://ui.adsabs.harvard.edu/abs/2022ApJ...938...80H}
}

@ARTICLE{Tanaka2024,
       author = {{Tanaka}, Kunihiko and {Mangum}, Jeffrey G. and {Viti}, Serena and {Mart{\'\i}n}, Sergio and {Harada}, Nanase and {Sakamoto}, Kazushi and {Muller}, Sebastien and {Yoshimura}, Yuki and {Nakanishi}, Kouichiro and {Herrero-Illana}, Rub{\'e}n and {Emig}, Kimberly L. and {M{\"u}hle}, S. and {Kaneko}, Hiroyuki and {Tosaki}, Tomoka and {Behrens}, Erica and {Rivilla}, V{\'\i}ctor M. and {Colzi}, Laura and {Nishimura}, Yuri and {Humire}, P.~K. and {Bouvier}, Mathilde and {Huang}, Ko-Yun and {Butterworth}, Joshua and {Meier}, David S. and {van der Werf}, Paul P.},
        title = "{Volume Density Structure of the Central Molecular Zone NGC 253 through ALCHEMI Excitation Analysis}",
      journal = {ApJ},
         year = 2024,
        month = jan,
       volume = {961},
       number = {1},
          eid = {18},
        pages = {18},
          doi = {10.3847/1538-4357/ad0e64},
archivePrefix = {arXiv},
       eprint = {2311.12106},
 primaryClass = {astro-ph.GA},
       adsurl = {https://ui.adsabs.harvard.edu/abs/2024ApJ...961...18T}
}

@ARTICLE{Butterworth2024,
       author = {{Butterworth}, J. and {Viti}, S. and {Van der Werf}, P.~P. and {Mangum}, J.~G. and {Mart{\'\i}n}, S. and {Harada}, N. and {Emig}, K.~L. and {Muller}, S. and {Sakamoto}, K. and {Yoshimura}, Y. and {Tanaka}, K. and {Herrero-Illana}, R. and {Colzi}, L. and {Rivilla}, V.~M. and {Huang}, K.~Y. and {Bouvier}, M. and {Behrens}, E. and {Henkel}, C. and {Yan}, Y.~T. and {Meier}, D.~S. and {Zhou}, D.},
        title = "{Molecular isotopologue measurements toward super star clusters and the relation to their ages in NGC253 with ALCHEMI}",
      journal = {arXiv},
         year = 2024,
        month = feb,
          eid = {arXiv:2402.10721},
        pages = {arXiv:2402.10721},
          doi = {10.48550/arXiv.2402.10721},
archivePrefix = {arXiv},
       eprint = {2402.10721},
 primaryClass = {astro-ph.GA},
       adsurl = {https://ui.adsabs.harvard.edu/abs/2024arXiv240210721B}
}

@ARTICLE{Bouvier2024,
       author = {{Bouvier}, M. and {Viti}, S. and {Behrens}, E. and {Butterworth}, J. and {Huang}, K. -Y. and {Mangum}, J.~G. and {Harada}, N. and {Mart{\'\i}n}, S. and {Rivilla}, V.~M. and {Muller}, S. and {Sakamoto}, K. and {Yoshimura}, Y. and {Tanaka}, K. and {Nakanishi}, K. and {Herrero-Illana}, R. and {Colzi}, L. and {Gorski}, M.~D. and {Henkel}, C. and {Humire}, P.~K. and {Meier}, D.~S. and {van der Werf}, P.~P. and {Yan}, Y.~T.},
        title = "{An ALCHEMI inspection of sulphur-bearing species towards the central molecular zone of NGC 253}",
      journal = {arXiv},
         year = 2024,
        month = may,
          eid = {arXiv:2405.08408},
        pages = {arXiv:2405.08408},
          doi = {10.48550/arXiv.2405.08408},
archivePrefix = {arXiv},
       eprint = {2405.08408},
 primaryClass = {astro-ph.GA},
       adsurl = {https://ui.adsabs.harvard.edu/abs/2024arXiv240508408B}
}

@ARTICLE{Schinnerer2024,
       author = {{Schinnerer}, E. and {Leroy}, A.~K.},
        title = "{Molecular Gas and the Star-Formation Process on Cloud Scales in Nearby Galaxies}",
      journal = {ARA\&A},
         year = 2024,
        month = sep,
       volume = {62},
       number = {1},
        pages = {369-436},
          doi = {10.1146/annurev-astro-071221-052651},
archivePrefix = {arXiv},
       eprint = {2403.19843},
 primaryClass = {astro-ph.GA},
       adsurl = {https://ui.adsabs.harvard.edu/abs/2024ARA&A..62..369S}
}

@ARTICLE{Harada2024ApJS,
       author = {{Harada}, Nanase and {Meier}, David S. and {Mart{\'\i}n}, Sergio and {Muller}, Sebastien and {Sakamoto}, Kazushi and {Saito}, Toshiki and {Gorski}, Mark D. and {Henkel}, Christian and {Tanaka}, Kunihiko and {Mangum}, Jeffrey G. and {Aalto}, Susanne and {Aladro}, Rebeca and {Bouvier}, Mathilde and {Colzi}, Laura and {Emig}, Kimberly L. and {Herrero-Illana}, Rub{\'e}n and {Huang}, Ko-Yun and {Kohno}, Kotaro and {K{\"o}nig}, Sabine and {Nakanishi}, Kouichiro and {Nishimura}, Yuri and {Takano}, Shuro and {Rivilla}, V{\'\i}ctor M. and {Viti}, Serena and {Watanabe}, Yoshimasa and {van der Werf}, Paul P. and {Yoshimura}, Yuki},
        title = "{The ALCHEMI Atlas: Principal Component Analysis Reveals Starburst Evolution in NGC 253}",
      journal = {ApJS},
         year = 2024,
        month = apr,
       volume = {271},
       number = {2},
          eid = {38},
        pages = {38},
          doi = {10.3847/1538-4365/ad1937},
archivePrefix = {arXiv},
       eprint = {2401.02578},
 primaryClass = {astro-ph.GA},
       adsurl = {https://ui.adsabs.harvard.edu/abs/2024ApJS..271...38H}
}

@ARTICLE{Bao2024,
       author = {{Bao}, Min and {Harada}, Nanase and {Kohno}, Kotaro and {Yoshimura}, Yuki and {Egusa}, Fumi and {Nishimura}, Yuri and {Tanaka}, Kunihiko and {Nakanishi}, Kouichiro and {Mart{\'\i}n}, Sergio and {Mangum}, Jeffrey G. and {Sakamoto}, Kazushi and {Muller}, S{\'e}bastien and {Bouvier}, Mathilde and {Colzi}, Laura and {Emig}, Kimberly L. and {Meier}, David S. and {Henkel}, Christian and {Humire}, Pedro and {Huang}, Ko-Yun and {Rivilla}, V{\'\i}ctor M. and {van der Werf}, Paul and {Viti}, Serena},
        title = "{Physical properties of the southwest outflow streamer in the starburst galaxy NGC 253 with ALCHEMI}",
      journal = {A\&A},
         year = 2024,
        month = jul,
       volume = {687},
          eid = {A43},
        pages = {A43},
          doi = {10.1051/0004-6361/202349050},
archivePrefix = {arXiv},
       eprint = {2404.04791},
 primaryClass = {astro-ph.GA},
       adsurl = {https://ui.adsabs.harvard.edu/abs/2024A&A...687A..43B}
}

@ARTICLE{Santa-Maria2023,
       author = {{Santa-Maria}, M.~G. and {Goicoechea}, J.~R. and {Pety}, J. and {Gerin}, M. and {Orkisz}, J.~H. and {Le Petit}, F. and {Einig}, L. and {Palud}, P. and {de Souza Magalhaes}, V. and {Be{\v{s}}li{\'c}}, I. and {Segal}, L. and {Bardeau}, S. and {Bron}, E. and {Chainais}, P. and {Chanussot}, J. and {Gratier}, P. and {Guzm{\'a}n}, V.~V. and {Hughes}, A. and {Languignon}, D. and {Levrier}, F. and {Lis}, D.~C. and {Liszt}, H.~S. and {Le Bourlot}, J. and {Oya}, Y. and {{\"O}berg}, K. and {Peretto}, N. and {Roueff}, E. and {Roueff}, A. and {Sievers}, A. and {Thouvenin}, P. -A. and {Yamamoto}, S.},
        title = "{HCN emission from translucent gas and UV-illuminated cloud edges revealed by wide-field IRAM 30 m maps of the Orion B GMC. Revisiting its role as a tracer of the dense gas reservoir for star formation}",
      journal = {A\&A},
         year = 2023,
        month = nov,
       volume = {679},
          eid = {A4},
        pages = {A4},
          doi = {10.1051/0004-6361/202346598},
archivePrefix = {arXiv},
       eprint = {2309.03186},
 primaryClass = {astro-ph.GA},
       adsurl = {https://ui.adsabs.harvard.edu/abs/2023A&A...679A...4S}
}

@ARTICLE{Eibensteiner2022,
       author = {{Eibensteiner}, C. and {Barnes}, A.~T. and {Bigiel}, F. and {Schinnerer}, E. and {Liu}, D. and {Meier}, D.~S. and {Usero}, A. and {Leroy}, A.~K. and {Rosolowsky}, E. and {Puschnig}, J. and {Lazar}, I. and {Pety}, J. and {Lopez}, L.~A. and {Emsellem}, E. and {Be{\v{s}}li{\'c}}, I. and {Querejeta}, M. and {Murphy}, E.~J. and {den Brok}, J. and {Schruba}, A. and {Chevance}, M. and {Glover}, S.~C.~O. and {Gao}, Y. and {Grasha}, K. and {Hassani}, H. and {Henshaw}, J.~D. and {Jimenez-Donaire}, M.~J. and {Klessen}, R.~S. and {Kruijssen}, J.~M.~D. and {Pan}, H. -A. and {Saito}, T. and {Sormani}, M.~C. and {Teng}, Y. -H. and {Williams}, T.~G.},
        title = "{A 2-3 mm high-resolution molecular line survey towards the centre of the nearby spiral galaxy NGC 6946}",
      journal = {A\&A},
         year = 2022,
        month = mar,
       volume = {659},
          eid = {A173},
        pages = {A173},
          doi = {10.1051/0004-6361/202142624},
archivePrefix = {arXiv},
       eprint = {2201.02209},
 primaryClass = {astro-ph.GA},
       adsurl = {https://ui.adsabs.harvard.edu/abs/2022A&A...659A.173E}
}

@ARTICLE{Behrens2024,
       author = {{Behrens}, Erica and {Mangum}, Jeffrey G. and {Viti}, Serena and {Holdship}, Jonathan and {Huang}, Ko-Yun and {Bouvier}, Mathilde and {Butterworth}, Joshua and {Eibensteiner}, Cosima and {Harada}, Nanase and {Martin}, Sergio and {Sakamoto}, Kazushi and {Muller}, Sebastien and {Tanaka}, Kunihiko and {Colzi}, Laura and {Henkel}, Christian and {Meier}, David S. and {Rivilla}, Victor M. and {van der Werf}, Paul P.},
        title = "{Neural Network Constraints on the Cosmic-Ray Ionization Rate and Other Physical Conditions in NGC 253 with ALCHEMI Measurements of HCN and HNC}",
      journal = {arXiv},
         year = 2024,
        month = sep,
          eid = {arXiv:2409.13821},
        pages = {arXiv:2409.13821},
          doi = {10.48550/arXiv.2409.13821},
archivePrefix = {arXiv},
       eprint = {2409.13821},
 primaryClass = {astro-ph.GA},
       adsurl = {https://ui.adsabs.harvard.edu/abs/2024arXiv240913821B}
}

@article{nautilus,
    author = {Lange, Johannes U},
    title = "{nautilus: boosting Bayesian importance nested sampling with deep learning}",
    journal = {Monthly Notices of the Royal Astronomical Society},
    volume = {525},
    number = {2},
    pages = {3181-3194},
    year = {2023},
    month = {08},
    doi = {10.1093/mnras/stad2441},
    url = {https://doi.org/10.1093/mnras/stad2441},
    eprint = {https://academic.oup.com/mnras/article-pdf/525/2/3181/51331635/stad2441.pdf},
}

@ARTICLE{Feroz2019,
       author = {{Feroz}, Farhan and {Hobson}, Michael P. and {Cameron}, Ewan and {Pettitt}, Anthony N.},
        title = "{Importance Nested Sampling and the MultiNest Algorithm}",
      journal = {The Open Journal of Astrophysics},
         year = 2019,
        month = nov,
       volume = {2},
       number = {1},
          eid = {10},
        pages = {10},
          doi = {10.21105/astro.1306.2144},
archivePrefix = {arXiv},
       eprint = {1306.2144},
 primaryClass = {astro-ph.IM},
       adsurl = {https://ui.adsabs.harvard.edu/abs/2019OJAp....2E..10F}
}

@ARTICLE{Shirley2015,
       author = {{Shirley}, Yancy L.},
        title = "{The Critical Density and the Effective Excitation Density of Commonly Observed Molecular Dense Gas Tracers}",
      journal = {PASP},
         year = 2015,
        month = mar,
       volume = {127},
       number = {949},
        pages = {299},
          doi = {10.1086/680342},
archivePrefix = {arXiv},
       eprint = {1501.01629},
 primaryClass = {astro-ph.IM},
       adsurl = {https://ui.adsabs.harvard.edu/abs/2015PASP..127..299S}
}

@ARTICLE{Reach2019,
       author = {{Reach}, William T. and {Tram}, Le Ngoc and {Richter}, Matthew and {Gusdorf}, Antoine and {DeWitt}, Curtis},
        title = "{Supernova Shocks in Molecular Clouds: Velocity Distribution of Molecular Hydrogen}",
      journal = {ApJ},
         year = 2019,
        month = oct,
       volume = {884},
       number = {1},
          eid = {81},
        pages = {81},
          doi = {10.3847/1538-4357/ab41f7},
archivePrefix = {arXiv},
       eprint = {1909.02079},
 primaryClass = {astro-ph.GA},
       adsurl = {https://ui.adsabs.harvard.edu/abs/2019ApJ...884...81R}
}

@ARTICLE{Tu2025,
       author = {{Tu}, Tian-Yu and {Chen}, Yang and {Liu}, Qian-Cheng},
        title = "{Mapping the Dense Molecular Gas toward 13 Supernova Remnants}",
      journal = {ApJ},
         year = 2025,
        month = jan,
       volume = {978},
       number = {1},
          eid = {83},
        pages = {83},
          doi = {10.3847/1538-4357/ad9390},
archivePrefix = {arXiv},
       eprint = {2411.09138},
 primaryClass = {astro-ph.GA},
       adsurl = {https://ui.adsabs.harvard.edu/abs/2025ApJ...978...83T}
}

@ARTICLE{Vaupre2014,
       author = {{Vaupr{\'e}}, S. and {Hily-Blant}, P. and {Ceccarelli}, C. and {Dubus}, G. and {Gabici}, S. and {Montmerle}, T.},
        title = "{Cosmic ray induced ionisation of a molecular cloud shocked by the W28 supernova remnant}",
      journal = {A\&A},
         year = 2014,
        month = aug,
       volume = {568},
          eid = {A50},
        pages = {A50},
          doi = {10.1051/0004-6361/201424036},
archivePrefix = {arXiv},
       eprint = {1407.0205},
 primaryClass = {astro-ph.GA},
       adsurl = {https://ui.adsabs.harvard.edu/abs/2014A&A...568A..50V}
}

@ARTICLE{Hernandez-Vera2023,
       author = {{Hern{\'a}ndez-Vera}, C. and {Guzm{\'a}n}, V.~V. and {Goicoechea}, J.~R. and {Maillard}, V. and {Pety}, J. and {Le Petit}, F. and {Gerin}, M. and {Bron}, E. and {Roueff}, E. and {Abergel}, A. and {Schirmer}, T. and {Carpenter}, J. and {Gratier}, P. and {Gordon}, K. and {Misselt}, K.},
        title = "{The extremely sharp transition between molecular and ionized gas in the Horsehead nebula}",
      journal = {A\&A},
         year = 2023,
        month = sep,
       volume = {677},
          eid = {A152},
        pages = {A152},
          doi = {10.1051/0004-6361/202347206},
archivePrefix = {arXiv},
       eprint = {2307.09540},
 primaryClass = {astro-ph.GA},
       adsurl = {https://ui.adsabs.harvard.edu/abs/2023A&A...677A.152H}
}

@ARTICLE{Xu2019,
       author = {{Xu}, Jin-Long and {Zavagno}, Annie and {Yu}, Naiping and {Liu}, Xiao-Lan and {Xu}, Ye and {Yuan}, Jinghua and {Zhang}, Chuan-Peng and {Zhang}, Si-Ju and {Zhang}, Guo-Yin and {Ning}, Chang-Chun and {Ju}, Bing-Gang},
        title = "{The effects of ionization feedback on star formation: a case study of the M 16 H II region}",
      journal = {A\&A},
         year = 2019,
        month = jul,
       volume = {627},
          eid = {A27},
        pages = {A27},
          doi = {10.1051/0004-6361/201935024},
archivePrefix = {arXiv},
       eprint = {1905.08030},
 primaryClass = {astro-ph.SR},
       adsurl = {https://ui.adsabs.harvard.edu/abs/2019A&A...627A..27X}
}

@ARTICLE{Peschken2023,
       author = {{Peschken}, N. and {Hanasz}, M. and {Naab}, T. and {W{\'o}lta{\'n}ski}, D. and {Gawryszczak}, A.},
        title = "{The phase structure of cosmic ray driven outflows in stream fed disc galaxies}",
      journal = {MNRAS},
         year = 2023,
        month = jul,
       volume = {522},
       number = {4},
        pages = {5529-5545},
          doi = {10.1093/mnras/stad1358},
archivePrefix = {arXiv},
       eprint = {2210.17328},
 primaryClass = {astro-ph.GA},
       adsurl = {https://ui.adsabs.harvard.edu/abs/2023MNRAS.522.5529P}
}

@ARTICLE{Li2020,
       author = {{Li}, Fei and {Wang}, Junzhi and {Fang}, Min and {Tan}, Qing-Hua and {Zhang}, Zhi-Yu and {Gao}, Yu and {Li}, Shanghuo},
        title = "{HCN 3-2 survey towards a sample of local galaxies}",
      journal = {PASJ},
         year = 2020,
        month = jun,
       volume = {72},
       number = {3},
          eid = {41},
        pages = {41},
          doi = {10.1093/pasj/psaa025},
archivePrefix = {arXiv},
       eprint = {2003.13009},
 primaryClass = {astro-ph.GA},
       adsurl = {https://ui.adsabs.harvard.edu/abs/2020PASJ...72...41L}
}

@ARTICLE{Gao2004,
       author = {{Gao}, Yu and {Solomon}, Philip M.},
        title = "{The Star Formation Rate and Dense Molecular Gas in Galaxies}",
      journal = {ApJ},
         year = 2004,
        month = may,
       volume = {606},
       number = {1},
        pages = {271-290},
          doi = {10.1086/382999},
archivePrefix = {arXiv},
       eprint = {astro-ph/0310339},
 primaryClass = {astro-ph},
       adsurl = {https://ui.adsabs.harvard.edu/abs/2004ApJ...606..271G}
}

@ARTICLE{Jones2023,
       author = {{Jones}, Gerwyn H. and {Clark}, Paul C. and {Glover}, Simon C.~O. and {Hacar}, Alvaro},
        title = "{On the density regime probed by HCN emission}",
      journal = {MNRAS},
         year = 2023,
        month = mar,
       volume = {520},
       number = {1},
        pages = {1005-1021},
          doi = {10.1093/mnras/stad202},
archivePrefix = {arXiv},
       eprint = {2112.05543},
 primaryClass = {astro-ph.GA},
       adsurl = {https://ui.adsabs.harvard.edu/abs/2023MNRAS.520.1005J}
}

@ARTICLE{Heckman1990,
       author = {{Heckman}, Timothy M. and {Armus}, Lee and {Miley}, George K.},
        title = "{On the Nature and Implications of Starburst-driven Galactic Superwinds}",
      journal = {ApJS},
         year = 1990,
        month = dec,
       volume = {74},
        pages = {833},
          doi = {10.1086/191522},
       adsurl = {https://ui.adsabs.harvard.edu/abs/1990ApJS...74..833H}
}

@ARTICLE{Lehnert1996,
       author = {{Lehnert}, Matthew D. and {Heckman}, Timothy M.},
        title = "{Ionized Gas in the Halos of Edge-on Starburst Galaxies: Evidence for Supernova-driven Superwinds}",
      journal = {ApJ},
         year = 1996,
        month = may,
       volume = {462},
        pages = {651},
          doi = {10.1086/177180},
       adsurl = {https://ui.adsabs.harvard.edu/abs/1996ApJ...462..651L}
}

@ARTICLE{Ceverino2009,
       author = {{Ceverino}, Daniel and {Klypin}, Anatoly},
        title = "{The Role of Stellar Feedback in the Formation of Galaxies}",
      journal = {ApJ},
         year = 2009,
        month = apr,
       volume = {695},
       number = {1},
        pages = {292-309},
          doi = {10.1088/0004-637X/695/1/292},
archivePrefix = {arXiv},
       eprint = {0712.3285},
 primaryClass = {astro-ph},
       adsurl = {https://ui.adsabs.harvard.edu/abs/2009ApJ...695..292C}
}

@ARTICLE{Hopkins2012,
       author = {{Hopkins}, Philip F. and {Quataert}, Eliot and {Murray}, Norman},
        title = "{Stellar feedback in galaxies and the origin of galaxy-scale winds}",
      journal = {MNRAS},
         year = 2012,
        month = apr,
       volume = {421},
       number = {4},
        pages = {3522-3537},
          doi = {10.1111/j.1365-2966.2012.20593.x},
archivePrefix = {arXiv},
       eprint = {1110.4638},
 primaryClass = {astro-ph.CO},
       adsurl = {https://ui.adsabs.harvard.edu/abs/2012MNRAS.421.3522H}
}

@ARTICLE{Agertz2013,
       author = {{Agertz}, Oscar and {Kravtsov}, Andrey V. and {Leitner}, Samuel N. and {Gnedin}, Nickolay Y.},
        title = "{Toward a Complete Accounting of Energy and Momentum from Stellar Feedback in Galaxy Formation Simulations}",
      journal = {ApJ},
         year = 2013,
        month = jun,
       volume = {770},
       number = {1},
          eid = {25},
        pages = {25},
          doi = {10.1088/0004-637X/770/1/25},
archivePrefix = {arXiv},
       eprint = {1210.4957},
 primaryClass = {astro-ph.CO},
       adsurl = {https://ui.adsabs.harvard.edu/abs/2013ApJ...770...25A}
}

@ARTICLE{Usero2015,
       author = {{Usero}, Antonio and {Leroy}, Adam K. and {Walter}, Fabian and {Schruba}, Andreas and {Garc{\'\i}a-Burillo}, Santiago and {Sandstrom}, Karin and {Bigiel}, Frank and {Brinks}, Elias and {Kramer}, Carsten and {Rosolowsky}, Erik and {Schuster}, Karl-Friedrich and {de Blok}, W.~J.~G.},
        title = "{Variations in the Star Formation Efficiency of the Dense Molecular Gas across the Disks of Star-forming Galaxies}",
      journal = {AJ},
         year = 2015,
        month = oct,
       volume = {150},
       number = {4},
          eid = {115},
        pages = {115},
          doi = {10.1088/0004-6256/150/4/115},
archivePrefix = {arXiv},
       eprint = {1506.00703},
 primaryClass = {astro-ph.GA},
       adsurl = {https://ui.adsabs.harvard.edu/abs/2015AJ....150..115U}
}

@ARTICLE{Neumann2024,
       author = {{Neumann}, Lukas and {Bigiel}, Frank and {Barnes}, Ashley T. and {Gallagher}, Molly J. and {Leroy}, Adam and {Usero}, Antonio and {Rosolowsky}, Erik and {Be{\v{s}}li{\'c}}, Ivana and {Boquien}, M{\'e}d{\'e}ric and {Cao}, Yixian and {Chevance}, M{\'e}lanie and {Colombo}, Dario and {Dale}, Daniel A. and {Eibensteiner}, Cosima and {Grasha}, Kathryn and {Henshaw}, Jonathan D. and {Jim{\'e}nez-Donaire}, Mar{\'\i}a J. and {Meidt}, Sharon and {Menon}, Shyam H. and {Murphy}, Eric J. and {Pan}, Hsi-An and {Querejeta}, Miguel and {Saito}, Toshiki and {Schinnerer}, Eva and {Stuber}, Sophia K. and {Teng}, Yu-Hsuan and {Williams}, Thomas G.},
        title = "{A 260 pc resolution ALMA map of HCN(1{\textendash}0) in the galaxy NGC 4321}",
      journal = {A\&A},
         year = 2024,
        month = nov,
       volume = {691},
          eid = {A121},
        pages = {A121},
          doi = {10.1051/0004-6361/202449496},
archivePrefix = {arXiv},
       eprint = {2406.12025},
 primaryClass = {astro-ph.GA},
       adsurl = {https://ui.adsabs.harvard.edu/abs/2024A&A...691A.121N}
}

@ARTICLE{Bemis2024,
       author = {{Bemis}, Ashley R. and {Wilson}, Christine D. and {Sharda}, Piyush and {Roberts}, Ian D. and {He}, Hao},
        title = "{Does the HCN/CO ratio trace the star-forming fraction of gas?: II. Variations in CO and HCN emissivity}",
      journal = {A\&A},
         year = 2024,
        month = dec,
       volume = {692},
          eid = {A146},
        pages = {A146},
          doi = {10.1051/0004-6361/202347879},
archivePrefix = {arXiv},
       eprint = {2410.00243},
 primaryClass = {astro-ph.GA},
       adsurl = {https://ui.adsabs.harvard.edu/abs/2024A&A...692A.146B}
}

@ARTICLE{Garcia-Rodriguez2023,
       author = {{Garc{\'\i}a-Rodr{\'\i}guez}, A. and {Usero}, A. and {Leroy}, A.~K. and {Bigiel}, F. and {Jim{\'e}nez-Donaire}, M.~J. and {Liu}, D. and {Querejeta}, M. and {Saito}, T. and {Schinnerer}, E. and {Barnes}, A. and {Belfiore}, F. and {Be{\v{s}}li{\'c}}, I. and {Cao}, Y. and {Chevance}, M. and {Dale}, D.~A. and {den Brok}, J.~S. and {Eibensteiner}, C. and {Garc{\'\i}a-Burillo}, S. and {Glover}, S.~C.~O. and {Klessen}, R.~S. and {Pety}, J. and {Puschnig}, J. and {Rosolowsky}, E. and {Sandstrom}, K. and {Sormani}, M.~C. and {Teng}, Y. -H. and {Williams}, T.~G.},
        title = "{Sub-kiloparsec empirical relations and excitation conditions of HCN and HCO$^{+}$ J = 3-2 in nearby star-forming galaxies}",
      journal = {A\&A},
         year = 2023,
        month = apr,
       volume = {672},
          eid = {A96},
        pages = {A96},
          doi = {10.1051/0004-6361/202244317},
archivePrefix = {arXiv},
       eprint = {2302.00450},
 primaryClass = {astro-ph.GA},
       adsurl = {https://ui.adsabs.harvard.edu/abs/2023A&A...672A..96G}
}

@ARTICLE{Garcia-Burillo2012,
       author = {{Garc{\'\i}a-Burillo}, S. and {Usero}, A. and {Alonso-Herrero}, A. and {Graci{\'a}-Carpio}, J. and {Pereira-Santaella}, M. and {Colina}, L. and {Planesas}, P. and {Arribas}, S.},
        title = "{Star-formation laws in luminous infrared galaxies. New observational constraints on models}",
      journal = {A\&A},
         year = 2012,
        month = mar,
       volume = {539},
          eid = {A8},
        pages = {A8},
          doi = {10.1051/0004-6361/201117838},
archivePrefix = {arXiv},
       eprint = {1111.6773},
 primaryClass = {astro-ph.CO},
       adsurl = {https://ui.adsabs.harvard.edu/abs/2012A&A...539A...8G}
}

@ARTICLE{Onus2018,
       author = {{Onus}, Adam and {Krumholz}, Mark R. and {Federrath}, Christoph},
        title = "{Numerical calibration of the HCN-star formation correlation}",
      journal = {MNRAS},
         year = 2018,
        month = sep,
       volume = {479},
       number = {2},
        pages = {1702-1710},
          doi = {10.1093/mnras/sty1662},
archivePrefix = {arXiv},
       eprint = {1801.09952},
 primaryClass = {astro-ph.GA},
       adsurl = {https://ui.adsabs.harvard.edu/abs/2018MNRAS.479.1702O}
}

@ARTICLE{Gong2025arXiv,
       author = {{Gong}, Y. and {Henkel}, C. and {Bop}, C.~T. and {Mangum}, J.~G. and {Behrens}, E. and {Du}, F.~J. and {Zhang}, S.~B. and {Martin}, S. and {Menten}, K.~M. and {Harada}, N. and {Bouvier}, M. and {Tang}, X.~D. and {Tanaka}, K. and {Viti}, S. and {Yan}, Y.~T. and {Yang}, W. and {Mao}, R.~Q. and {Quan}, D.~H.},
        title = "{Shock-induced HCNH+ abundance enhancement in the heart of the starburst galaxy NGC 253 unveiled by ALCHEMI}",
      journal = {arXiv},
         year = 2025,
        month = feb,
          eid = {arXiv:2502.20894},
        pages = {arXiv:2502.20894},
          doi = {10.48550/arXiv.2502.20894},
archivePrefix = {arXiv},
       eprint = {2502.20894},
 primaryClass = {astro-ph.GA},
       adsurl = {https://ui.adsabs.harvard.edu/abs/2025arXiv250220894G}
}

@article{Kishikawa2024,
    author = {Kishikawa, Ryo and Harada, Nanase and Saito, Toshiki and Aalto, Susanne and Colzi, Laura and Gorski, Mark and Henkel, Christian and Mangum, Jeffrey G and Martín, Sergio and Muller, Sebastian and Nishimura, Yuri and Rivilla, Víctor M and Sakamoto, Kazushi and van der Werf, Paul and Viti, Serena},
    title = {Components of star formation in NGC 253: Non-negative matrix factorization analysis with the ALCHEMI integrated intensity images},
    journal = {Publications of the Astronomical Society of Japan},
    volume = {77},
    number = {1},
    pages = {1-20},
    year = {2024},
    month = {11},
    issn = {2053-051X},
    doi = {10.1093/pasj/psae095},
    url = {https://doi.org/10.1093/pasj/psae095},
    eprint = {https://academic.oup.com/pasj/article-pdf/77/1/1/60906348/psae095.pdf},
}

@ARTICLE{Rybak2022,
       author = {{Rybak}, M. and {Hodge}, J.~A. and {Greve}, T.~R. and {Riechers}, D. and {Lamperti}, I. and {van Marrewijk}, J. and {Walter}, F. and {Wagg}, J. and {van der Werf}, P.~P.},
        title = "{PRUSSIC. I. A JVLA survey of HCN, HCO$^{+}$, and HNC (1-0) emission in z {\ensuremath{\sim}} 3 dusty galaxies: Low dense-gas fractions in high-redshift star-forming galaxies}",
      journal = {A\&A},
         year = 2022,
        month = nov,
       volume = {667},
          eid = {A70},
        pages = {A70},
          doi = {10.1051/0004-6361/202243894},
archivePrefix = {arXiv},
       eprint = {2207.06967},
 primaryClass = {astro-ph.GA},
       adsurl = {https://ui.adsabs.harvard.edu/abs/2022A&A...667A..70R}
}

@ARTICLE{Oteo2017,
       author = {{Oteo}, I. and {Zhang}, Z. -Y. and {Yang}, C. and {Ivison}, R.~J. and {Omont}, A. and {Bremer}, M. and {Bussmann}, S. and {Cooray}, A. and {Cox}, P. and {Dannerbauer}, H. and {Dunne}, L. and {Eales}, S. and {Furlanetto}, C. and {Gavazzi}, R. and {Gao}, Y. and {Greve}, T.~R. and {Nayyeri}, H. and {Negrello}, M. and {Neri}, R. and {Riechers}, D. and {Tunnard}, R. and {Wagg}, J. and {Van der Werf}, P.},
        title = "{High Dense Gas Fraction in Intensely Star-forming Dusty Galaxies}",
      journal = {ApJ},
         year = 2017,
        month = dec,
       volume = {850},
       number = {2},
          eid = {170},
        pages = {170},
          doi = {10.3847/1538-4357/aa8ee3},
archivePrefix = {arXiv},
       eprint = {1701.05901},
 primaryClass = {astro-ph.GA},
       adsurl = {https://ui.adsabs.harvard.edu/abs/2017ApJ...850..170O}
}

@ARTICLE{Spilker2014,
       author = {{Spilker}, J.~S. and {Marrone}, D.~P. and {Aguirre}, J.~E. and {Aravena}, M. and {Ashby}, M.~L.~N. and {B{\'e}thermin}, M. and {Bradford}, C.~M. and {Bothwell}, M.~S. and {Brodwin}, M. and {Carlstrom}, J.~E. and {Chapman}, S.~C. and {Crawford}, T.~M. and {de Breuck}, C. and {Fassnacht}, C.~D. and {Gonzalez}, A.~H. and {Greve}, T.~R. and {Gullberg}, B. and {Hezaveh}, Y. and {Holzapfel}, W.~L. and {Husband}, K. and {Ma}, J. and {Malkan}, M. and {Murphy}, E.~J. and {Reichardt}, C.~L. and {Rotermund}, K.~M. and {Stalder}, B. and {Stark}, A.~A. and {Strandet}, M. and {Vieira}, J.~D. and {Wei{\ss}}, A. and {Welikala}, N.},
        title = "{The Rest-frame Submillimeter Spectrum of High-redshift, Dusty, Star-forming Galaxies}",
      journal = {ApJ},
         year = 2014,
        month = apr,
       volume = {785},
       number = {2},
          eid = {149},
        pages = {149},
          doi = {10.1088/0004-637X/785/2/149},
archivePrefix = {arXiv},
       eprint = {1403.1667},
 primaryClass = {astro-ph.GA},
       adsurl = {https://ui.adsabs.harvard.edu/abs/2014ApJ...785..149S}
}

@ARTICLE{Patra2025,
       author = {{Patra}, Sudeshna and {Evans}, Neal J. and {Kim}, Kee-Tae and {Heyer}, Mark and {Giannetti}, Andrea and {Elia}, Davide and {Jose}, Jessy and {Kauffmann}, Jens and {Samal}, Manash R. and {Karska}, Agata and {Das}, Swagat R. and {Lee}, Gyuho and {Park}, Geumsook},
        title = "{Variation of Dense Gas Mass{\textendash}Luminosity Conversion Factor with Metallicity in the Milky Way}",
      journal = {\apj},
         year = 2025,
        month = apr,
       volume = {983},
       number = {2},
          eid = {133},
        pages = {133},
          doi = {10.3847/1538-4357/adbf8d},
archivePrefix = {arXiv},
       eprint = {2503.07931},
 primaryClass = {astro-ph.GA},
       adsurl = {https://ui.adsabs.harvard.edu/abs/2025ApJ...983..133P}
}

@ARTICLE{Aalto2002,
       author = {{Aalto}, S. and {Polatidis}, A.~G. and {H{\"u}ttemeister}, S. and {Curran}, S.~J.},
        title = "{CN and HNC line emission in IR luminous galaxies}",
      journal = {\aap},
         year = 2002,
        month = jan,
       volume = {381},
        pages = {783-794},
          doi = {10.1051/0004-6361:20011514},
archivePrefix = {arXiv},
       eprint = {astro-ph/0111323},
 primaryClass = {astro-ph},
       adsurl = {https://ui.adsabs.harvard.edu/abs/2002A&A...381..783A}
}

@ARTICLE{Imanishi2023,
       author = {{Imanishi}, Masatoshi and {Baba}, Shunsuke and {Nakanishi}, Kouichiro and {Izumi}, Takuma},
        title = "{Dense Molecular Gas Properties of the Central Kiloparsec of Nearby Ultraluminous Infrared Galaxies Constrained by ALMA Three Transition-line Observations}",
      journal = {\apj},
         year = 2023,
        month = jun,
       volume = {950},
       number = {1},
          eid = {75},
        pages = {75},
          doi = {10.3847/1538-4357/acc388},
archivePrefix = {arXiv},
       eprint = {2303.08178},
 primaryClass = {astro-ph.GA},
       adsurl = {https://ui.adsabs.harvard.edu/abs/2023ApJ...950...75I}
}

@ARTICLE{Saykally1976,
       author = {{Saykally}, R.~J. and {Szanto}, P.~G. and {Anderson}, T.~G. and {Woods}, R.~C.},
        title = "{The microwave spectrum of hydrogen isocyanide.}",
      journal = {\apjl},
         year = 1976,
        month = mar,
       volume = {204},
        pages = {L143-L145},
          doi = {10.1086/182074},
       adsurl = {https://ui.adsabs.harvard.edu/abs/1976ApJ...204L.143S}
}

@ARTICLE{Ahrens2002,
       author = {{Ahrens}, Volker and {Lewen}, Frank and {Takano}, Shuro and {Winnewisser}, Gisbert and {Urban}, {\v{S}}tep{\'a}n and {Negirev}, A.~A. and {Koroliev}, A.~N.},
        title = "{Sub-Doppler Saturation Spectroscopy of HCN up to 1 THz and Detection of J = 3 {\textemdash}> 2 (4{\textemdash}> 3) Emission from TMC1}",
      journal = {Zeitschrift Naturforschung Teil A},
         year = 2002,
        month = aug,
       volume = {57},
       number = {8},
        pages = {669-681},
          doi = {10.1515/zna-2002-0806},
       adsurl = {https://ui.adsabs.harvard.edu/abs/2002ZNatA..57..669A}
}

@ARTICLE{Bouvier2025arXiv,
       author = {{Bouvier}, M. and {Viti}, S. and {Mangum}, J.~G. and {Eibensteiner}, C. and {Behrens}, E. and {Rivilla}, V.~M. and {L{\'o}pez-Gallifa}, {\'A}. and {Mart{\'\i}n}, S. and {Harada}, N. and {Muller}, S. and {Colzi}, L. and {Sakamoto}, K.},
        title = "{Complex Organic Molecules towards the central molecular zone of NGC 253}",
      journal = {arXiv},
         year = 2025,
        month = apr,
          eid = {arXiv:2504.19631},
        pages = {arXiv:2504.19631},
          doi = {10.48550/arXiv.2504.19631},
archivePrefix = {arXiv},
       eprint = {2504.19631},
 primaryClass = {astro-ph.GA},
       adsurl = {https://ui.adsabs.harvard.edu/abs/2025arXiv250419631B}
}

@article{astropy:2013,
Adsurl = {http://adsabs.harvard.edu/abs/2013A%26A...558A..33A},
Archiveprefix = {arXiv},
Author = {{Astropy Collaboration} and {Robitaille}, T.~P. and {Tollerud}, E.~J. and {Greenfield}, P. and {Droettboom}, M. and {Bray}, E. and {Aldcroft}, T. and {Davis}, M. and {Ginsburg}, A. and {Price-Whelan}, A.~M. and {Kerzendorf}, W.~E. and {Conley}, A. and {Crighton}, N. and {Barbary}, K. and {Muna}, D. and {Ferguson}, H. and {Grollier}, F. and {Parikh}, M.~M. and {Nair}, P.~H. and {Unther}, H.~M. and {Deil}, C. and {Woillez}, J. and {Conseil}, S. and {Kramer}, R. and {Turner}, J.~E.~H. and {Singer}, L. and {Fox}, R. and {Weaver}, B.~A. and {Zabalza}, V. and {Edwards}, Z.~I. and {Azalee Bostroem}, K. and {Burke}, D.~J. and {Casey}, A.~R. and {Crawford}, S.~M. and {Dencheva}, N. and {Ely}, J. and {Jenness}, T. and {Labrie}, K. and {Lim}, P.~L. and {Pierfederici}, F. and {Pontzen}, A. and {Ptak}, A. and {Refsdal}, B. and {Servillat}, M. and {Streicher}, O.},
Doi = {10.1051/0004-6361/201322068},
Eid = {A33},
Eprint = {1307.6212},
Journal = {A\&A},
Month = oct,
Pages = {A33},
Primaryclass = {astro-ph.IM},
Title = {{Astropy: A community Python package for astronomy}},
Volume = 558,
Year = 2013}

@ARTICLE{astropy:2018,
       author = {{Astropy Collaboration} and {Price-Whelan}, A.~M. and
         {Sip{\H{o}}cz}, B.~M. and {G{\"u}nther}, H.~M. and {Lim}, P.~L. and
         {Crawford}, S.~M. and {Conseil}, S. and {Shupe}, D.~L. and
         {Craig}, M.~W. and {Dencheva}, N. and {Ginsburg}, A. and {Vand
        erPlas}, J.~T. and {Bradley}, L.~D. and {P{\'e}rez-Su{\'a}rez}, D. and
         {de Val-Borro}, M. and {Aldcroft}, T.~L. and {Cruz}, K.~L. and
         {Robitaille}, T.~P. and {Tollerud}, E.~J. and {Ardelean}, C. and
         {Babej}, T. and {Bach}, Y.~P. and {Bachetti}, M. and {Bakanov}, A.~V. and
         {Bamford}, S.~P. and {Barentsen}, G. and {Barmby}, P. and
         {Baumbach}, A. and {Berry}, K.~L. and {Biscani}, F. and {Boquien}, M. and
         {Bostroem}, K.~A. and {Bouma}, L.~G. and {Brammer}, G.~B. and
         {Bray}, E.~M. and {Breytenbach}, H. and {Buddelmeijer}, H. and
         {Burke}, D.~J. and {Calderone}, G. and {Cano Rodr{\'\i}guez}, J.~L. and
         {Cara}, M. and {Cardoso}, J.~V.~M. and {Cheedella}, S. and {Copin}, Y. and
         {Corrales}, L. and {Crichton}, D. and {D'Avella}, D. and {Deil}, C. and
         {Depagne}, {\'E}. and {Dietrich}, J.~P. and {Donath}, A. and
         {Droettboom}, M. and {Earl}, N. and {Erben}, T. and {Fabbro}, S. and
         {Ferreira}, L.~A. and {Finethy}, T. and {Fox}, R.~T. and
         {Garrison}, L.~H. and {Gibbons}, S.~L.~J. and {Goldstein}, D.~A. and
         {Gommers}, R. and {Greco}, J.~P. and {Greenfield}, P. and
         {Groener}, A.~M. and {Grollier}, F. and {Hagen}, A. and {Hirst}, P. and
         {Homeier}, D. and {Horton}, A.~J. and {Hosseinzadeh}, G. and {Hu}, L. and
         {Hunkeler}, J.~S. and {Ivezi{\'c}}, {\v{Z}}. and {Jain}, A. and
         {Jenness}, T. and {Kanarek}, G. and {Kendrew}, S. and {Kern}, N.~S. and
         {Kerzendorf}, W.~E. and {Khvalko}, A. and {King}, J. and {Kirkby}, D. and
         {Kulkarni}, A.~M. and {Kumar}, A. and {Lee}, A. and {Lenz}, D. and
         {Littlefair}, S.~P. and {Ma}, Z. and {Macleod}, D.~M. and
         {Mastropietro}, M. and {McCully}, C. and {Montagnac}, S. and
         {Morris}, B.~M. and {Mueller}, M. and {Mumford}, S.~J. and {Muna}, D. and
         {Murphy}, N.~A. and {Nelson}, S. and {Nguyen}, G.~H. and
         {Ninan}, J.~P. and {N{\"o}the}, M. and {Ogaz}, S. and {Oh}, S. and
         {Parejko}, J.~K. and {Parley}, N. and {Pascual}, S. and {Patil}, R. and
         {Patil}, A.~A. and {Plunkett}, A.~L. and {Prochaska}, J.~X. and
         {Rastogi}, T. and {Reddy Janga}, V. and {Sabater}, J. and
         {Sakurikar}, P. and {Seifert}, M. and {Sherbert}, L.~E. and
         {Sherwood-Taylor}, H. and {Shih}, A.~Y. and {Sick}, J. and
         {Silbiger}, M.~T. and {Singanamalla}, S. and {Singer}, L.~P. and
         {Sladen}, P.~H. and {Sooley}, K.~A. and {Sornarajah}, S. and
         {Streicher}, O. and {Teuben}, P. and {Thomas}, S.~W. and
         {Tremblay}, G.~R. and {Turner}, J.~E.~H. and {Terr{\'o}n}, V. and
         {van Kerkwijk}, M.~H. and {de la Vega}, A. and {Watkins}, L.~L. and
         {Weaver}, B.~A. and {Whitmore}, J.~B. and {Woillez}, J. and
         {Zabalza}, V. and {Astropy Contributors}},
        title = "{The Astropy Project: Building an Open-science Project and Status of the v2.0 Core Package}",
      journal = {AJ},
         year = 2018,
        month = sep,
       volume = {156},
       number = {3},
          eid = {123},
        pages = {123},
          doi = {10.3847/1538-3881/aabc4f},
archivePrefix = {arXiv},
       eprint = {1801.02634},
 primaryClass = {astro-ph.IM},
       adsurl = {https://ui.adsabs.harvard.edu/abs/2018AJ....156..123A}
}

@ARTICLE{astropy:2022,
       author = {{Astropy Collaboration} and {Price-Whelan}, Adrian M. and {Lim}, Pey Lian and {Earl}, Nicholas and {Starkman}, Nathaniel and {Bradley}, Larry and {Shupe}, David L. and {Patil}, Aarya A. and {Corrales}, Lia and {Brasseur}, C.~E. and {N{"o}the}, Maximilian and {Donath}, Axel and {Tollerud}, Erik and {Morris}, Brett M. and {Ginsburg}, Adam and {Vaher}, Eero and {Weaver}, Benjamin A. and {Tocknell}, James and {Jamieson}, William and {van Kerkwijk}, Marten H. and {Robitaille}, Thomas P. and {Merry}, Bruce and {Bachetti}, Matteo and {G{"u}nther}, H. Moritz and {Aldcroft}, Thomas L. and {Alvarado-Montes}, Jaime A. and {Archibald}, Anne M. and {B{'o}di}, Attila and {Bapat}, Shreyas and {Barentsen}, Geert and {Baz{'a}n}, Juanjo and {Biswas}, Manish and {Boquien}, M{'e}d{'e}ric and {Burke}, D.~J. and {Cara}, Daria and {Cara}, Mihai and {Conroy}, Kyle E. and {Conseil}, Simon and {Craig}, Matthew W. and {Cross}, Robert M. and {Cruz}, Kelle L. and {D'Eugenio}, Francesco and {Dencheva}, Nadia and {Devillepoix}, Hadrien A.~R. and {Dietrich}, J{"o}rg P. and {Eigenbrot}, Arthur Davis and {Erben}, Thomas and {Ferreira}, Leonardo and {Foreman-Mackey}, Daniel and {Fox}, Ryan and {Freij}, Nabil and {Garg}, Suyog and {Geda}, Robel and {Glattly}, Lauren and {Gondhalekar}, Yash and {Gordon}, Karl D. and {Grant}, David and {Greenfield}, Perry and {Groener}, Austen M. and {Guest}, Steve and {Gurovich}, Sebastian and {Handberg}, Rasmus and {Hart}, Akeem and {Hatfield-Dodds}, Zac and {Homeier}, Derek and {Hosseinzadeh}, Griffin and {Jenness}, Tim and {Jones}, Craig K. and {Joseph}, Prajwel and {Kalmbach}, J. Bryce and {Karamehmetoglu}, Emir and {Ka{l}uszy{'n}ski}, Miko{l}aj and {Kelley}, Michael S.~P. and {Kern}, Nicholas and {Kerzendorf}, Wolfgang E. and {Koch}, Eric W. and {Kulumani}, Shankar and {Lee}, Antony and {Ly}, Chun and {Ma}, Zhiyuan and {MacBride}, Conor and {Maljaars}, Jakob M. and {Muna}, Demitri and {Murphy}, N.~A. and {Norman}, Henrik and {O'Steen}, Richard and {Oman}, Kyle A. and {Pacifici}, Camilla and {Pascual}, Sergio and {Pascual-Granado}, J. and {Patil}, Rohit R. and {Perren}, Gabriel I. and {Pickering}, Timothy E. and {Rastogi}, Tanuj and {Roulston}, Benjamin R. and {Ryan}, Daniel F. and {Rykoff}, Eli S. and {Sabater}, Jose and {Sakurikar}, Parikshit and {Salgado}, Jes{'u}s and {Sanghi}, Aniket and {Saunders}, Nicholas and {Savchenko}, Volodymyr and {Schwardt}, Ludwig and {Seifert-Eckert}, Michael and {Shih}, Albert Y. and {Jain}, Anany Shrey and {Shukla}, Gyanendra and {Sick}, Jonathan and {Simpson}, Chris and {Singanamalla}, Sudheesh and {Singer}, Leo P. and {Singhal}, Jaladh and {Sinha}, Manodeep and {Sip{H{o}}cz}, Brigitta M. and {Spitler}, Lee R. and {Stansby}, David and {Streicher}, Ole and {{{S}}umak}, Jani and {Swinbank}, John D. and {Taranu}, Dan S. and {Tewary}, Nikita and {Tremblay}, Grant R. and {Val-Borro}, Miguel de and {Van Kooten}, Samuel J. and {Vasovi{'c}}, Zlatan and {Verma}, Shresth and {de Miranda Cardoso}, Jos{'e} Vin{'i}cius and {Williams}, Peter K.~G. and {Wilson}, Tom J. and {Winkel}, Benjamin and {Wood-Vasey}, W.~M. and {Xue}, Rui and {Yoachim}, Peter and {Zhang}, Chen and {Zonca}, Andrea and {Astropy Project Contributors}},
        title = "{The Astropy Project: Sustaining and Growing a Community-oriented Open-source Project and the Latest Major Release (v5.0) of the Core Package}",
      journal = {ApJ},
         year = 2022,
        month = aug,
       volume = {935},
       number = {2},
          eid = {167},
        pages = {167},
          doi = {10.3847/1538-4357/ac7c74},
archivePrefix = {arXiv},
       eprint = {2206.14220},
 primaryClass = {astro-ph.IM},
       adsurl = {https://ui.adsabs.harvard.edu/abs/2022ApJ...935..167A}
}

@article{corner,
      doi = {10.21105/joss.00024},
      url = {https://doi.org/10.21105/joss.00024},
      year  = {2016},
      month = {jun},
      publisher = {The Open Journal},
      volume = {1},
      number = {2},
      pages = {24},
      author = {Daniel Foreman-Mackey},
      title = {corner.py: Scatterplot matrices in Python},
      journal = {The Journal of Open Source Software}
    }

@misc{tensorflow2015-whitepaper,
title={ {TensorFlow}: Large-Scale Machine Learning on Heterogeneous Systems},
url={https://www.tensorflow.org/},
note={Software available from tensorflow.org},
author={
    Mart\'{i}n~Abadi and
    Ashish~Agarwal and
    Paul~Barham and
    Eugene~Brevdo and
    Zhifeng~Chen and
    Craig~Citro and
    Greg~S.~Corrado and
    Andy~Davis and
    Jeffrey~Dean and
    Matthieu~Devin and
    Sanjay~Ghemawat and
    Ian~Goodfellow and
    Andrew~Harp and
    Geoffrey~Irving and
    Michael~Isard and
    Yangqing Jia and
    Rafal~Jozefowicz and
    Lukasz~Kaiser and
    Manjunath~Kudlur and
    Josh~Levenberg and
    Dandelion~Man\'{e} and
    Rajat~Monga and
    Sherry~Moore and
    Derek~Murray and
    Chris~Olah and
    Mike~Schuster and
    Jonathon~Shlens and
    Benoit~Steiner and
    Ilya~Sutskever and
    Kunal~Talwar and
    Paul~Tucker and
    Vincent~Vanhoucke and
    Vijay~Vasudevan and
    Fernanda~Vi\'{e}gas and
    Oriol~Vinyals and
    Pete~Warden and
    Martin~Wattenberg and
    Martin~Wicke and
    Yuan~Yu and
    Xiaoqiang~Zheng},
  year={2015},
}
\bibliographystyle{aasjournalv7}

\restartappendixnumbering\renewcommand{\theHfigure}{A\arabic{figure}}
\appendix 
\label{sec:appendix}

\section{Additional Figures}\label{sec:addFigs}

In the following, we present additional figures to supplement those in the main text. In Figure~\ref{fig:allCorners}, we show the full corner plots for our models constrained by the control combinations of HCN and HNC transitions (the $J=1-0$, $2-1$, $3-2$, and $4-3$ for both molecules) for the low-$n$-$\zeta$ and low-SNR regions, as the control result for the high-$n$-$\zeta$ region is shown in Figure \ref{fig:synthControl}.

To show examples of what the full 1- and 2-D posterior distributions for models constrained by each of the five transition test combinations may look like, we provide two corner plots for each of the five combination groupings. Though not every transition combination within a particular group will yield similar results, we show examples that are roughly representative of the constraining combinations in each group. The posterior distributions obtained from single transitions ($\color{Sienna3}\blacklozenge$) are shown in Figure~\ref{fig:singleTcorners}, matched HCN and HNC transitions (\textcolor{SpringGreen4}{\ding{53}}) are shown in Figure~\ref{fig:matchedCorners}, single molecule combinations ($\color{RoyalBlue3}\blacksquare$) are shown in Figure~\ref{fig:singleMolCorners}, all transitions of one molecule supplemented with one transition from the other (\textcolor{Purple4}{\CIRCLE}) are shown in Figure~\ref{fig:allMol+1_corners}, and single transitions removed ($\color{VioletRed1}\blacktriangle$) are shown in Figure~\ref{fig:1TremovedCorners}.

We supplement Figure \ref{fig:synthControl} showing the parameter results from measured versus synthetic data (Section \ref{sec:synth} with additional figures for test cases constrained by just the HCN $1-0$ transition (Figure \ref{fig:synthHCN10}) and just the four HNC transitions (Figure \ref{fig:synthHNConly}).}

We also provide versions of Figure~\ref{fig:reg26_1d} for the high-$n$-$\zeta$ region (Figure~\ref{fig:reg53_1d}) and for the the low-SNR region (Figure~\ref{fig:reg51_1d}).

\begin{figure*}
    \gridline{
    \leftfig{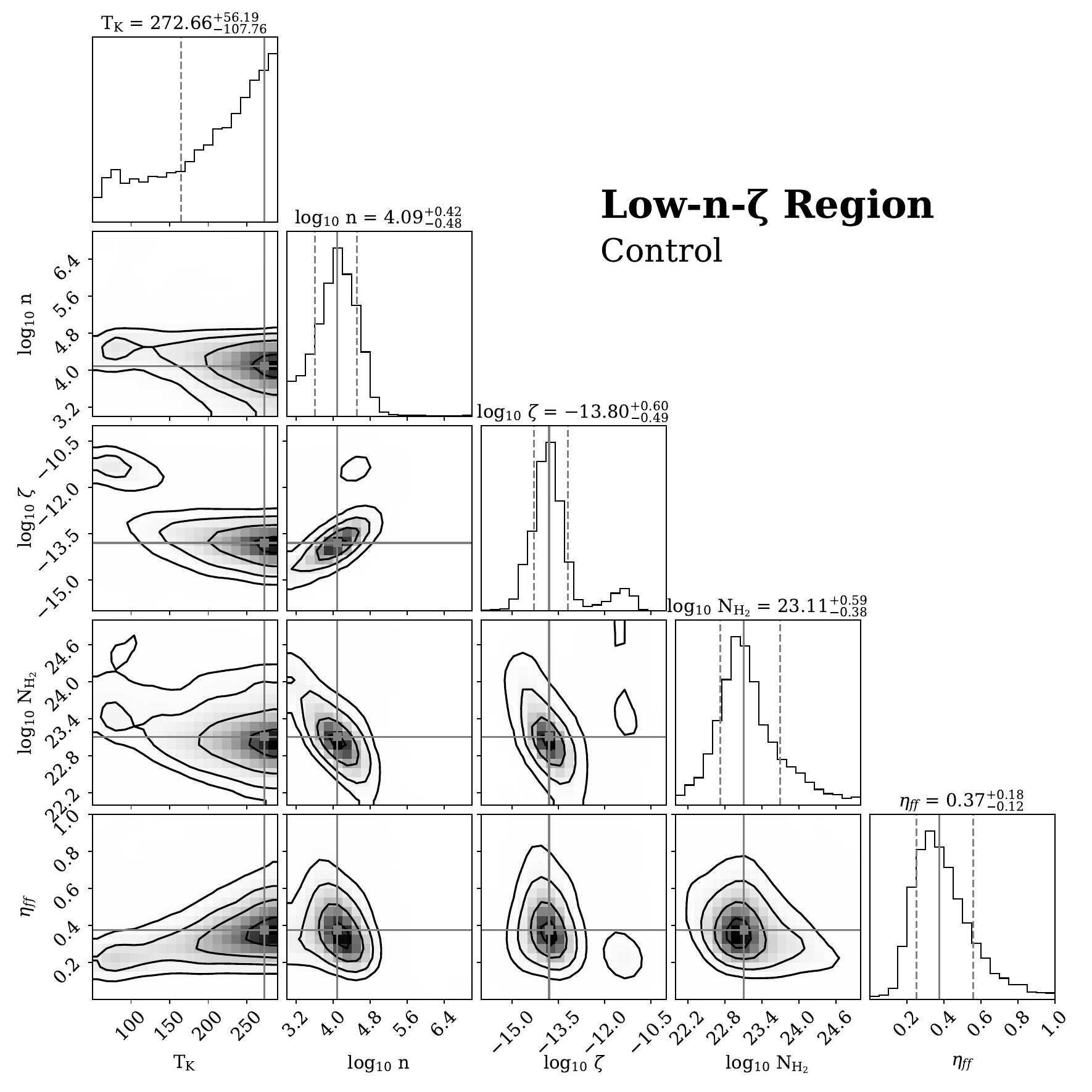}{0.5\textwidth}{(a)}
    \rightfig{region53_control_paperFig.pdf}{0.5\textwidth}{(b)}
    }
    \gridline{\fig{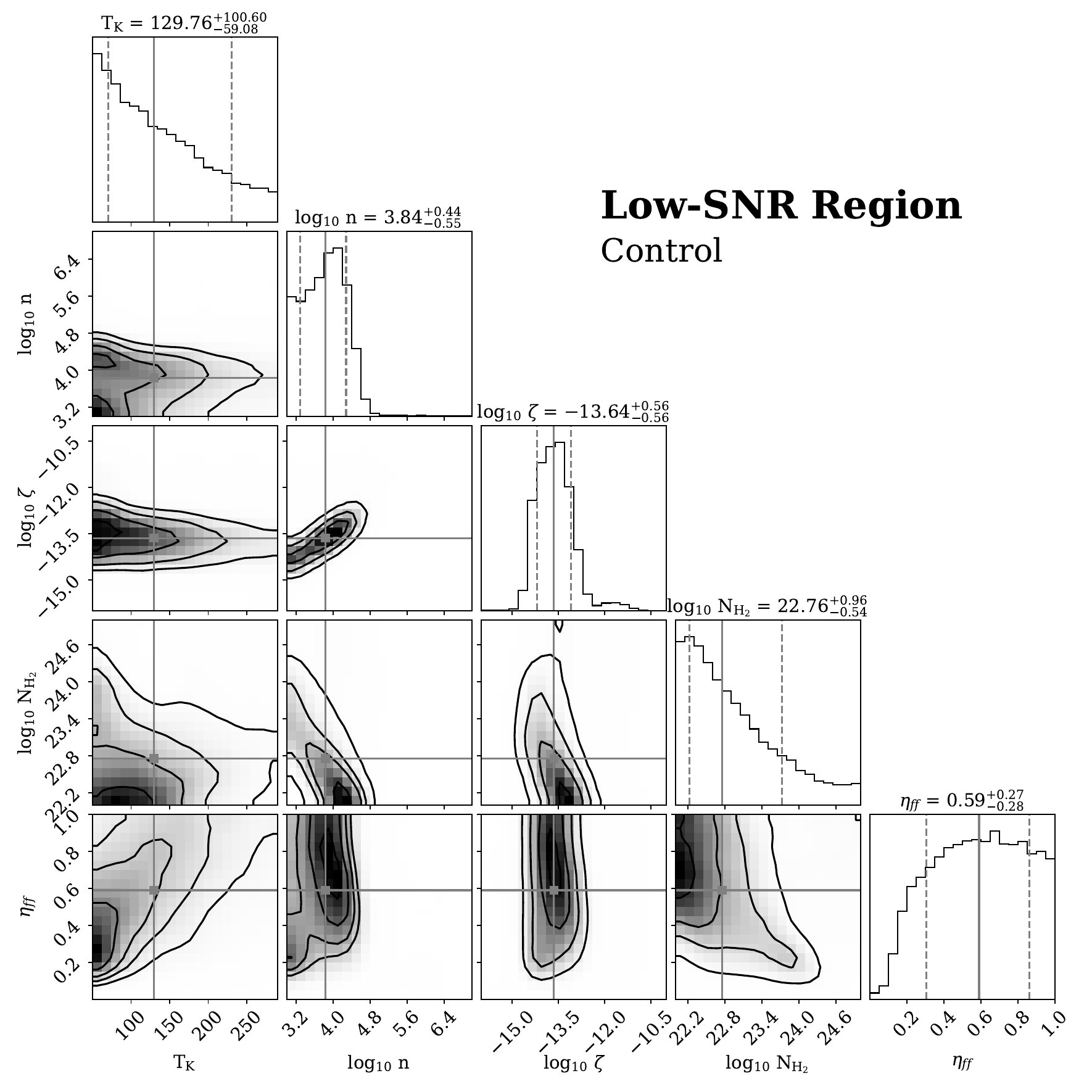}{0.5\textwidth}{(c)}}
    \caption{Posterior distributions for the low-$n$-$\zeta$ region (top left), high-$n$-$\zeta$ region (top right), and low-SNR region (bottom) using the control set of HCN and HNC transitions.}
    \label{fig:allCorners}
\end{figure*}

\begin{figure*}
    \gridline{
    \leftfig{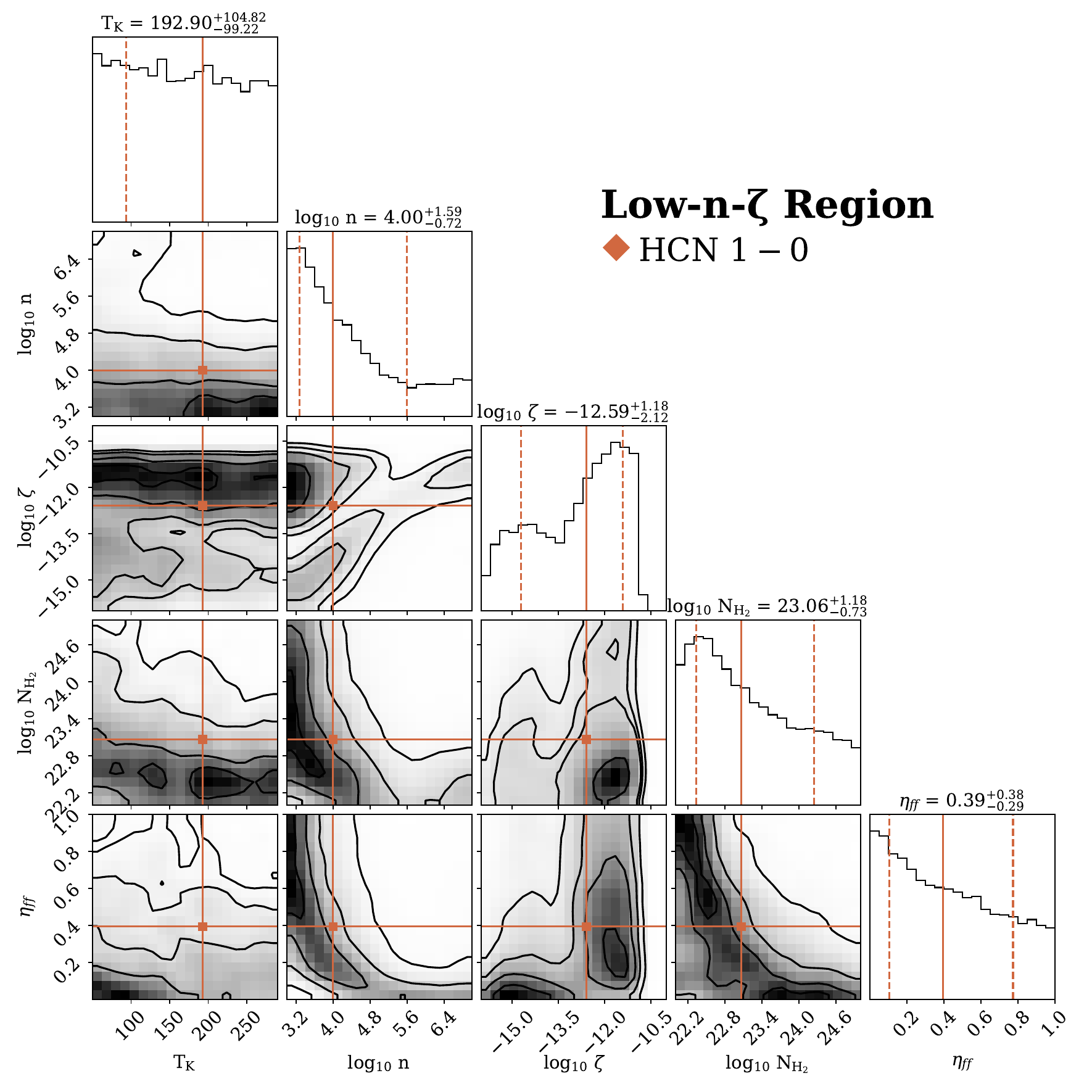}{0.5\textwidth}{(a)}
    \rightfig{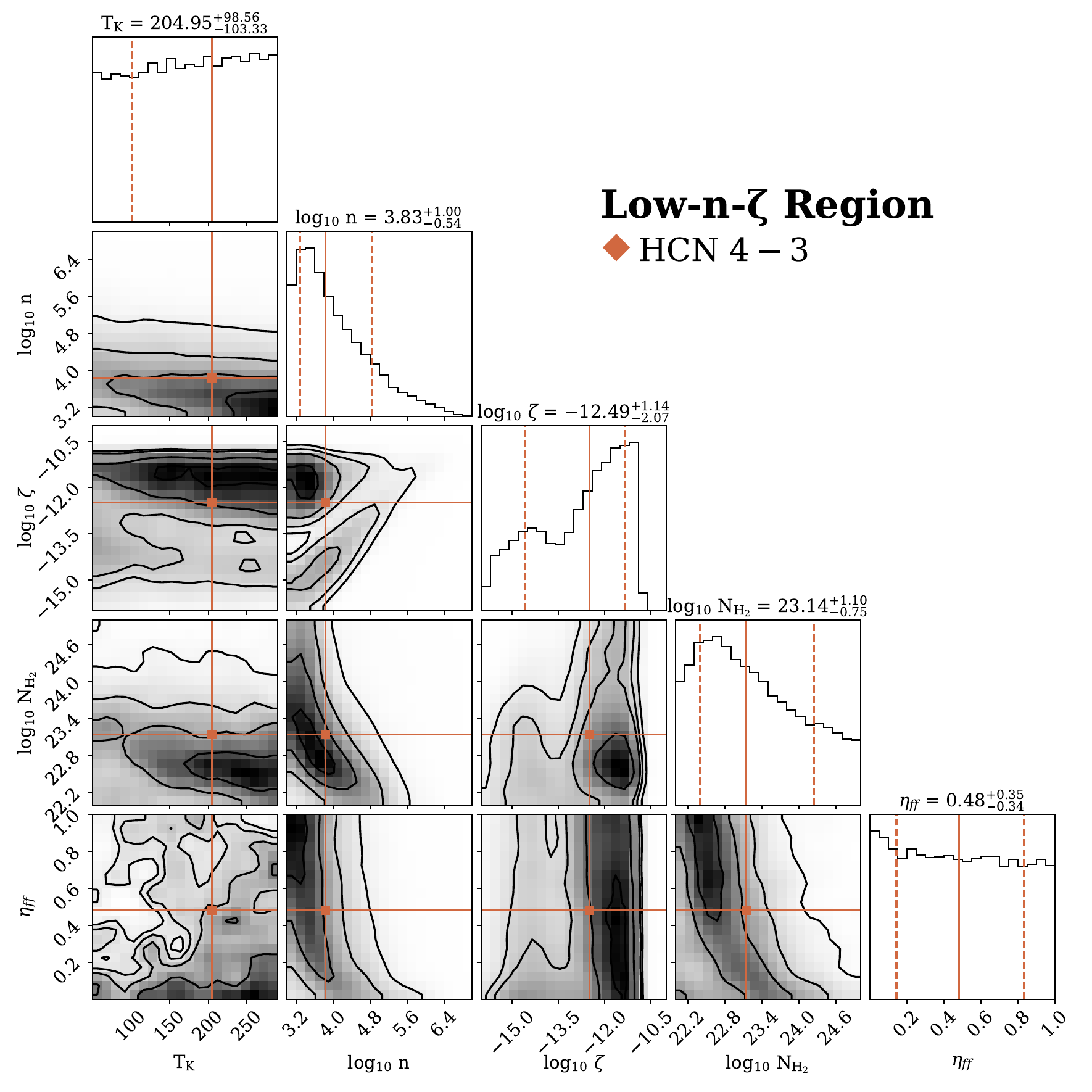}{0.5\textwidth}{(b)}
    }
    \caption{Corner plots for two examples where we constrain our models with single transitions in the low-$n$-$\zeta$ region: \emph{(a)} HCN $1=0$ and \emph{(b)} HCN $4-3$.}
    \label{fig:singleTcorners}
\end{figure*}

\begin{figure*}
    \gridline{
    \leftfig{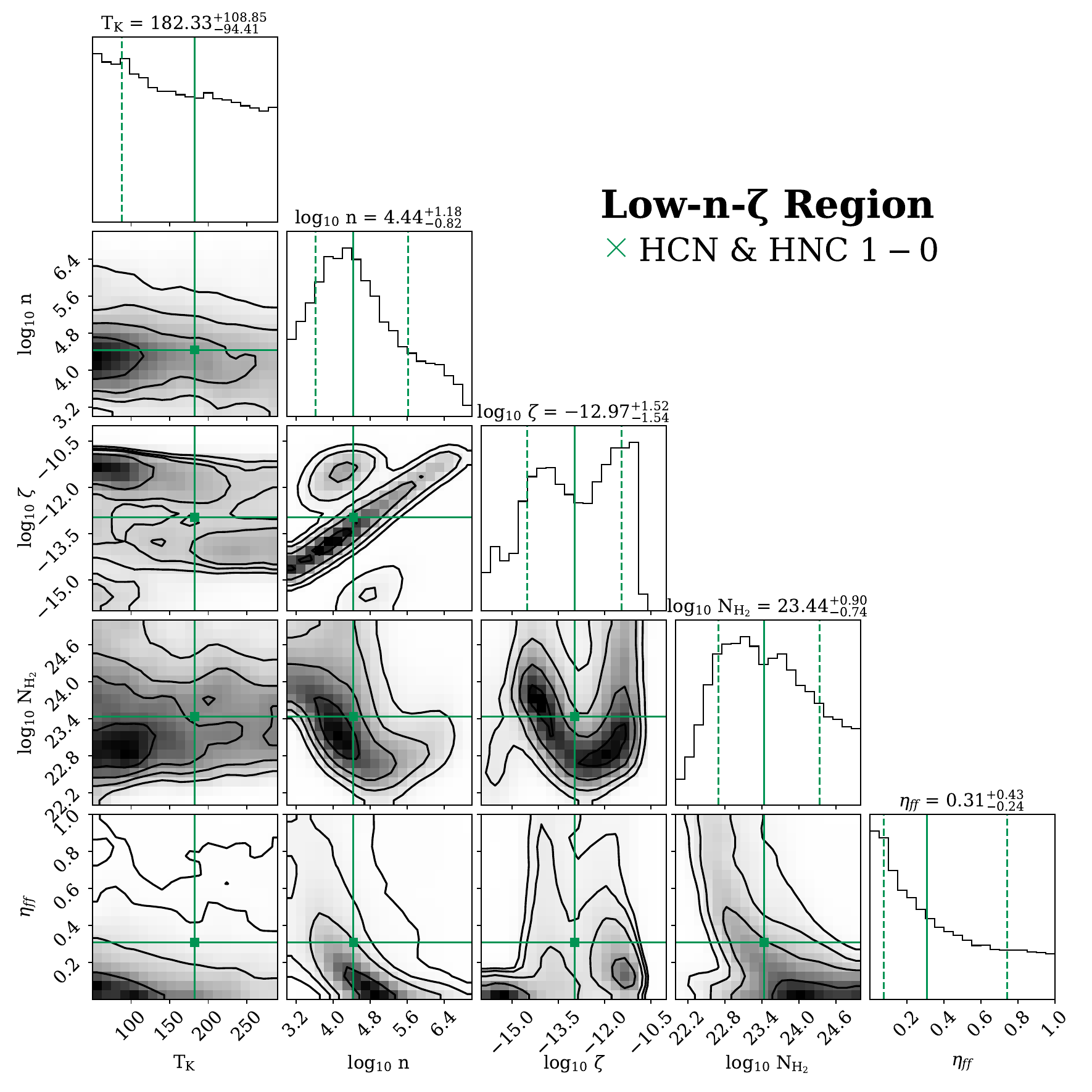}{0.5\textwidth}{(a)}
    \rightfig{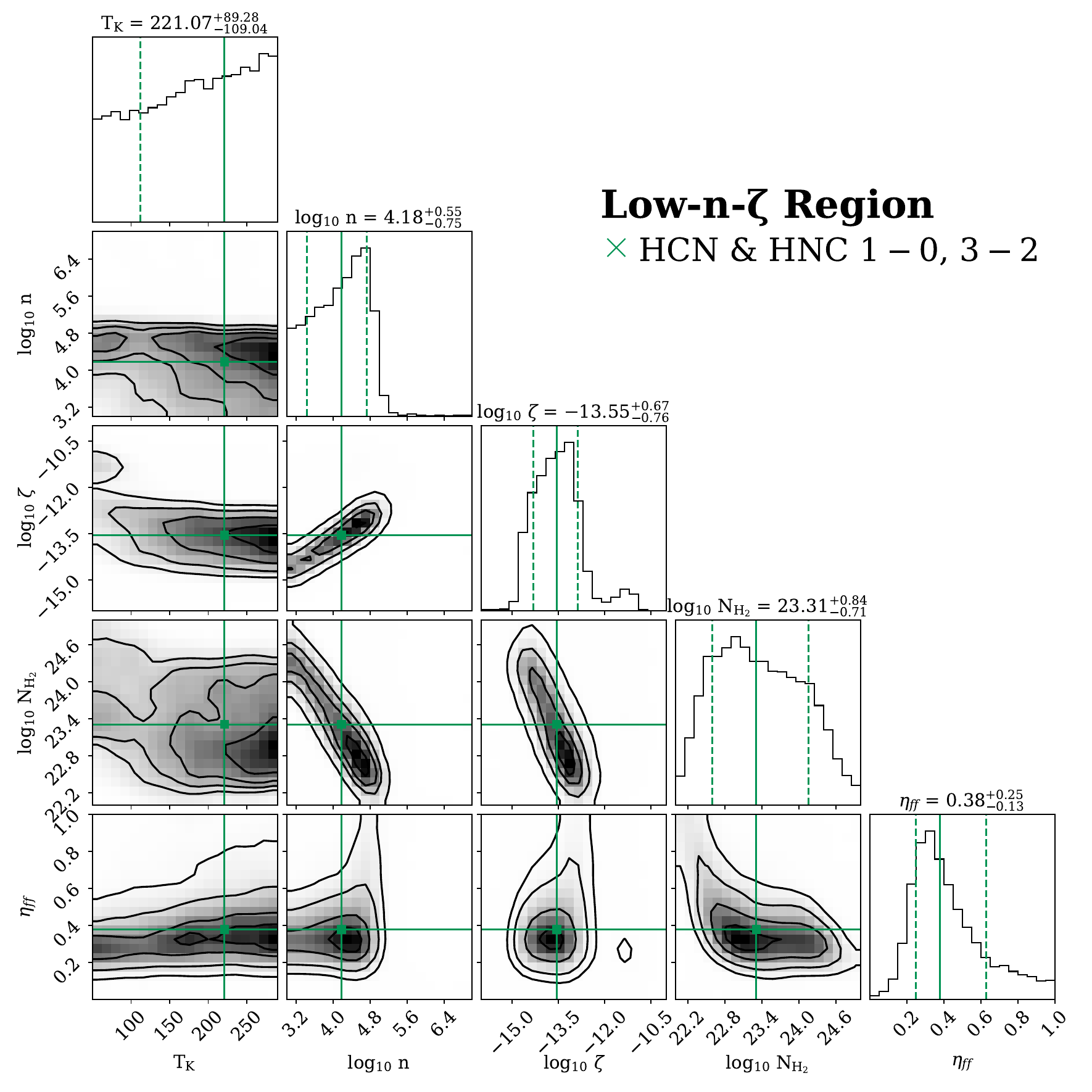}{0.5\textwidth}{(b)}
    }
    \caption{Same as in Figure~\ref{fig:singleTcorners} but for two examples of matched HCN and HNC transitions: \emph{(a)}: HCN and HNC $1-0$ and \emph{(b)} HCN and HNC $1-0$ and $3-2$.}
    \label{fig:matchedCorners}
\end{figure*}

\begin{figure*}
    \gridline{
    \leftfig{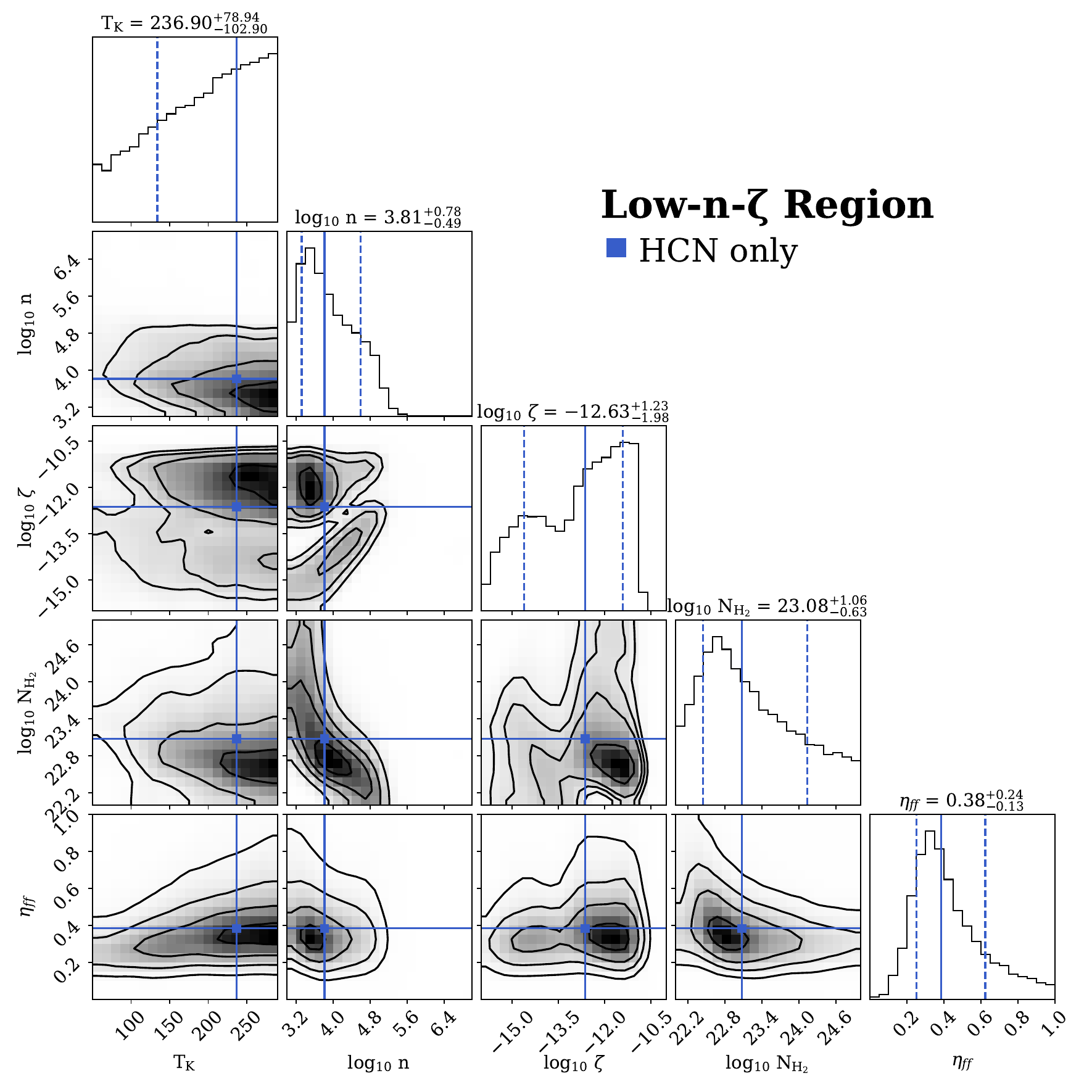}{0.5\textwidth}{(a)}
    \rightfig{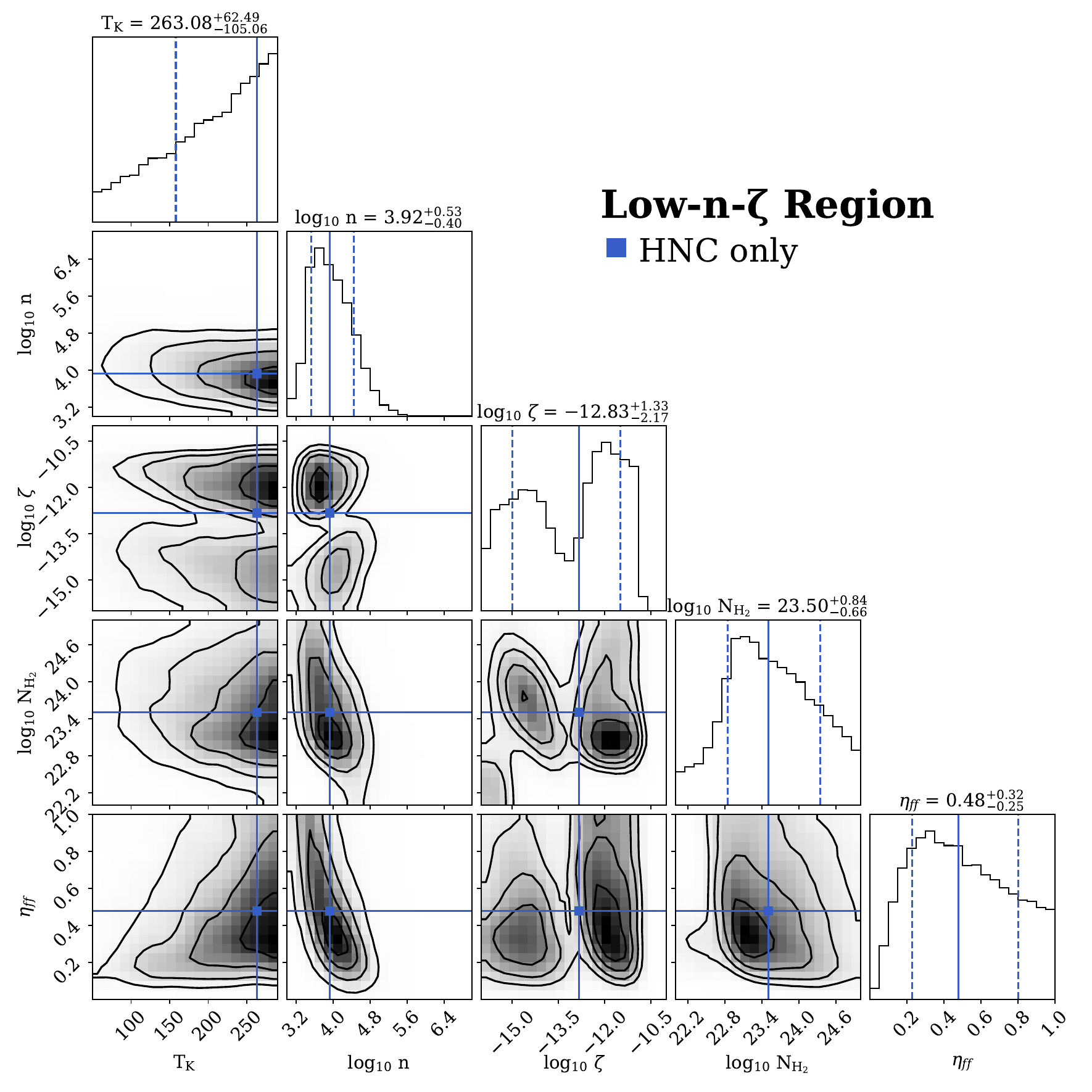}{0.5\textwidth}{(b)}
    }
    \caption{Same as in Figures~\ref{fig:singleTcorners} and \ref{fig:matchedCorners} but for two examples of combinations where we use all the transitions of a single molecule: \emph{(a)}: HCN only and \emph{(b)} HNC only.}
    \label{fig:singleMolCorners}
\end{figure*}

\begin{figure*}
    \gridline{
    \leftfig{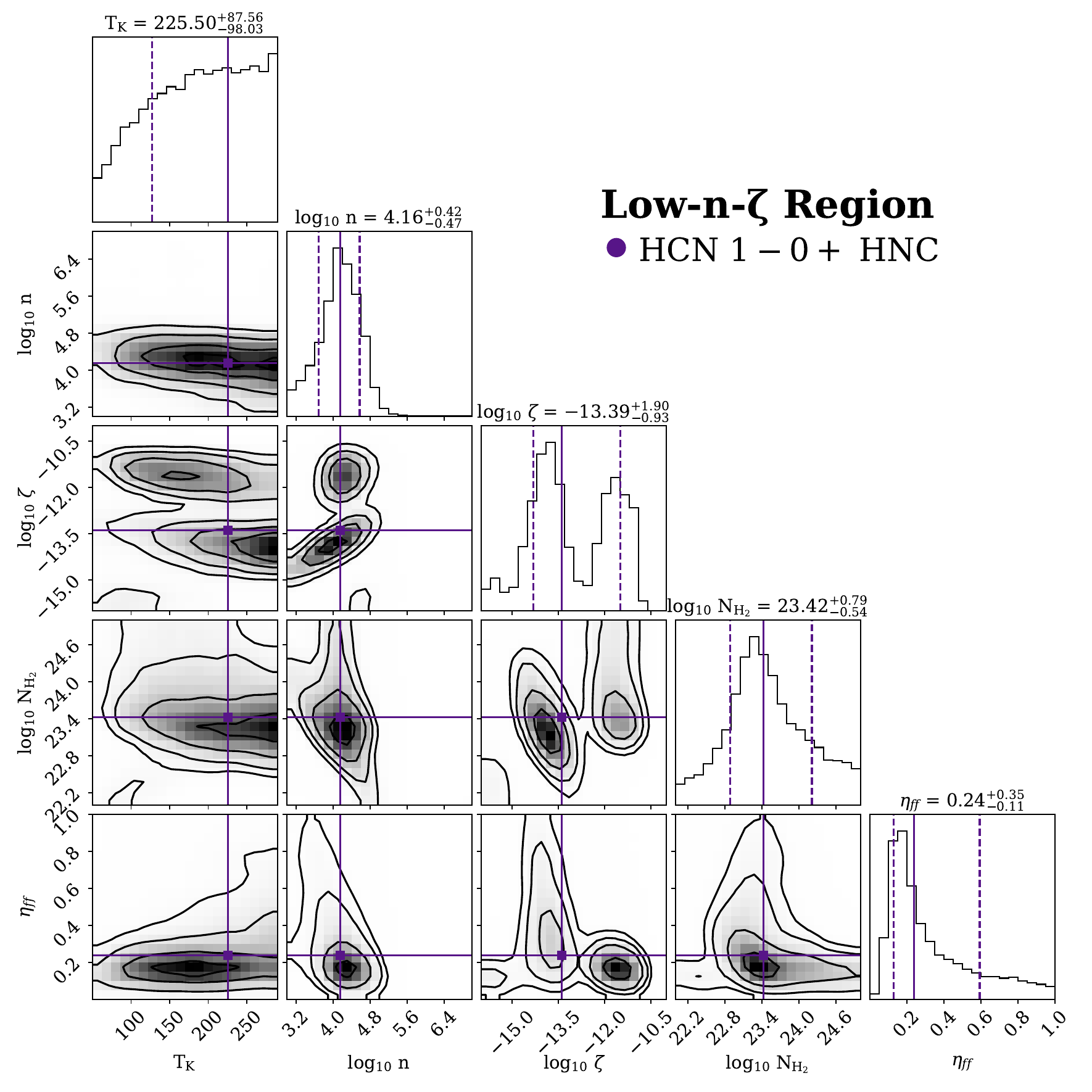}{0.5\textwidth}{(a)}
    \rightfig{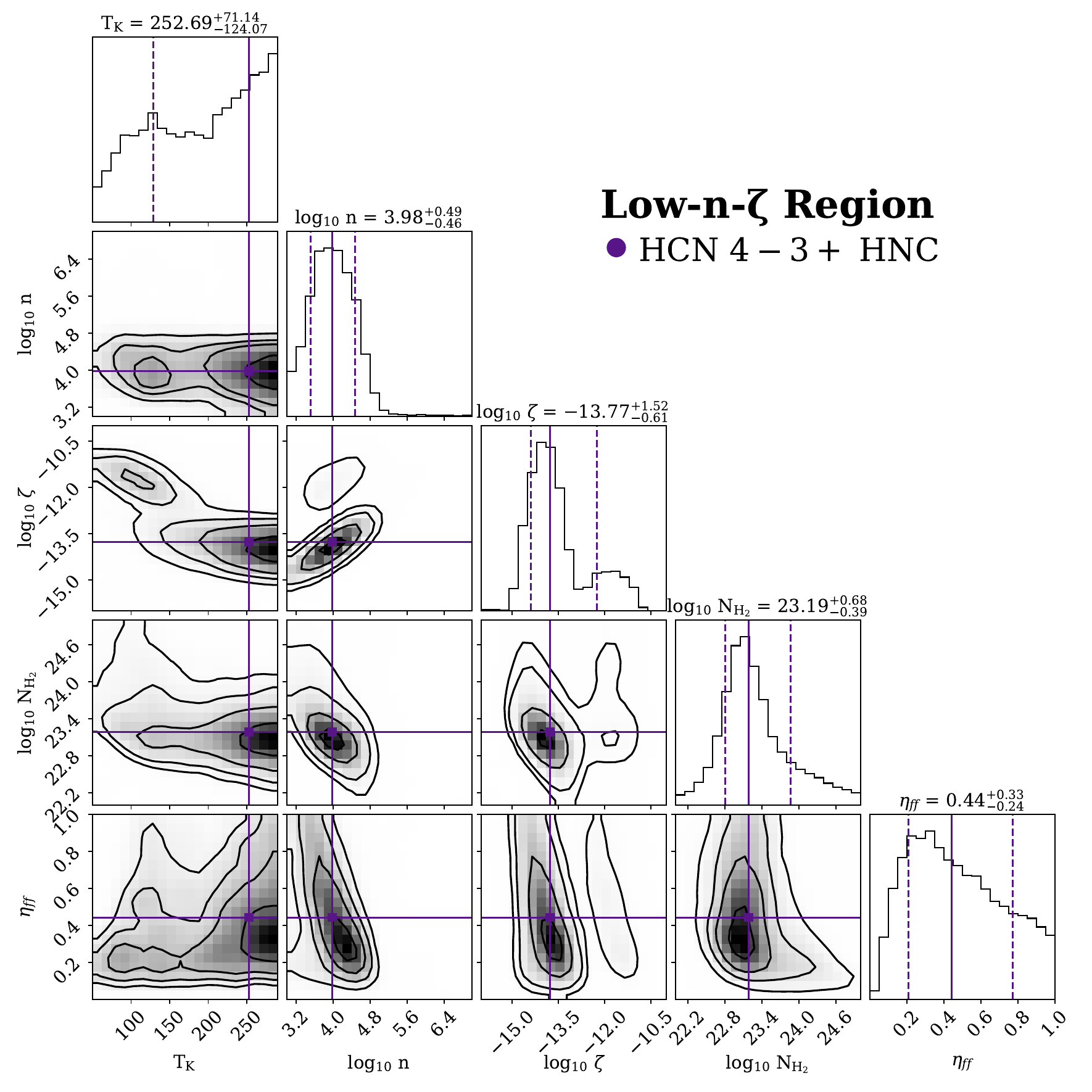}{0.5\textwidth}{(b)}
    }
    \caption{Same as in Figures~\ref{fig:singleTcorners}, \ref{fig:matchedCorners}, and \ref{fig:singleMolCorners} but for two examples of combinations where we use all the transitions of one molecule supplemented with a single transition of the opposing molecule: \emph{(a)}: HCN $1-0$ + all HNC \emph{(b)} HCN $4-3$ + all HNC.}
    \label{fig:allMol+1_corners}
\end{figure*}

\begin{figure*}
    \gridline{
    \leftfig{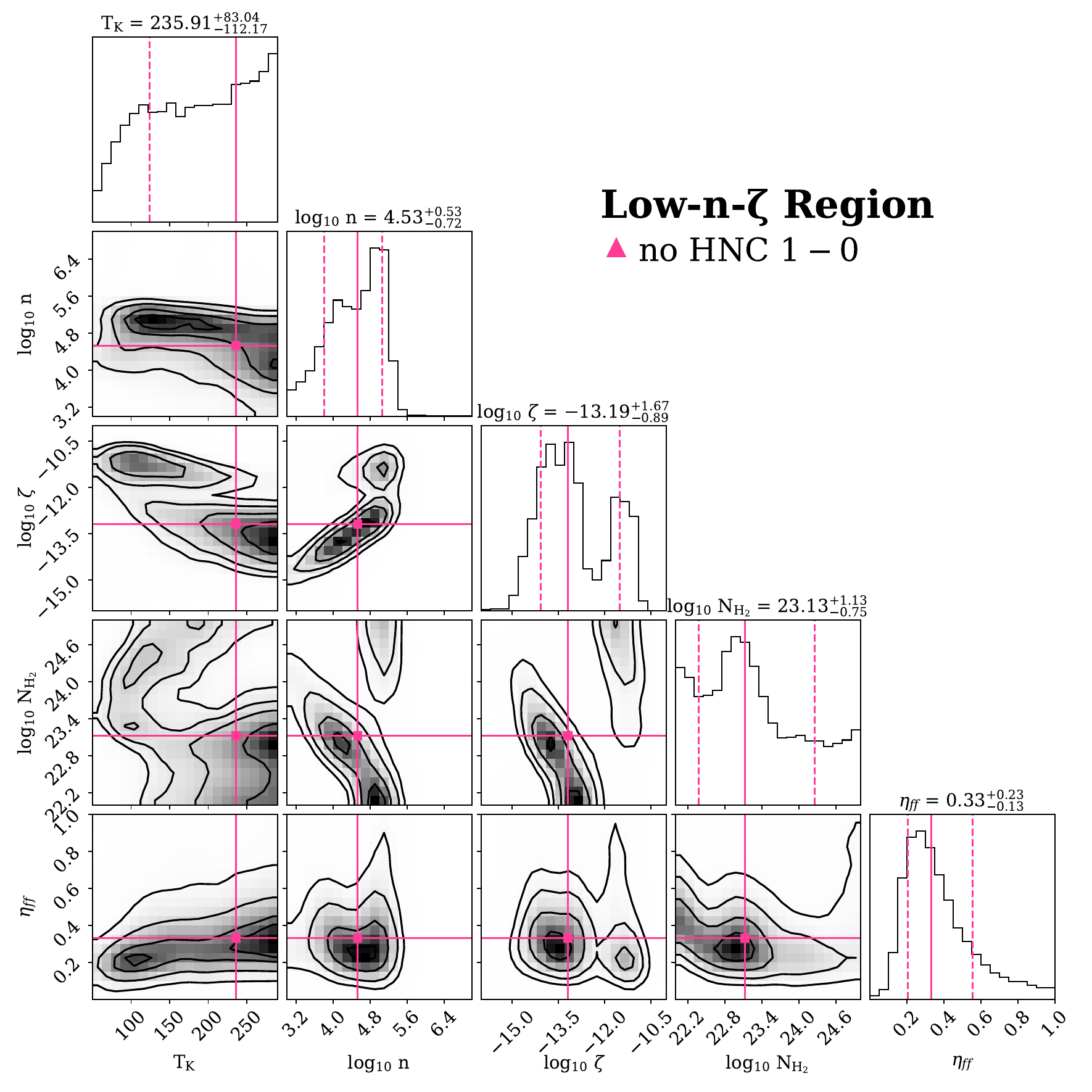}{0.5\textwidth}{(a)}
    \rightfig{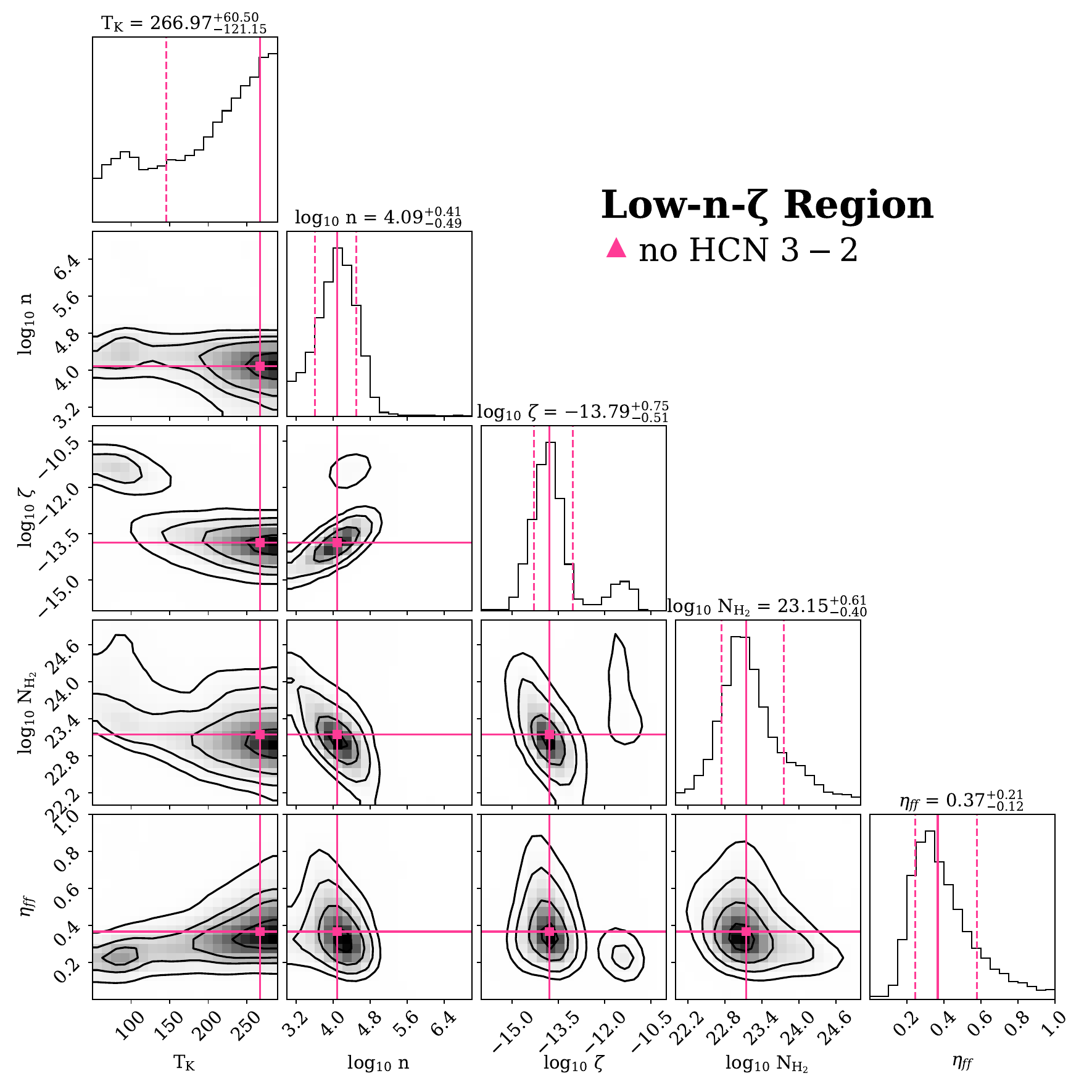}{0.5\textwidth}{(b)}
    }
    \caption{Same as in Figures~\ref{fig:singleTcorners}, \ref{fig:matchedCorners}, \ref{fig:singleMolCorners}, and \ref{fig:allMol+1_corners} but for two examples of combinations where we remove a single transition from the set of 8: \emph{(a)}: no HNC $1-0$  \emph{(b)} no HCN $3-2$.}
    \label{fig:1TremovedCorners}
\end{figure*}

\begin{figure*}
\centering 
    \gridline{
    \leftfig{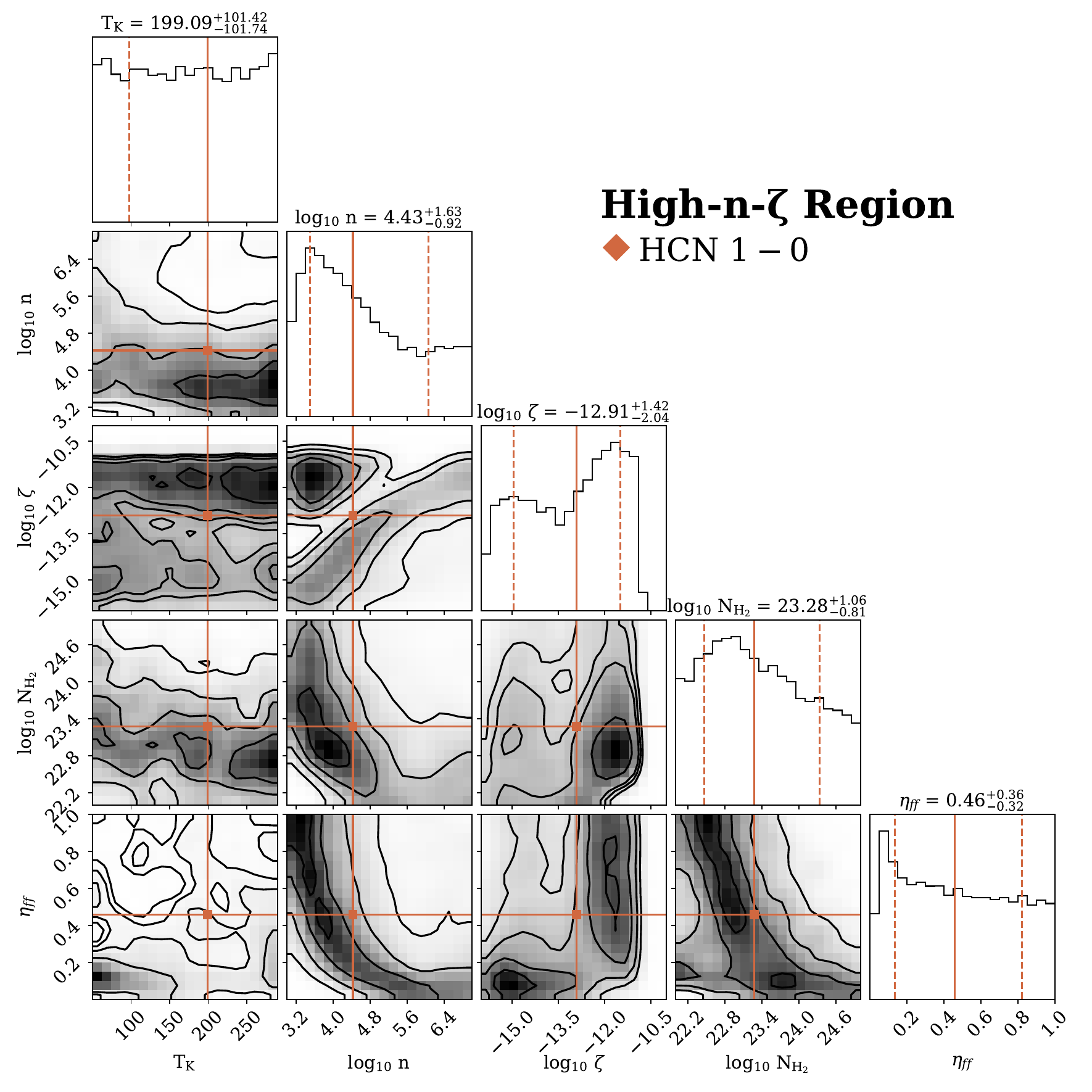}{0.5\textwidth}{}
    \rightfig{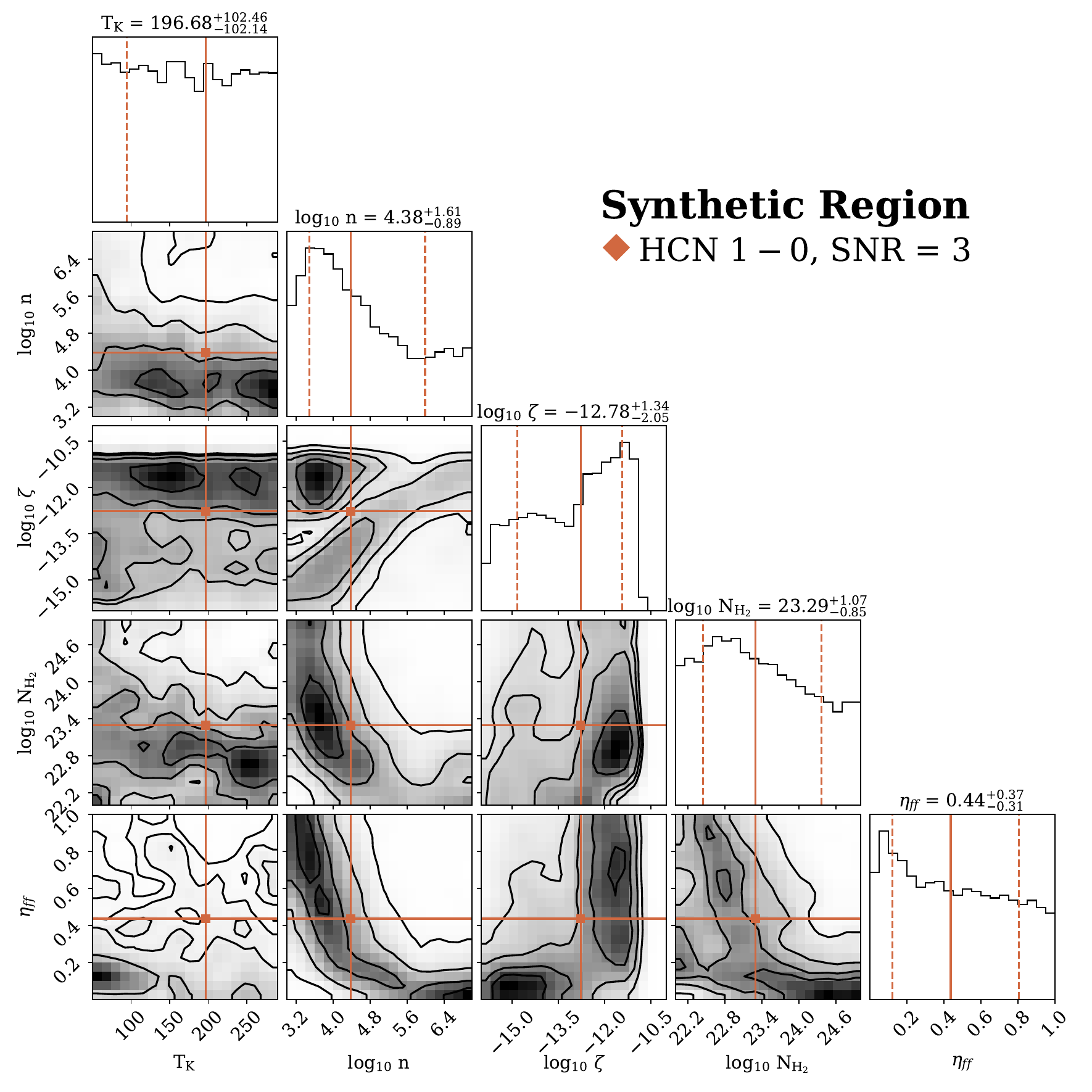}{0.5\textwidth}{}
    }
    \gridline{
    \fig{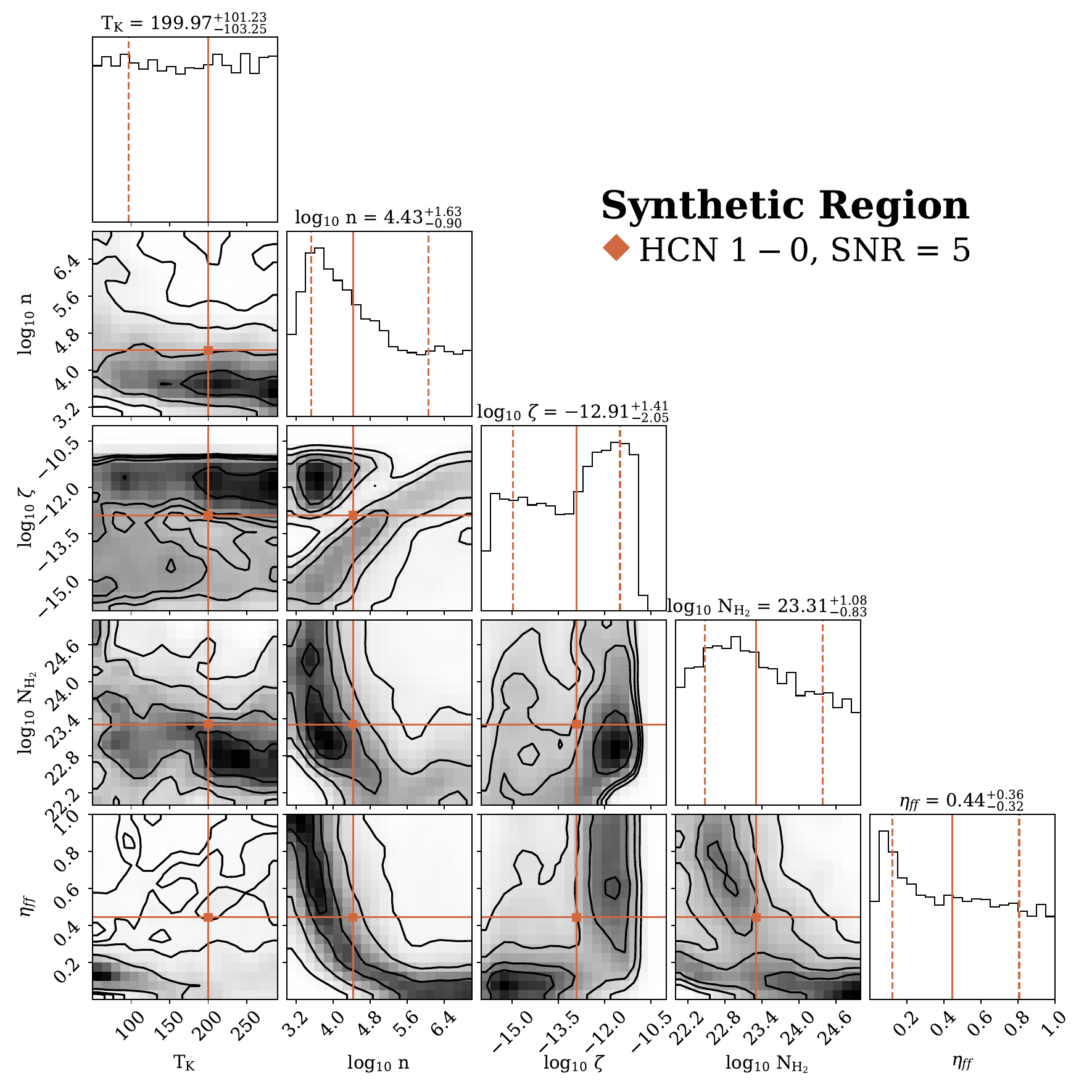}{0.5\textwidth}{}
    }
    \vspace{-8mm}
    \caption{Bayesian inference results for the models constrained solely by HCN $1-0$ for the high-$n$-$\zeta$ (top left) and synthetic regions with SNR = 3 (top right) and SNR = 5 (bottom).}
    \label{fig:synthHCN10}
\end{figure*}

\begin{figure*}
\centering 
    \gridline{
    \leftfig{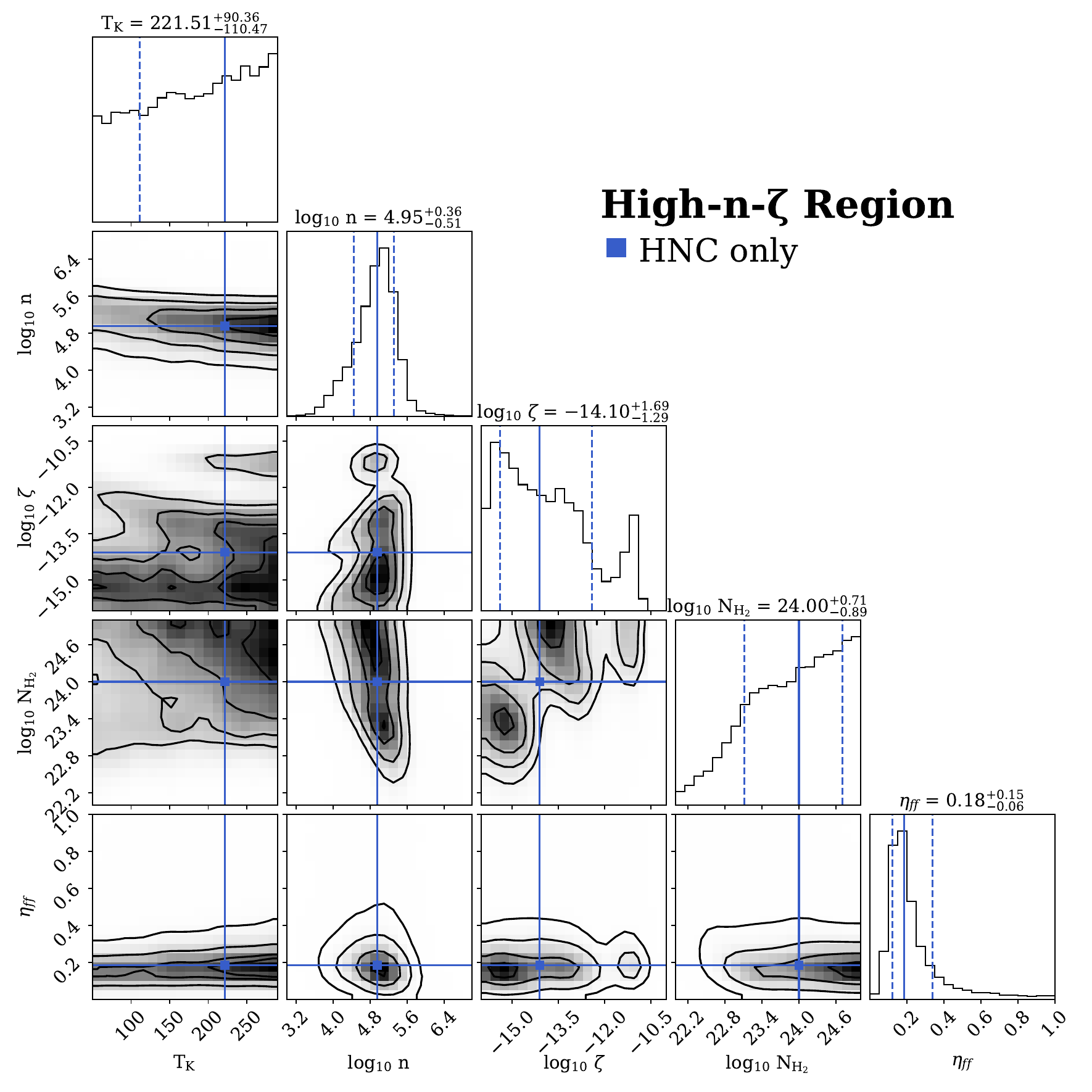}{0.5\textwidth}{}
    \rightfig{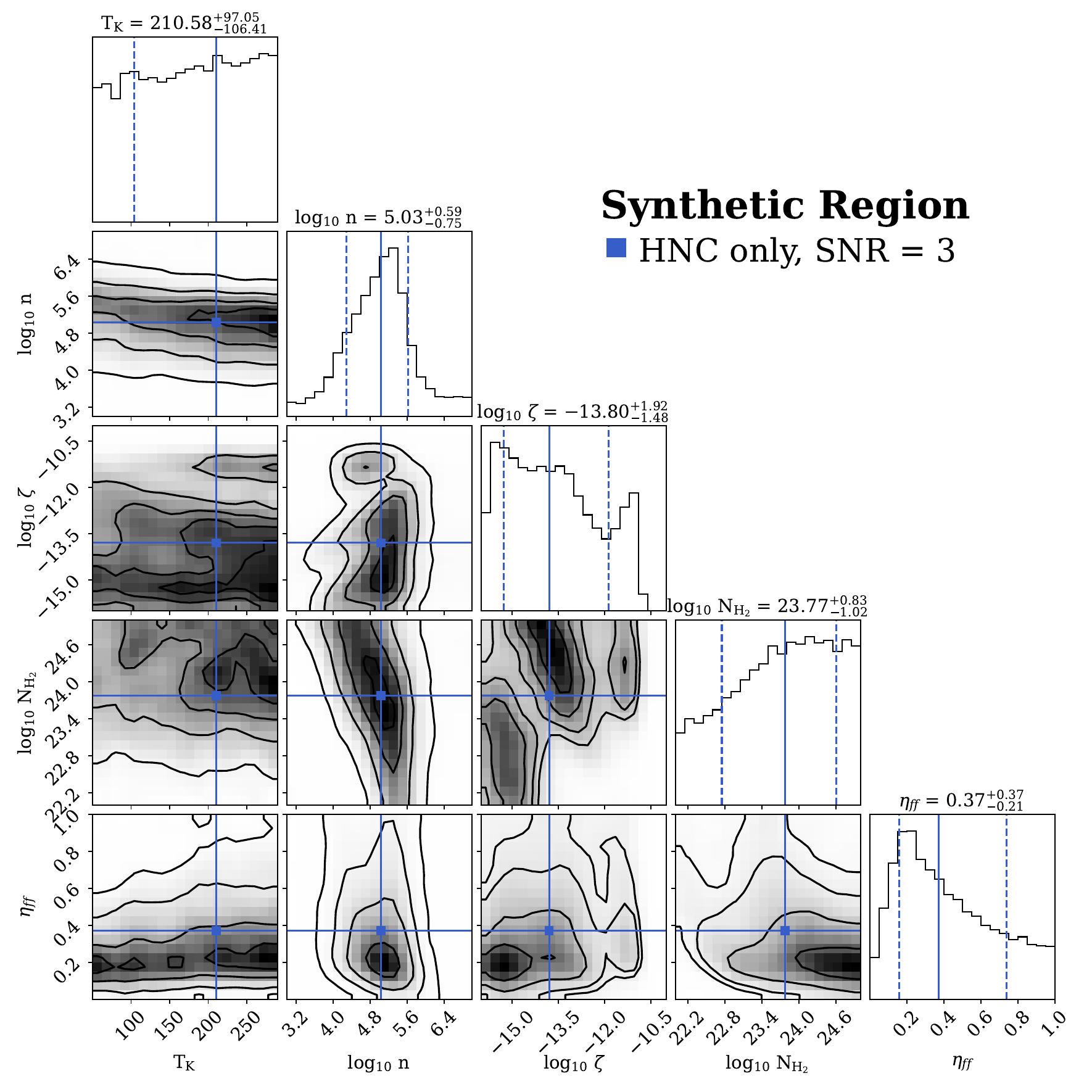}{0.5\textwidth}{}
    }
    \gridline{
    \fig{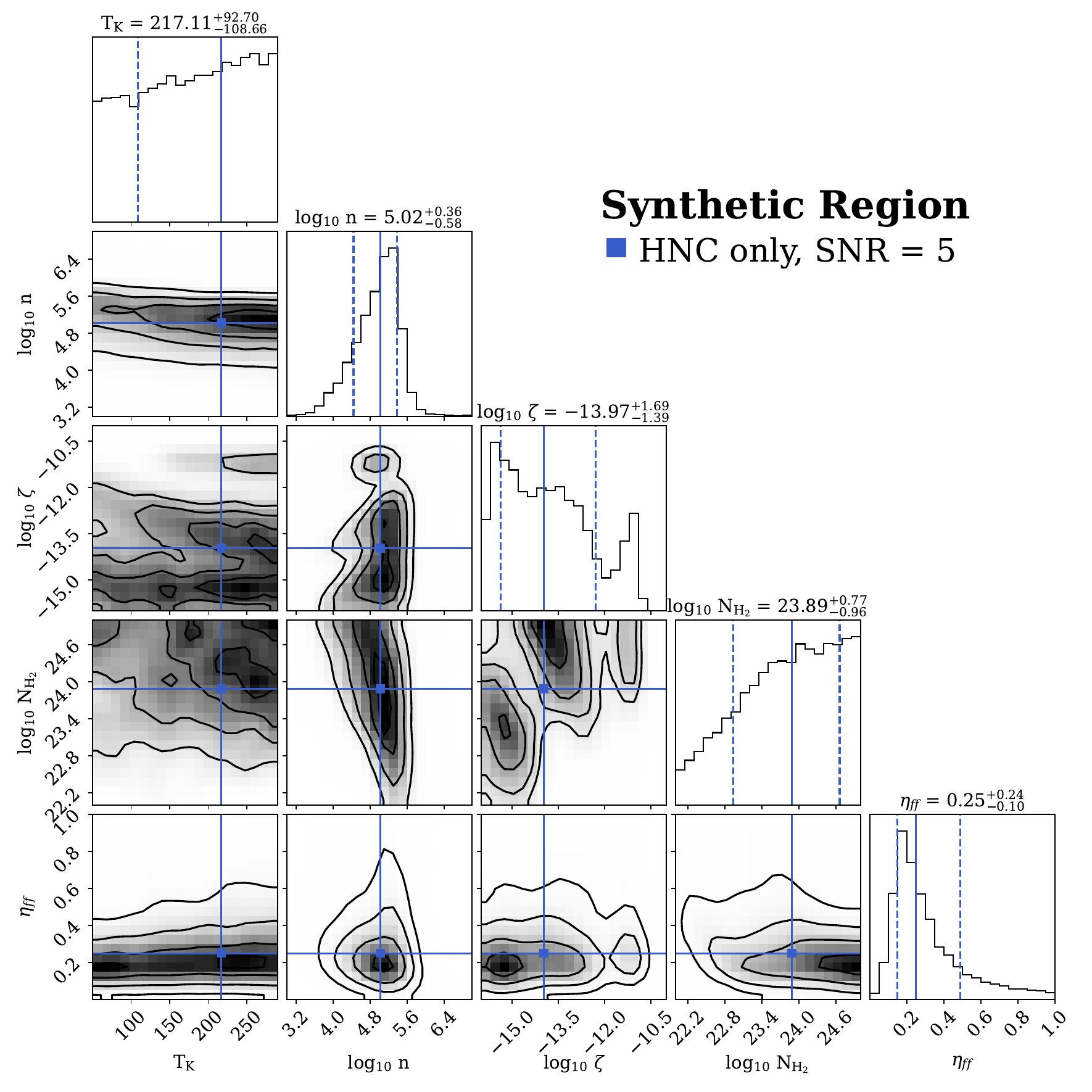}{0.5\textwidth}{}
    }
    \vspace{-8mm}
    \caption{Bayesian inference results for the models constrained solely by HNC transitions for the high-$n$-$\zeta$ (top left) and synthetic regions for SNR = 3 (top right) and SNR = 5 (bottom).}
    \label{fig:synthHNConly}
\end{figure*}

\begin{figure*}
    \centering
    \includegraphics[width=0.95\textwidth]{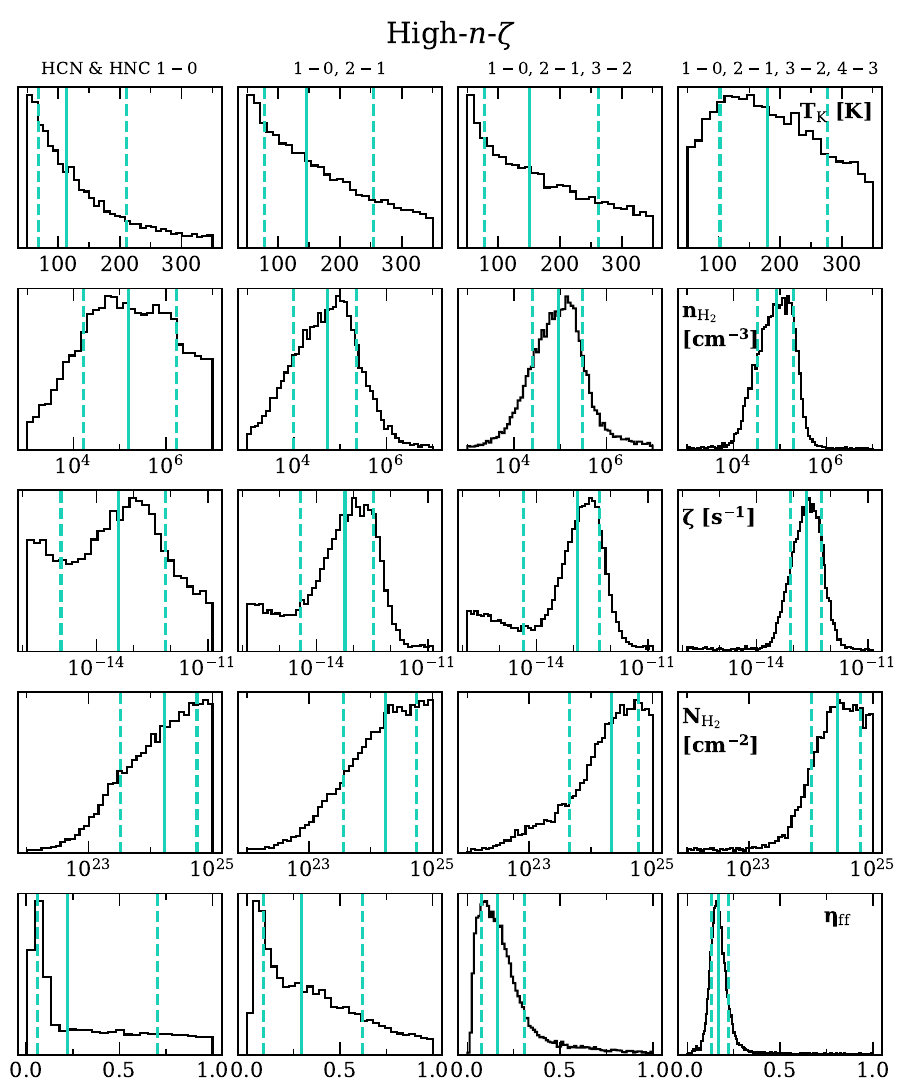}
    \caption{Same as in Figure \ref{fig:reg26_1d} but for the high-$n$-$\zeta$ region.}
    \label{fig:reg53_1d}
\end{figure*}

\begin{figure*}
    \centering
    \includegraphics[width=0.95\textwidth]{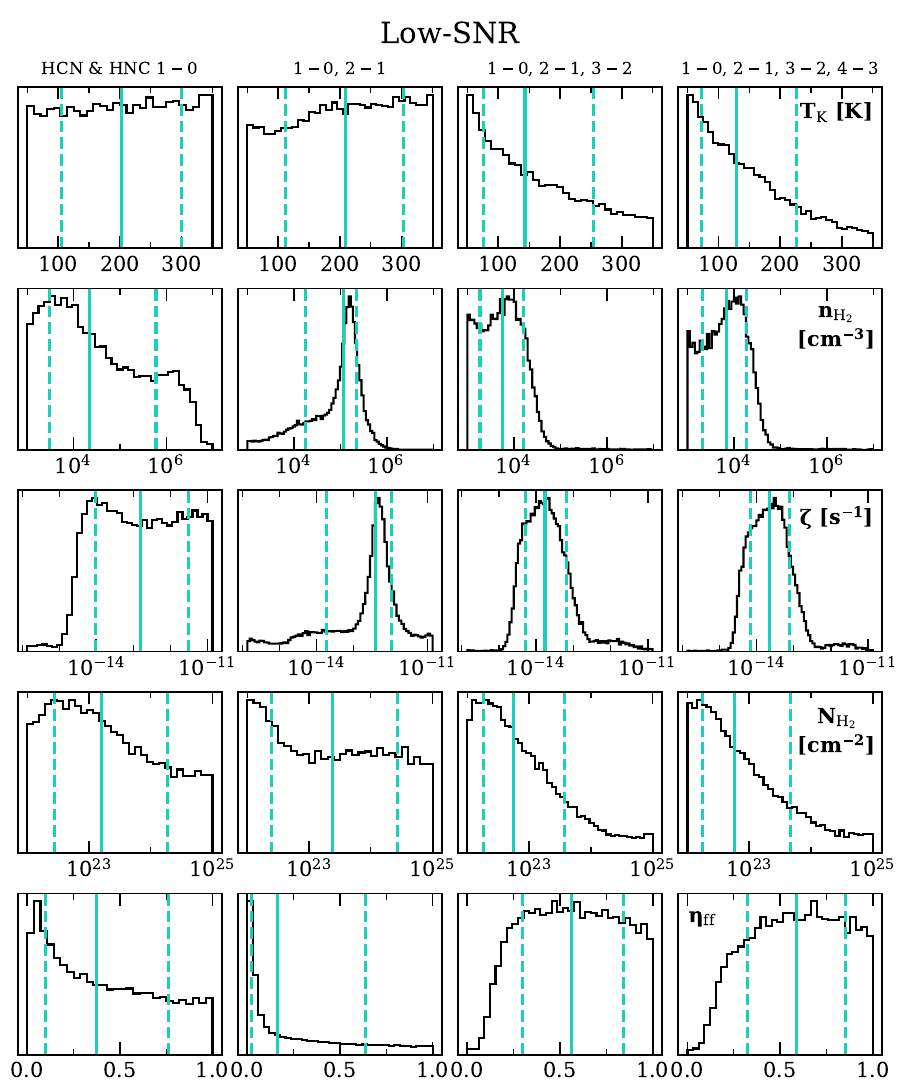}
    \caption{Same as in Figure \ref{fig:reg26_1d} and \ref{fig:reg53_1d} but for the low-SNR region.}
    \label{fig:reg51_1d}
\end{figure*}

\begin{figure*}
    \centering
    \includegraphics[width=0.8\textwidth]{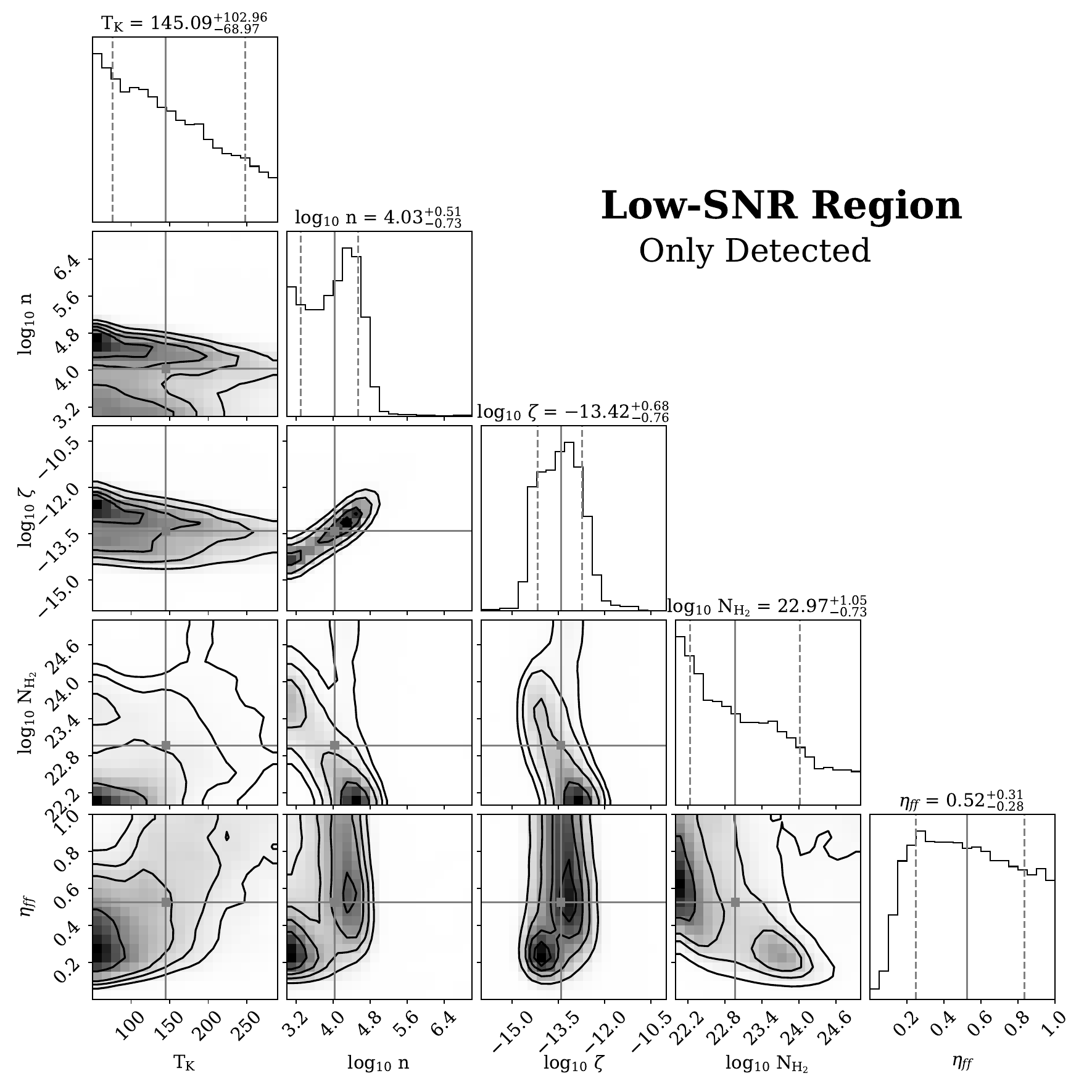}
    \caption{Inference results for low-SNR region where models were constrained with only the 5 detected transitions (no upper limits). These results are very similar to those seen in the low-SNR panel of Figure~\ref{fig:allCorners}, indicating that the 3$\sigma$ upper limits do not have a large impact on the parameter inference results.}
    \label{fig:reg51_onlyDet}
\end{figure*}



\end{CJK*}
\end{document}